\documentclass[12pt]{iopart}
\makeatletter
\renewcommand\@appendixstar{\@@par
 \ifnumbysec 
 \@addtoreset{table}{section}
 \@addtoreset{figure}{section}\fi
 \setcounter{section}{0}
 \setcounter{subsection}{0}
 \setcounter{subsubsection}{0}
 \setcounter{equation}{0}
 \setcounter{figure}{0}
 \setcounter{table}{0}
 \def\thesection{\Alph{section}} % this line has been \def\thesection{Appendix \Alph{section}} before
 \def\theequation{\ifnumbysec
      \Alph{section}.\arabic{equation}\else
      \Alph{section}\arabic{equation}\fi}
 \def\thetable{\ifnumbysec
      \Alph{section}\arabic{table}\else
      A\arabic{table}\fi}
 \def\thefigure{\ifnumbysec
      \Alph{section}\arabic{figure}\else
      A\arabic{figure}\fi}}
\makeatother

\usepackage{iopams} 
\expandafter\let\csname equation*\endcsname\relax

\expandafter\let\csname endequation*\endcsname\relax

\usepackage{amsmath} 
\usepackage{graphicx}% Include figure files
\usepackage{dcolumn}% Align table columns on decimal point
\usepackage{bm}% bold math
\usepackage[markings, customcolors]{hf-tikz}
\usepackage{circuitikz}
\usetikzlibrary{shapes}
\usepackage{caption}
\usepackage{subcaption}
\usepackage{booktabs}
\usepackage{xcolor}
\usepackage{fancybox}
\usepackage{hyperref}
\usepackage{listings}
\usepackage{braket}

\definecolor{DarkGreen}{RGB}{0,75,0}

\tikzset{cross/.style={cross out, draw=black, thick, minimum size=2*(#1-\pgflinewidth), inner sep=0pt, outer sep=0pt},
cross/.default={5 pt}}

\tikzset{lattice/.pic={
\draw [thick, color=black!60!blue] (-4, 0)--(0, -4);
\draw [thick, color=black!60!blue] (0, -4)--(4, 0);
\draw [thick, color=black!60!blue] (4, 0)--(0, 4);
\draw [thick, color=black!60!blue] (0, 4)--(-4, 0);
\draw [thick, color=black!60!blue] (-3,1)--(1, -3);
\draw [thick, color=black!60!blue] (-2,2)--(2, -2);
\draw [thick, color=black!60!blue] (-1,3)--(3, -1);
\draw [thick, color=black!60!blue] (-3,-1)--(1, 3);
\draw [thick, color=black!60!blue] (-2,-2)--(2, 2);
\draw [thick, color=black!60!blue] (-1,-3)--(3, 1);
\filldraw [color=black!80!blue]   (0, 4) circle (0.3cm);
\filldraw [color=black!80!blue]   (0, -4) circle (0.3cm);
\foreach \x in {0,...,4}
\filldraw  [color=black!80!blue] (-4+\x*2,0) circle (0.3cm);
\foreach \x in {0,...,3}
\filldraw  [color=black!80!blue] (-3+\x*2,1) circle (0.3cm);
\foreach \x in {0,...,3}
\filldraw  [color=black!80!blue] (-3+\x*2,-1) circle (0.3cm);
\foreach \x in {0,...,2}
\filldraw  [color=black!80!blue] (-2+\x*2,2) circle (0.3cm);
\foreach \x in {0,...,2}
\filldraw  [color=black!80!blue] (-2+\x*2,-2) circle (0.3cm);
\foreach \x in {0,...,1}
\filldraw  [color=black!80!blue] (-1+\x*2,3) circle (0.3cm);
\foreach \x in {0,...,1}
\filldraw  [color=black!80!blue] (-1+\x*2,-3) circle (0.3cm);
}
}
\tikzset{quantumCircuit/.pic={
\foreach \x in {0,1}
\foreach \y in {0,1,2}
\draw[thick] (0+2*\x,0-\y*2)--(1+2*\x,0-\y*2);
\foreach \y in {0,1,2}
\draw[thick] (4,0-\y*2)--(4.5,0-\y*2);
\foreach \x in {0,...,1}
\foreach \y in {0,...,2}
\draw[thick] (1+2*\x,-0.5-\y*2) rectangle (2+2*\x,0.5-\y*2);
\node at (1.5,0) {\tiny{$U_{11}$}};
\node at (1.5+2,0) {\tiny{$U_{12}$}};
\node at (1.5,-2) {\tiny{$U_{21}$}};
\node at (1.5+2,-2) {\tiny{$U_{22}$}};
\node at (1.5,-4) {\tiny{$U_{31}$}};
\node at (1.5+2,-4) {\tiny{$U_{32}$}};
}
}
\tikzset{latticeFeynman/.pic={
\foreach \y in {0,...,3}
\draw [ultra thick, color=orange] (0,0-\y*2)--(6, 0-\y*2);
\foreach \x in {0,...,3}
\draw [ultra thick, black!50!blue] (0+2*\x,0)--(0+2*\x, -6);
\foreach \y in {0,1,2}
\draw [ultra thick, black!50!blue] (0,-2-\y*2)--(2,0-\y*2)--(4,-2-\y*2)--(6,0-\y*2);
\foreach \x in {0,...,3}
\foreach \y in {0,...,3}
\filldraw [color=black!80!blue]   (0+\x*2, 0-\y*2) circle (0.3cm);
}
}

\begin{document}

\title{Hamiltonian quantum computing with superconducting qubits}

\author{A. Ciani$^{1,4}$, B. M. Terhal$^{2,3,4}$, D. P. DiVincenzo$^{1,3,4}$}
\address{$^1$ Institute for Quantum Information, RWTH Aachen University,                                D-52056 Aachen,                        
  Germany}
\address{$^2$ QuTech, Delft University of Technology, P.O. Box 5046, 2600 GA
    Delft,
  The Netherlands}
  \address{$^3$ Peter Gr\"{u}nberg Institute, Theoretical Nanoelectronics,
    Forschungszentrum J\"{u}lich,
  D-52425
  J\"{u}lich,
  Germany}
\address{$^4$ J\"{u}lich-Aachen Research Alliance (JARA),
    Fundamentals of Future Information Technologies,
  D-52425
  J\"{u}lich,
  Germany}

%\address{IOP Publishing, Temple Circus, Temple Way, Bristol BS1 6HG, UK}
%\ead{submissions@iop.org}
%\vspace{10pt}
%\begin{indented}
%\item[]August 2017
%\end{indented}

\begin{abstract}
We consider how the Hamiltonian Quantum Computing scheme introduced in [New Journal of Physics, vol. 18, p. 023042, 2016] can be implemented using a 2D array of superconducting transmon qubits. We show how the scheme requires  the engineering of strong attractive cross-Kerr and weak flip-flop or hopping interactions and we detail how this can be achieved. Our proposal uses a new electric circuit for obtaining the attractive cross-Kerr coupling between transmons via a dipole-like element. We discuss and numerically analyze the forward motion and execution of the computation and its dependence on coupling strengths and their variability. We extend [New Journal of Physics, vol. 18, p. 023042, 2016] by explicitly showing how to construct a direct Toffoli gate, thus establishing computational universality via the Hadamard and Toffoli gate or via controlled-Hadamard, Hadamard and CNOT. 
\end{abstract}

\maketitle

\tableofcontents

\section{Introduction}
In a seminal paper of the early days of quantum information, Feynman introduced a model capable of performing an arbitrary quantum computation with a time-independent Hamiltonian \cite{feynman1986}. In this time-independent approach to quantum computing, the system is prepared in an initial state, it evolves under the action of a Hamiltonian for a certain time and is finally measured to extract the result of the computation. In Feynman's model a fundamental role is played by a quantum clock system whose state changes upon the application of gates. While Feynman's approach to quantum computing has not received much attention at the experimental level compared to the circuit model, its importance from a theoretical point of view has long been recognized. In particular, Feynman's model has been used to analyze the QMA-completeness of the $k$-local Hamiltonian problem \cite{kitaevBook}, and to show a formal equivalence between adiabatic and circuit-based quantum computation\cite{ahronovAdiabatic2007, lidarmMizelProof}. 

One of the challenges in a practical implementation of the Feynman Hamiltonian is the presence of the clock system, which requires high-weight and non-local interactions in space. An alternative to the concept of a global clock is a model where each information-carrying particle has its own local clock. This idea of an asynchronous cellular automaton was first formulated by Margolus \cite{Margolus90parallelquantum}, and analyzed in much greater detail in Refs.~\cite{janzing2007, mizel2001, mizel2004, breuckmannTerhal, gossetTerhal, terhalLloyd}. The idea has been formalized under the name `space-time circuit-to-Hamiltonian' construction in Ref.~\cite{breuckmannTerhal}. 
In this model the position of each particle with an internal, information-carrying, state represents its local clock. In order to implement a computation involving multi-qubit gates it is necessary to achieve coordination between the local clocks, \i.e. clock times need to align for particles to interact. The need for alignment requires an attractive interaction between the particles.

There are alternative constructions showing universal Hamiltonian quantum computation in which mobile multiple interacting particles do not require particles to move together. In these constructions particles interact as wavepackets in scattering regions \cite{CGW:walk,childs+:switch}. Ref.~\cite{englund:walk} has considered how to implement such alternative multi-particle walk with an ultra-cold bosonic atom system.

The appeal of a Hamiltonian approach to quantum computing is that it does not require active driving fields to enact logic: information carriers are moving through gates in space instead of time-dependent gates being applied to stationary qubits.  It means that a realization of this approach is much closer to the idea of analog quantum simulation \cite{georgescuNori}, but with the benefit of allowing for universal computation. 
In a 2D lattice realization, information is entered on the side of the lattice while the passive interactions in the bulk of the lattice are engineered to implement a chosen 1D quantum circuit, \i.e. a quantum circuit with nearest-neighbor gates between qubits on a 1D line. 

This computational model also lends itself well to quantum learning since gates executed in regions on the lattice can simply depend on angle parameters which can be altered from one run of the computation to the next. \par 

In this paper we consider the scheme for Hamiltonian quantum computing on a two-dimensional lattice developed by Lloyd and Terhal in Ref.~\cite{terhalLloyd}. The scheme is most easily formulated in terms of particles carrying an internal spin degree of freedom, that can sit on the sites of the lattice. The lattice grid is depicted in Fig.~\ref{basicGrid}. To each particle we associate a horizontal track composed of different sites on which the particle can hop. When hopping takes place a corresponding unitary gate is applied to the internal spin degree of freedom. The coordination of the motion is achieved by means of strong attractive interactions between particles on adjacent tracks. These attractive terms are associated with the edges of the lattice depicted in Fig. \ref{basicGrid}. As shown in \cite{terhalLloyd} the problem can be mapped to qubits using a dual rail encoding. In this mapping, the necessary interactions to be engineered between the qubits are strong interactions at the edges which induce excitations to move together and weaker hopping or flip-flop interactions across the plaquettes.  

In this paper, we show how the scheme of Lloyd and Terhal can form the basis of a very concrete architecture using a planar array of transmon qubits. We show how the scheme requires strong attractive cross-Kerr and weak hopping interactions and how one could go about putting these together. Before going into details, we summarize the overall structure and features of our proposed architecture in the next Section \ref{sec:overview}. 

%Universal computation in this model can be achieved using the Hadamard gate, the CNOT gate as well as the controlled-Hadamard or the Toffoli gate, but at no extra cost the model allows for real single-qubit rotations $U(\theta)$ and controlled-$U(\theta)$ gates. In addition, we extend the Lloyd-Terhal scheme by explicitly discussing how to construct a Toffoli gate and in general any controlled-controlled unitary between neighbouring particles. 

%Besides computational universality, the model  interesting physics questions, see the discussion in Section \ref{transmon:noise} on the coherent propagation of the computational wavefront and the emergence of classical time which can be studied by realizing the model.
%BMT added a paragraph above

\begin{figure}
\centering
\begin{tikzpicture}[scale=0.4]
\pic [scale=0.4] at (0,0) {lattice};
\draw [ultra thick, color=red!70!black] (0,4)--(-4,0)--(0,-4);
\foreach \x in {0,..., 4}
\filldraw [color=red!70!black]   (0-\x*1, 4-\x*1) circle (0.3cm);
\foreach \x in {0,..., 4}
\filldraw [color=red!70!black]   (-4+\x*1, 0-\x*1) circle (0.3cm);
\end{tikzpicture}
\caption{Example of a small rotated grid where particles can hop over $N_{\mathrm{track}}=9$ horizontal tracks. Red dots denote the sites that are occupied by the particles: shown is an initial configuration.}
\label{basicGrid}
\end{figure}
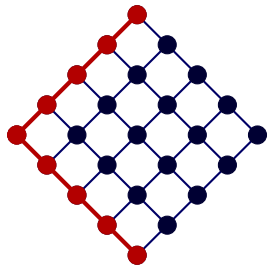 
\begin{figure}
\centering
\includegraphics[scale=0.06]{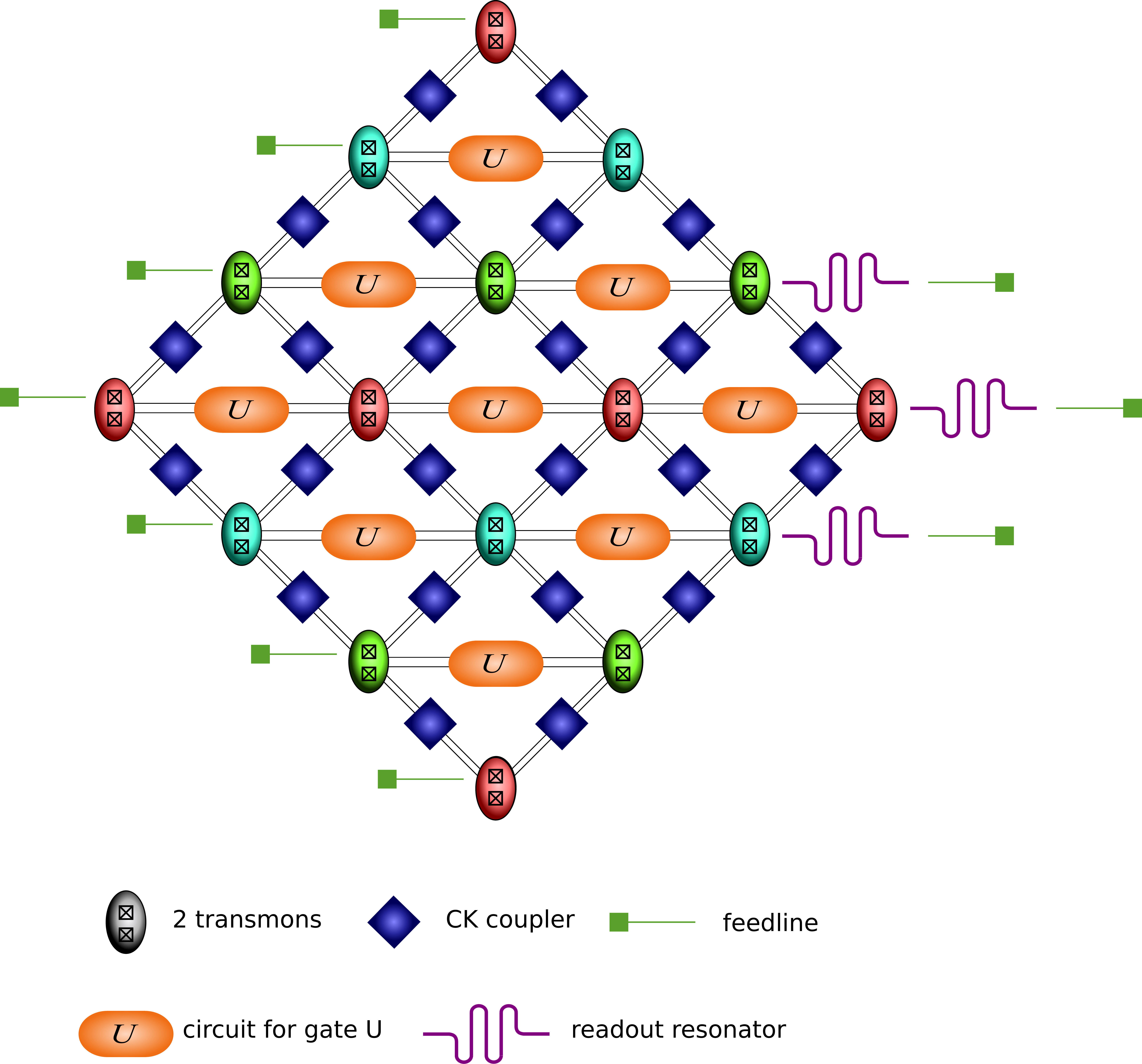}
\caption{Global layout concept for the Hamiltonian quantum computing scheme with superconducting transmon qubits. A layout with $N_{\rm track}=7$ horizontal tracks is shown. The microwave feedlines on the left prepare excitations in a subset of transmons on the left (one excitation per transmon pair) to set an initial bitstring of the computation. The bulk of the grid is used to realize gates such as $U=I$, $U=H$ (Fig.~\ref{circuitGates}) or $U={\rm CNOT}$, controlled-Hadamard or a Toffoli gate. If a CNOT or controlled-Hadamard is to be executed, the couplers in a region of the lattice are modified as shown in Fig.~\ref{fig:CNOT-CKerr}. The blue interactions represent doubled cross-Kerr (CK) interactions, see Fig.~\ref{strongZZcoup}. }.\label{layoutConcept}
\end{figure}

\begin{figure}[htb]
\centering
\begin{subfigure}[t]{0.5\textwidth}
\includegraphics[scale=0.15]{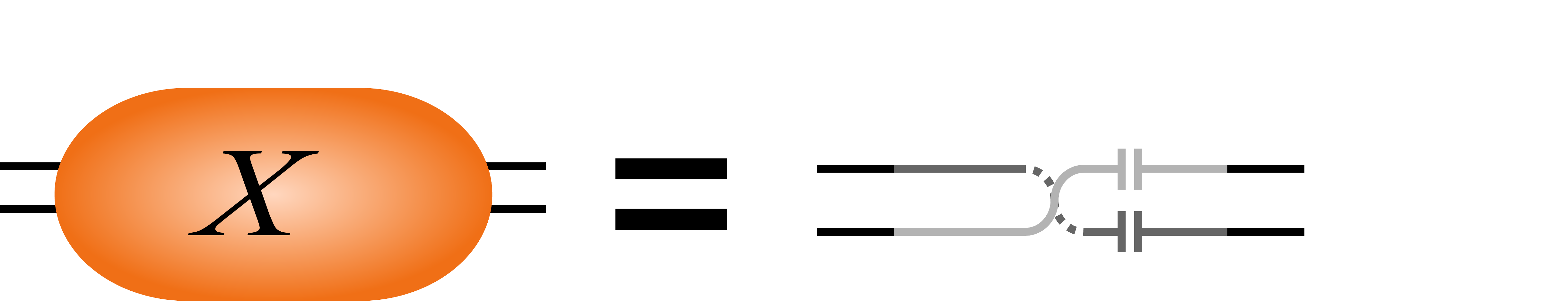}
\caption{X gate circuit.}
\end{subfigure}
\begin{subfigure}[t]{0.5\textwidth}
\includegraphics[scale=0.15]{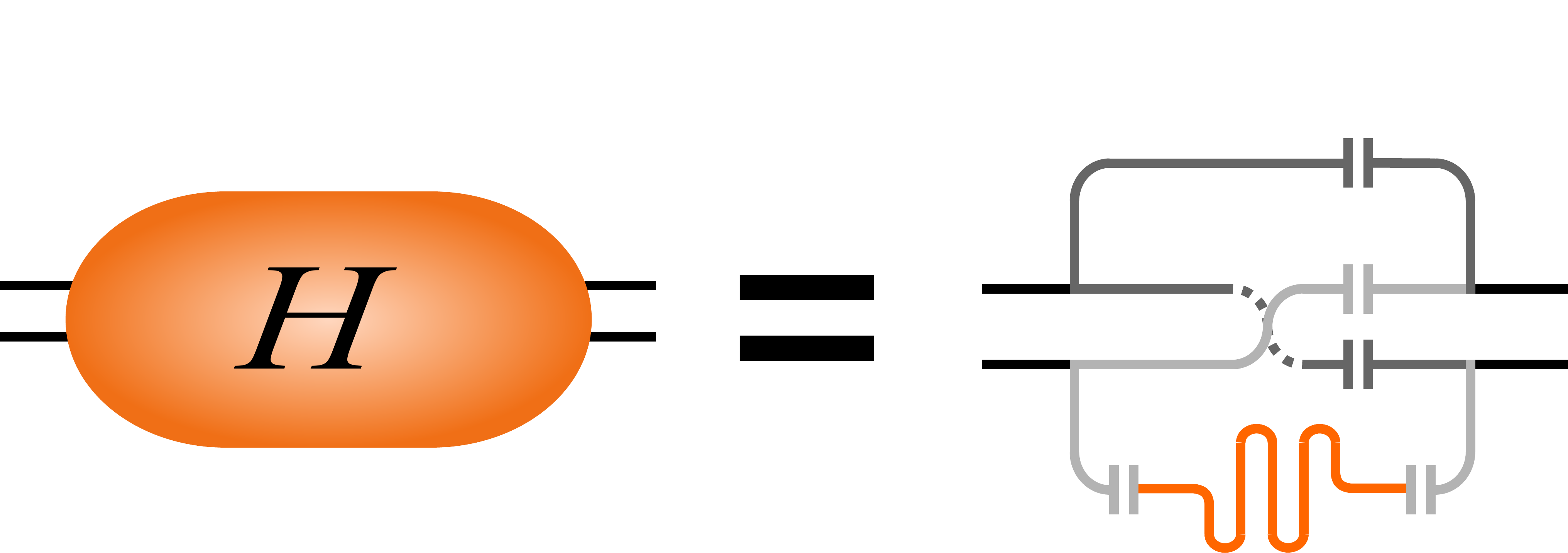}
\caption{Hadamard gate circuit.}
\label{Hgate}
\end{subfigure}
\caption{Examples of quantum electric circuits for implementing single-qubit gates, constructed via direct or resonator-mediated capacitive couplers. }
\label{circuitGates}
\end{figure}

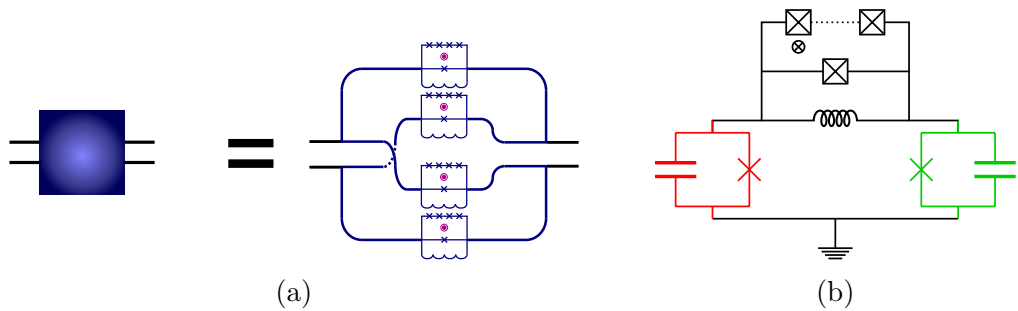
\begin{figure}[htb]
\centering
\begin{subfigure}[t]{0.5 \textwidth}
\centering
\includegraphics[scale=0.17]{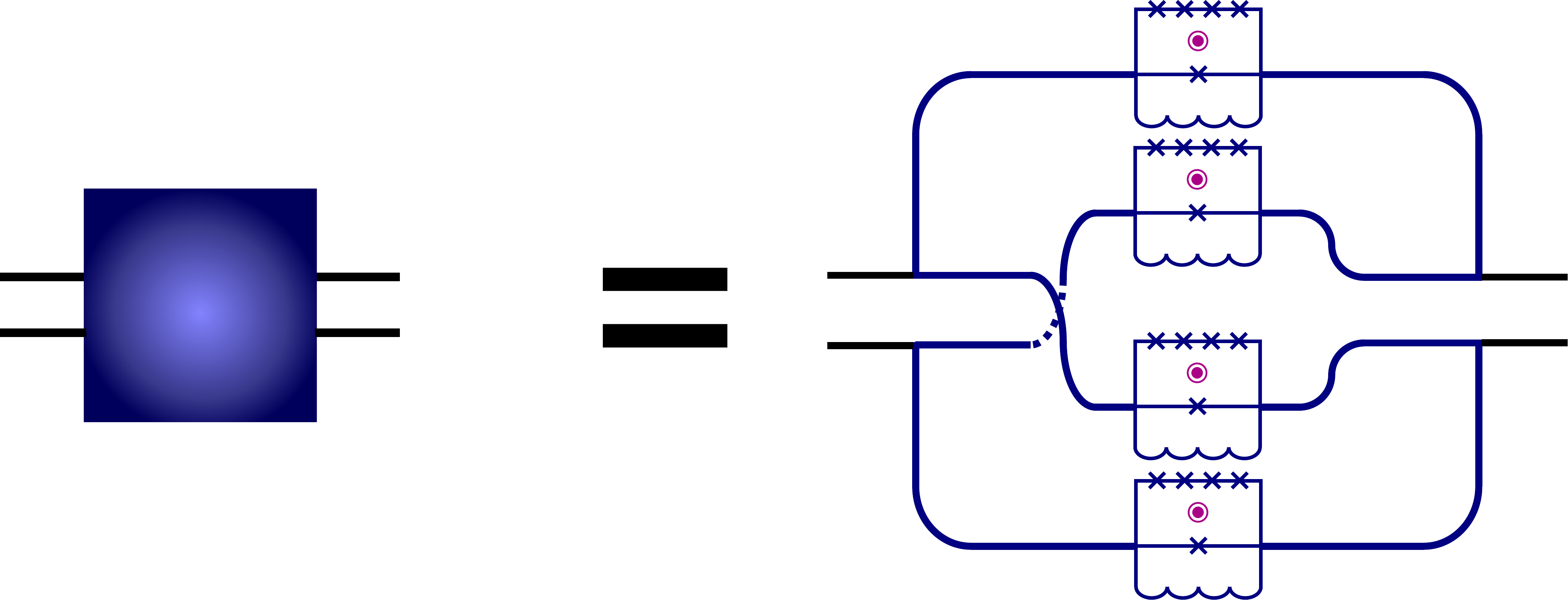}
\subcaption{}
\end{subfigure}
\begin{subfigure}[t]{0.4 \textwidth}
\centering
\resizebox{0.8\textwidth}{!}{
\begin{circuitikz}[scale=0.5, american voltages]
\draw[line width=1.0] (0,0)
to [short] (-5,0);
\draw[line width=1.0, color=red] (-5,0)
to [short] (-5, 0.5)
to [short] (-6.5, 0.5)
to [C] (-6.5, 3.5)
to [short] (-3.5, 3.5)
to [short] (-3.5, 0.5)
to [short] (-5, 0.5);
\draw[line width=3.0, black] (-3.5, 2) node[cross=7.5,color=red,rotate=0, line width=1.0] {};
\draw[line width=1.0] (0,0)
to [short] (5,0);
\draw[line width=1.0, color=green!80!black] (5,0)
to [short] (5, 0.5)
to [short] (6.5, 0.5)
to [C] (6.5, 3.5)
to [short] (3.5, 3.5)
to [short] (3.5, 0.5)
to [short] (5, 0.5);
\draw[line width=3.0, black] (3.5, 2) node[cross=7.5,color=green!80!black,rotate=0, line width=1.0] {};
%array
\draw [line width=1.0] (-5,3.5)--(-5, 4)--(-3, 4)--(-3, 8)--(-2, 8);
\draw [line width=1.0, color=red] (-5,3.5)--(-5, 4);
\draw [line width=1.0] (-2, 7.5) rectangle (-1, 8.5);
\draw[line width=3.0, black] (-1.5, 8) node[cross=7.5,color=black,rotate=0, line width=1.0] {};
\draw [line width=1.0] (5,3.5)--(5, 4)--(3, 4)--(3, 8)--(2, 8);
\draw [line width=1.0, color=green!80!black] (5,3.5)--(5, 4);
\draw [line width=1.0] (2, 7.5) rectangle (1, 8.5);
\draw[line width=3.0, black] (1.5, 8) node[cross=7.5,color=black,rotate=0, line width=1.0] {};
\draw [line width=1.0, dotted] (-1, 8)--(1,8);
%junction
\draw [line width=1.0] (-3, 6)--(-0.5, 6);
\draw [line width=1.0] (-0.5, 5.5) rectangle (0.5, 6.5);
\draw[line width=3.0, black] (0, 6) node[cross=7.5,color=black,rotate=0, line width=1.0] {};
\draw [line width=1.0] (3, 6)--(0.5, 6);
%inductance
\draw [line width=1.0, color=black] (-3, 4) 
to [short] (-2, 4)
to [L, line width=0.6, color=black] (2,4)
to [short] (3,4);
\draw[line width=1.0] (0,0)
to (0,-0.5) node [ground]{};
\draw[line width=1.0] (-1.5, 7) circle (7pt);
\draw[line width=3.0, black] (-1.5, 7) node[cross=3.5,color=black,rotate=0, line width=1.0] {};
\end{circuitikz}
}
\subcaption{}
\label{fig::CKcirc}
\end{subfigure}
\caption{(a) Schematic circuit for realizing the doubled {\em attractive} cross-Kerr interactions between two sets of transmon pairs, one pair at each site. The two input (output) lines denote one pair. The cross-Kerr interaction is doubled in the sense that it is a sum of four cross-Kerr interactions between the pairs.  (b) Coupling element that realizes the cross-Kerr interaction between two transmons on different tracks. The coupler is represented in more detail in Fig.~\ref{capFig}, and analyzed in Sec. \ref{transmonImpl}. }
\label{strongZZcoup}
\end{figure}

\begin{figure}[htb]
\centering
\includegraphics[scale=0.11]{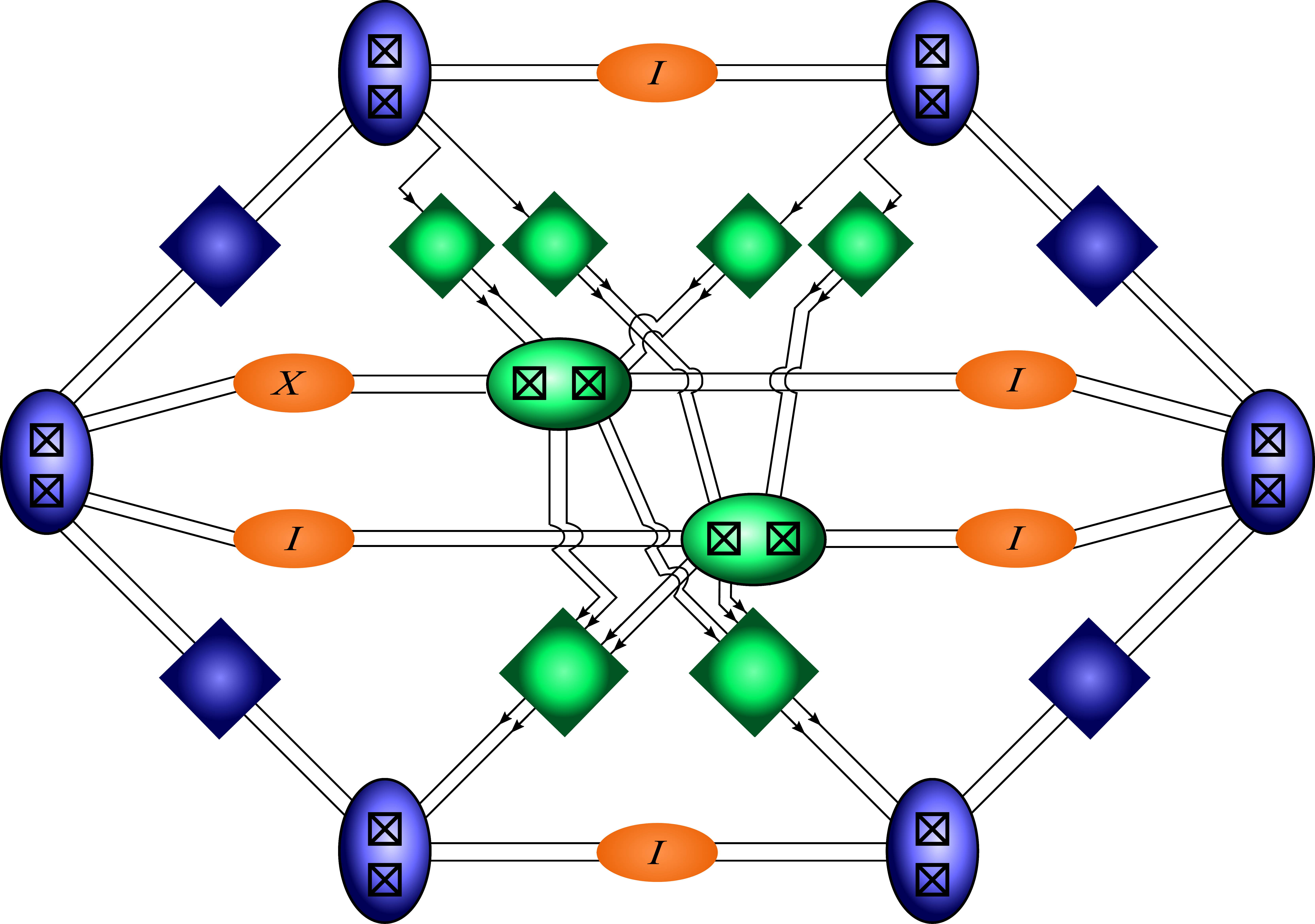}
\caption{Coupling scheme for a CNOT region: all green edges in Fig.~\ref{cnotFig} are represented by green cross-Kerr couplers. The figure represents the two plaquettes that need to be modified in order to implement a CNOT, as detailed in Subsec. \ref{sec:mql}. We have a total of $16$ transmons and a total of $6$ elements applying unitaries like in Fig. \ref{circuitGates}. The green cross-Kerr couplers have a variable number of input and output lines: between {\em each} input and output line there should exist an attractive cross-Kerr interaction formed by the electric circuit in Fig.~\ref{fig::CKcirc}. This dense coupling, in particular between the transmons on the bottom track and the middle track, makes the CNOT region the most complicated element of our proposal.}
\label{fig:CNOT-CKerr}
\end{figure}

\subsection{Overview of Transmon Qubit Implementation Proposal}
\label{sec:overview}

Consider the following layout sketch shown in Fig.~\ref{layoutConcept}. On each site of the lattice we have a pair of transmon qubits \cite{koch2007}:
% the ground state according to our convention is |11>, which is a consequence of how sigma_z is defined in quantum info vs quantum optics.
the ground-state $\ket{00}$ denotes the absence of a qubit information carrier, while the single-excitation states $\ket{01}$ and $\ket{10}$ represent the qubit. This means that at a certain site only one of the two transmons is in the excited state. Furthermore, on each horizontal track there is at most a single transmon excitation carrying the qubit information, all other transmon qubits are in the $\ket{0}$ state. From here onwards the word track refers to horizontal tracks.

%While it is relatively easy to design a circuit-QED element for an interaction by itself, it is generally difficult to generate all interactions together without creating spurious terms. 
We consider a system of grounded transmon qubits, in which each interaction is implemented by a coupling element, which might be direct or indirect, in a modular way. In addition, as a design principle, we try to design a system that is as passive as possible, meaning that we do not use active pulses or parametric drives to engineer the interactions. However, we will allow the presence of constant flux biases. 

The scheme does not require any active control apart from initialization and readout and works in the following way. The feedline on the left of Fig.~\ref{layoutConcept} put transmon excitations on the left sites representing some computational state $\ket{x}$ with $x$ representing some bitstring. The system evolves under the action of an engineered Hamiltonian that implements a quantum circuit. After a certain time the qubits on the right side of the lattice are measured via a standard dispersive readout. This read-out is indicated by the read-out resonator + feedlines on the right in Fig.~\ref{layoutConcept}.
We discuss the need for measuring qubits in a larger measurement {\em region} and the idea of employing a MERA-like trap region in Section\ref{sec:timing-issues}.

The two transmon qubits which make up a pair should be identical in their dressed frequency \footnote{Dressed frequency means their frequency in the presence of all couplers.} so that the qubit carrier space is degenerate. Furthermore,  all transmon qubits on the same track have, in principle, the same dressed frequency.

Transmon qubits on adjacent tracks are detuned: in Fig.~\ref{layoutConcept} we have chosen to use three frequencies, shown as three different colors, a pattern which can be repeated throughout the lattice.  This detuning between qubits on adjacent tracks helps in mitigating unwanted cross-talk between the tracks. Such cross-talk is an unwanted side-effect of the circuit which induces the cross-Kerr couplings which we design to minimize.  In Table \ref{tab:numbers} we list all relevant energy scales for the couplers and interactions that are used in Fig.~\ref{layoutConcept}. 

%BMT g-> J, but careful about J_x..
% J_{CK}  (as compared to Delta)
% frequency-diff adjacent tracks order of magnitude 500Mh-1 GHz
% variation in freq on a track, O(Delta)= 0.7 E_c= O(100)MHz
% Frequency band= 1.5 GHz in this scheme
% g/2pi= O(1-10)MHz. -> J
% T_1 versus computation time. xi_loss=Number of tracks x decay time (1/T_1) x time of comp
% T_comp \sim Depth x g    xi=n^2  x gamma/g = n^2 x 
% xi_loss=1% T_1=50 x 10^-6.  g/2pi=5 x 10^6

%Sites associated to different tracks can have different colors, but sites on the same track must have the same color, which means that the transmons on the same track need to have the same on-site energy. 
%Consequently, the transmons in the bulk of the track have all the same frequencies, while those at the edges have slightly different frequencies because of the different renormalization coming from the attractive interactions at the edges. 

\begin{center}
\begin{tabular}{p{6cm} | p{4cm}}
\toprule
Flip-flop coupling $J$ & $ 2 \pi \times 10 \,  \mathrm{MHz}$ \\
$J(\sigma_+^1 \sigma_-^2+\sigma_-^1 \sigma_+^2$) & \\
\midrule
Transmon cross-Kerr coupling $\Delta$ & $2 \pi \times 100 \, \mathrm{MHz}$ \\
($-\Delta a^{\dagger} a b^{\dagger} b$) \\
\midrule
Transmon freq. detuning between & $ \gtrapprox  \, 500 \, \mathrm{MHz}$ \\
adj. tracks  \\
\midrule
Total transmon freq. bandwidth & $\approx \, 1.5 \, \mathrm{GHz} $ \\
\bottomrule 
\end{tabular}
 \captionof{table}{Estimates of typical parameters achievable in our implementation. }
\label{tab:numbers}
\end{center}

All gates are generated by direct capacitive couplings between the transmon qubits and capacitive couplings via an intermediate resonator. The coupling mechanism behind all these gates are weak $J/2\pi=O(1)-O(10)$MHz flip-flop (also referred to as hopping) couplings which move the transmon excitations along the track.
In Fig.~\ref{circuitGates} we see an example of a bit-flip $X$ and a Hadamard $H$ gate. The Hadamard gate requires a mediated flip-flop interaction via an off-resonantly coupled resonator, so as to obtain a sign change in the coupling parameter associated with the Hadamard gate, \i.e. the transition from $\ket{1}$ to $\ket{1}$ has amplitude $-1/\sqrt{2}$. By changing the strength of such flip-flop couplers, for example by tuning the frequency of the resonator, one can get any real single-qubit rotation $U(\theta)$.

The most challenging interactions to achieve with transmon qubits are the strong attractive cross-Kerr interactions on the edges. When two transmon qubits are viewed as anharmonic oscillators with annihilation operators $a$ and $b$, these cross-Kerr interactions are of the form $- \Delta a^{\dagger}a b^{\dagger} b$.

One of the main problems in obtaining these interactions is that they have a tendency to come together with unwanted flip-flop and/or cross-talk interactions. These flip-flop interactions would move information-carrying excitations from one track to another which is unwanted. Secondly, two successive intertrack flip-flops can induce a logical $X$ error by effectively moving the excitation from one transmon in a pair to the other transmon of the pair. Since we keep transmon qubits on adjacent tracks detuned, the first-order effect of residual flip-flop interactions is small. However, a sequence of two intertrack hops is a resonant process, so the strength of such terms needs to be small.

We propose a new element capable of generating large cross-Kerr interactions, while keeping the cross-talk moderate. The coupler is composed of an array of a few junctions  in parallel with a smaller Josephson junction and it is schematically depicted in Fig.~\ref{strongZZcoup}. The coupler requires a constant flux bias. Similar elements, but for (different) simulation purposes, have been proposed in Refs.~\cite{richerPop,snail} and realized in \cite{Kounalakis2018}, \cite{wallraff:observation}. In these last papers, an attractive cross-Kerr of strength $\Delta= 2 \pi \times O(7-10){\rm MHz}$ was measured which is below what we believe could be achieved with our coupler. For all such couplers, one expects that the maximal strength of the cross-Kerr interaction will be limited by the anharmonicity of the coupled modes, see also e.g. \cite{nigg:bb}, implying that anharmonic modes are needed to build cross-Kerr interactions.

In Fig.~\ref{strongZZcoup}, we see that at each normal edge we need four couplers, which is a consequence of the fact that in the dual rail encoding we need two transmons per site. In Fig.~\ref{fig:CNOT-CKerr} we show the couplers for a CNOT gate. The implementation of controlled unitaries, such as the CNOT, requires the splitting of an intermediate site into two sites, where the particle is routed depending on the state of the control. Thus, on the intermediate site we will have a total of four transmon qubits as can be seen from Fig. \ref{fig:CNOT-CKerr} (green sites).

The correct forward motion of the computation is realized in the limit $J/\Delta \ll 1$: in this limit, excitations only move when they do not incur energy penalties due to the strong 
cross-Kerr interaction, implying that they occupy nearest-neighbor lattice sites. 
The idea is that one can also selectively use the cross-Kerr interaction to determine where one excitation moves depending on the presence of another excitation, and thus what controlled state-change the qubit-carrying-excitation undergoes. This is the idea behind the CNOT, Controlled-Hadamard and Toffoli gates. The need to work in the limit $J/\Delta \ll 1$ while the computation time scales as $\sim 1/J$ requires the strongest possible $\Delta$, in turn allowing for the largest possible $J$ to execute the computation before excitations get lost via $T_1$-relaxation. Let's assume a typical transmon $T_1=50 \, \mu s$. The computation time $T_{\rm comp}$ scales as $1/J \times {\rm HoppingDepth}$ where HoppingDepth measures the maximum $\#$ site-to-site hoppings to execute some 1D nearest-neighbor circuit. If we consider an $N \times N$ lattice, then the probability to lose any of the transmon excitations scales as the number of excitations, namely $2N-1$, times their probability to decay in time $T_{\rm comp}$, proportional to $\approx T_{\rm comp}/T_1$. If we fix the total loss probability to be $0.1$ and take a HoppingDepth of 20 (a CNOT requires 3), one obtains $N=15$. This rough estimate shows how much computation can be executed in our proposed scheme given current transmon relaxation times.

The ideal model does not require the doubled cross-Kerr couplings to be identical throughout the lattice. While in principle the strength $\Delta$ should be the same for all interactions between two adjacent tracks, it can be different for different pairs of tracks without affecting the motion of the computational wavefront, as long as $J/\Delta$ is small. In addition, small variability in $\Delta$ for cross-Kerr interactions between tracks may not directly harm the computation. We discuss the interesting effect of static disorder on qubit frequency, hopping and cross-Kerr interaction in Section \ref{sec:error} and \ref{transmon:noise}.

A simpler, early-implementation, version of our model would be one without any computational logic. In this model one studies the dynamics of the transmon excitations. Each track consists of a single transmon qubit (not a pair representing a dual-rail encoded qubit) and transmon excitations would be supplied at the side of the lattice. The goal would be to experimentally observe how the combination of strong cross-Kerr interaction combined with weak hopping gives rise to a wavefront of excitations, forming a connected string which is propagating over the lattice.  The engineering of this simple Hamiltonian is close to that of a Bose-Hubbard model with strong cross-Kerr interactions on edges, self-Kerr anharmonicity for transmon qubits and weak capacitively-induced hopping across plaquettes.

\subsection{Overview of Paper}
The paper is organized as follows. In Sec.~\ref{reviewSec} we review the model presented in Ref.~\cite{terhalLloyd}. After briefly recalling the idea of the CNOT implementation in Sec.~\ref{subsecCNOT}, we show how to construct a direct Toffoli gate in Sec.~\ref{subsec:Tof} (with technical details in Appendix \ref{sec:tof}). In Sec. \ref{dualRailSec} we review the effect of the dual-rail encoding, leading to the coupling scheme shown in Fig.~\ref{layoutConcept}. In Sec.~\ref{sec:error} we numerically analyze errors in the correct forward propagation of the computation including the effects of variability in couplers and frequencies for small system sizes.

In Sec.~\ref{transmonImpl} we give details of the proposed implementation with transmon qubits describing the basic coupling elements that we sketched in Fig.~\ref{circuitGates} and Fig.~\ref{strongZZcoup}. In Section \ref{subsec::CKstrength} we discuss the strength of unwanted interactions mediated by the cross-Kerr couplers between three transmon qubits.

In Section \ref{transmon:noise} we discuss a challenge inherent in the realization of this computing architecture concerning the read-out of the computation. In Appendix \ref{FKPqc} we discuss a 2D lattice version of the Feynman-Kitaev Hamiltonian as an alternative to the Lloyd-Terhal scheme. We show how multi-qubit gates can be done in this scheme with ideas similar to those in Sec. \ref{sec:mql}. Appendix \ref{secControlledHadamard} shows how to use the simpler controlled-Hadamard gate to make a Toffoli gate. We end the paper with some open questions in Sec.~\ref{conclSec}.

In all Sections but Section \ref{transmonImpl} we set $\hbar=1$.

\section{Review of the Lloyd-Terhal scheme}
\label{reviewSec}

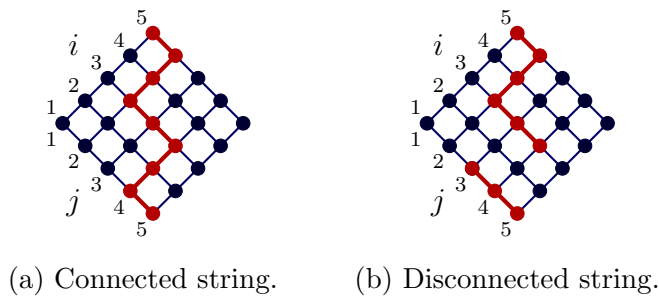
\begin{figure}
\centering
\begin{subfigure}[t]{0.3\textwidth}
\centering
\begin{tikzpicture}[scale=0.3]
\draw [thick, color=black!60!blue] (-4, 0)--(0, -4);
\draw [thick, color=black!60!blue] (0, -4)--(4, 0);
\draw [thick,color=black!60!blue] (4, 0)--(0, 4);
\draw [thick, color=black!60!blue] (0, 4)--(-4, 0);
\draw [thick, color=black!60!blue] (-3,1)--(1, -3);
\draw [thick, color=black!60!blue] (-2,2)--(2, -2);
\draw [thick, color=black!60!blue] (-1,3)--(3, -1);
\draw [thick, color=black!60!blue] (-3,-1)--(1, 3);
\draw [thick, color=black!60!blue] (-2,-2)--(2, 2);
\draw [thick, color=black!60!blue] (-1,-3)--(3, 1);
\filldraw [color=black!80!blue]   (-4, 0) circle (0.3cm);
\filldraw [color=black!80!blue]   (-2, 0) circle (0.3cm);
\filldraw [color=red!70!black]   (0, 0) circle (0.3cm);
\filldraw [color=black!80!blue]   (2, 0) circle (0.3cm);
\filldraw [color=black!80!blue]   (4, 0) circle (0.3cm);
\filldraw [color=black!80!blue]   (-3, 1) circle (0.3cm);
\filldraw [color=red!70!black]   (-1, 1) circle (0.3cm);
\filldraw [color=black!80!blue]   (+1, 1) circle (0.3cm);
\filldraw [color=black!80!blue]   (3, 1) circle (0.3cm);
\filldraw [color=black!80!blue]   (-3, -1) circle (0.3cm);
\filldraw [color=black!80!blue]   (-1, -1) circle (0.3cm);
\filldraw [color=red!70!black]   (+1, -1) circle (0.3cm);
\filldraw [color=black!80!blue]   (3, -1) circle (0.3cm);
\filldraw [color=black!80!blue]   (-2, 2) circle (0.3cm);
\filldraw [color=red!70!black]   (0, 2) circle (0.3cm);
\filldraw [color=black!80!blue]   (2, 2) circle (0.3cm);
\filldraw [color=black!80!blue]   (-2, -2) circle (0.3cm);
\filldraw [color=red!70!black]   (0, -2) circle (0.3cm);
\filldraw [color=black!80!blue]   (2, -2) circle (0.3cm);
\filldraw [color=black!80!blue]   (-1, 3) circle (0.3cm);
\filldraw [color=red!70!black]   (1, 3) circle (0.3cm);
\filldraw [color=red!70!black]   (-1, -3) circle (0.3cm);
\filldraw [color=black!80!blue]   (1, -3) circle (0.3cm);
\filldraw [color=red!70!black]   (0, 4) circle (0.3cm);
\filldraw [color=red!70!black]   (0, -4) circle (0.3cm);
\draw [ultra thick, color=red!70!black] (0, 4)--(1, 3);
\draw [ultra thick, color=red!70!black] (1, 3)--(0, 2);
\draw [ultra thick, color=red!70!black] (0, 2)--(-1, 1);
\draw [ultra thick, color=red!70!black] (-1, 1)--(0, 0);
\draw [ultra thick, color=red!70!black] (0, 0)--(1, -1);
\draw [ultra thick, color=red!70!black] (1, -1)--(0, -2);
\draw [ultra thick, color=red!70!black] (0, -2)--(-1, -3);
\draw [ultra thick, color=red!70!black] (-1, -3)--(0, -4);
\node at (-3.5, 3.5) {$i$};
\node at (-3.5, -3.5) {$j$};
\foreach \x in {1,...,5}
\node at (-4.5+\x-1, 0.7+\x-1) {\scriptsize{$\x$}};
\foreach \x in {1,...,5}
\node at (-4.5+\x-1, -0.7-\x+1) {\scriptsize{$\x$}};
\end{tikzpicture}
\subcaption{Connected string.}
\label{figConnected}
\end{subfigure}
\begin{subfigure}[t]{0.3\textwidth}
\centering
\begin{tikzpicture}[scale=0.3]
\draw [thick, color=black!60!blue] (-4, 0)--(0, -4);
\draw [thick, color=black!60!blue] (0, -4)--(4, 0);
\draw [thick, color=black!60!blue] (4, 0)--(0, 4);
\draw [thick, color=black!60!blue] (0, 4)--(-4, 0);
\draw [thick, color=black!60!blue] (-3,1)--(1, -3);
\draw [thick, color=black!60!blue] (-2,2)--(2, -2);
\draw [thick, color=black!60!blue] (-1,3)--(3, -1);
\draw [thick, color=black!60!blue] (-3,-1)--(1, 3);
\draw [thick, color=black!60!blue] (-2,-2)--(2, 2);
\draw [thick, color=black!60!blue] (-1,-3)--(3, 1);
\filldraw [color=black!80!blue]   (-4, 0) circle (0.3cm);
\filldraw [color=black!80!blue]   (-2, 0) circle (0.3cm);
\filldraw [color=red!70!black]   (0, 0) circle (0.3cm);
\filldraw [color=black!80!blue]   (2, 0) circle (0.3cm);
\filldraw [color=black!80!blue]   (4, 0) circle (0.3cm);
\filldraw [color=black!80!blue]   (-3, 1) circle (0.3cm);
\filldraw [color=red!70!black]   (-1, 1) circle (0.3cm);
\filldraw [color=black!80!blue]   (+1, 1) circle (0.3cm);
\filldraw [color=black!80!blue]   (3, 1) circle (0.3cm);
\filldraw [color=black!80!blue]   (-3, -1) circle (0.3cm);
\filldraw [color=black!80!blue]   (-1, -1) circle (0.3cm);
\filldraw [color=red!70!black]   (+1, -1) circle (0.3cm);
\filldraw [color=black!80!blue]   (3, -1) circle (0.3cm);
\filldraw [color=black!80!blue]   (-2, 2) circle (0.3cm);
\filldraw [color=red!70!black]   (0, 2) circle (0.3cm);
\filldraw [color=black!80!blue]   (2, 2) circle (0.3cm);
\filldraw [color=black!80!blue]   (-2, -2) circle (0.3cm);
\filldraw [color=red!70!black]   (-2, -2) circle (0.3cm);
\filldraw [color=black!80!blue]   (2, -2) circle (0.3cm);
\filldraw [color=black!80!blue]   (-1, 3) circle (0.3cm);
\filldraw [color=red!70!black]   (1, 3) circle (0.3cm);
\filldraw [color=red!70!black]   (-1, -3) circle (0.3cm);
\filldraw [color=black!80!blue]   (1, -3) circle (0.3cm);
\filldraw [color=red!70!black]   (0, 4) circle (0.3cm);
\filldraw [color=black!80!blue]   (0, -2) circle (0.3cm);
\filldraw [color=red!70!black]   (0, -4) circle (0.3cm);
\draw [ultra thick, color=red!70!black] (0, 4)--(1, 3);
\draw [ultra thick, color=red!70!black] (1, 3)--(0, 2);
\draw [ultra thick, color=red!70!black] (0, 2)--(-1, 1);
\draw [ultra thick, color=red!70!black] (-1, 1)--(0, 0);
\draw [ultra thick, color=red!70!black] (0, 0)--(1, -1);
\draw [ultra thick, color=red!70!black] (-2, -2)--(0, -4);
\node at (-3.5, 3.5) {$i$};
\node at (-3.5, -3.5) {$j$};
\foreach \x in {1,...,5}
\node at (-4.5+\x-1, 0.7+\x-1) {\scriptsize{$\x$}};
\foreach \x in {1,...,5}
\node at (-4.5+\x-1, -0.7-\x+1) {\scriptsize{$\x$}};
\end{tikzpicture}
\subcaption{Disconnected string.}
\end{subfigure}
\caption{Examples of connected and disconnected strings of particles. The red dots denote the position of the particles.}
\label{figLattice}
\end{figure}

In this section we review some elements of the scheme for quantum computing with a time-independent Hamiltonian proposed in Ref \cite{terhalLloyd}. The scheme is closely related to the one proposed in Ref \cite{janzing2007} and the one analyzed in \cite{gossetTerhal, breuckmannTerhal}. The main difference between the scheme in Ref. \cite{terhalLloyd} and the ones of Refs. \cite{janzing2007} and \cite{gossetTerhal} is the way gates are implemented.
The model is most simply explained in terms of particles with spin-1/2 moving on tracks: in Section \ref{dualRailSec} we discuss the dual-rail representation in terms of transmon qubits. We start by considering the rotated $N \times N$ lattice in Fig.~\ref{figLattice} with $N_{\rm track}=2N-1$ tracks. A site is denoted by $(i, j)$, with $i, j \in \{1,2, \dots, N\}$. Importantly, we will consider that at each time there is \emph{only} one particle per track \footnote{Note that a track is not identified by a constant index $i$ or $j$, but by a constant difference $i-j$ in our notation.}. This is ensured by initializing the system in a state with this property and by ensuring that the Hamiltonian preserves it. 

The quantum information is encoded in the spin degree of freedom of the particle.  Particles are allowed to hop horizontally from one site to the next one (and vice-versa), --we say that the particles move on tracks--, and when hopping takes place a single-qubit gate can be applied to the spin degree of freedom. We can thus associate a single-qubit gate to each plaquette of the lattice in Fig. \ref{figLattice}.  If our purpose were to simulate a quantum circuit with only single-qubit gates, then independent hopping dynamics for each qubit degree of computation would clearly suffice.  
One way of realizing two-qubit gates is to ensure that particles move coherently together, as a string of particles \cite{janzing2007, gossetTerhal}. If we let the spin-degree of freedom of one particle influence the whereabouts and the single-qubit logic on another particle, spin-controlled single-qubit gates can be realized.
An efficient forward computation can be achieved also in the context of quantum walk on a line, an idea that we will use in the discussion of the alternative lattice model in Appendix \ref{FKPqc}. \par 
The dynamics of the system is designed in such a way that if the system starts in a state in which the particles are connected as a string in Fig.~\ref{figConnected}, it will evolve only into states in which the particles remain connected. This last property will not be guaranteed exactly by the Hamiltonian, but perturbatively. We will refer to states with connected particles as valid or connected strings. Note that if these properties are satisfied, assuming that the system starts in the connected string in which all particles are on the left of the lattice, a particular connected string univocally identifies the gates that have been executed. The particular wedge-like geometry of the rotated lattice may seem artificial when the goal is to execute a nearest-neighbour circuit. However, this geometry can be shown to guarantee an efficient forward motion of the computation on the left-half of the lattice where the number of connected strings is an increasing function of lattice depth \cite{gossetTerhal, terhalLloyd}. The motion of the string is also depicted in Fig.~\ref{fig:screenWF}.
 \par

In the following part of this section we focus on the implementation of quantum circuits with only single-qubit gates, leaving the discussion of two- and three-qubit gates to the section \ref{sec:mql}. \par 

Before introducing Hamiltonians, let us give some mathematical definitions. 
 We define the particle number operator at a site $(i,j)$ as $\bm{n}[i,j]=\sum_{s=0,1} n_{s}[i,j]$, where $n_{s}[i,j]$ is the number operator for particle at site $(i,j)$ in internal spin state $s=0,1$. The operators $n_s[i, j]$ can be written in terms of creation and annihilation operators for a particle at site $(i, j)$ and internal state $s$, \i.e., $n_s[i,j]=a_{s}^{\dagger}[i,j] a_s[i,j]$. To describe the CNOT and Toffoli gate, we use split-sites labeled as $(i,j,k)$, $k=0,1$ so that $n_{s}[i,j,k]= a_{s}^{\dagger}[i,j,k]a_{s}[i,j,k]$, $\bm{n}[i,j,k]=\sum_{s=0}^1 n_{s}[i,j,k]$ and $\bm{n}[i,j]=\sum_{k=0}^1\bm{n}[i,j,k]$.
We call the set of edges $E$ and the set of plaquettes $P$. A site $(i, j)$ will occasionally be denoted with a compact symbol $\mu$ or $\nu$. We will identify an edge $e$ by the sites it connects and write $e=(\mu, \nu)$. \par 

The previously sketched ideas translate to a Hamiltonian
\begin{equation}
H= H_{\rm valid}+V_{\rm hop}.
\label{basicH}
\end{equation} 
where $H_{\rm valid}$ is defined as
\begin{equation}
H_{\rm valid}= -\Delta \sum_{(\mu, \nu) \in E} \bm{n}[\mu] \bm{n}[\nu],
\label{HstringTL}
\end{equation}
where $\Delta >0$. Consider the spectrum of $H_{\rm valid}$ in the particle number basis. The groundspace of $H_{\rm valid}$, in the sector with one excitation per track, is degenerate with eigenvalue $E_0=-(N_{\rm track}-1) \Delta$ and composed of all possible connected strings.  The number of connected strings on the lattice is ${2 (N-1) \choose N-1}$. Thus, including the spin-degree of freedom the groundspace is $2^{N_{\rm track}}{2(N-1) \choose N-1}$-dimensional. The first excited subspace is formed by the subspace of strings which are disconnected at a single site. These strings have an energy gap of $\Delta$ above the groundspace, \i.e., $E_1-E_0= \Delta$. In general, $H_{\rm valid}$ has $N_{\rm track}-2$ eigen-subspaces with energy $E_k-E_0=k \Delta$, where $k \in \{0,1, \dots, N_{\rm track}-1 \}$ denotes the number of points where the string is broken. 
Note that this Hamiltonian is fully degenerate with respect to the spin degree of freedom. However, no harm is done by having additional single-site terms of the form $\omega {\bf n}[\mu]$ in the Hamiltonian as long as $\omega$ is the same along the entire track (as there is a only single particle per track). In addition, the attractive intertrack strengths $\Delta$ in $H_{\rm valid}$ can be different for one pair of tracks as compared to another pair.

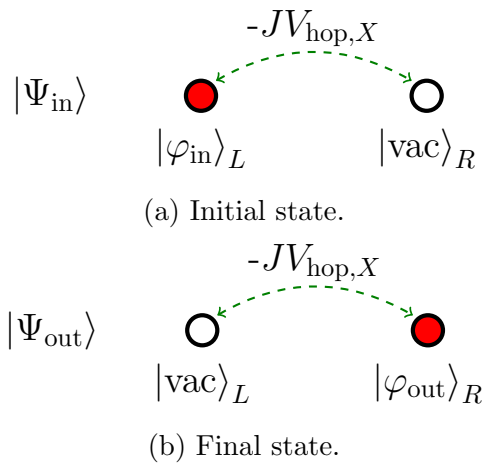
\begin{figure}[htb]
\centering
\begin{subfigure}[t]{0.5\textwidth}
\centering
\begin{tikzpicture}
\filldraw [red] (0,0) circle (0.2 cm);
\draw [ultra thick] (0,0) circle (0.2 cm);
\draw [ultra thick] (3,0) circle (0.2 cm);
\node at (0, -0.7) {\large{$\ket{\varphi_{\rm in}}_L$}};
\node at (3, -0.7) {\large{$\ket{\rm vac}_R$}};
\node at (-2, 0) {\large{$\ket{\Psi_{\rm in}}$}};
\draw [thick, color=green!50!black,  dashed, <->] (0.2, 0.2) to [bend left] (3-0.2,0+0.2);
\node at (1.5, 0.9) {\large{-$J V_{{\rm hop}, X}$}};
\end{tikzpicture}
\caption{Initial state.}
\label{singleQGa}
\end{subfigure}
\begin{subfigure}[t]{0.5\textwidth}
\centering
\begin{tikzpicture}
\filldraw [red] (3,0) circle (0.2 cm);
\draw [ultra thick] (0,0) circle (0.2 cm);
\draw [ultra thick] (3,0) circle (0.2 cm);
\node at (3, -0.7) {\large{$\ket{\varphi_{\rm out}}_R$}};
\node at (0, -0.7) {\large{$\ket{\rm vac}_L$}};
\node at (-2, 0) {\large{$\ket{\Psi_{\rm out}}$}};
\draw [thick, color=green!50!black,  dashed, <->] (0.2, 0.2) to [bend left] (3-0.2,0+0.2);
\node at (1.5, 0.9) {\large{-$J V_{{\rm hop}, X}$}};
\end{tikzpicture}
\caption{Final state.}
\label{singleQGb}
\end{subfigure}
\caption{Example of hopping executing a single-qubit $X$-gate.}
\label{singleQG}
\end{figure}

The hopping Hamiltonian $V_{\rm hop}$ is responsible for the dynamics of the system and for the application of the gates. Let us suppose that we want to run a quantum circuit composed only of single-qubit gates. In particular, we associate a single-qubit gate, denoted by $U_p$, with each plaquette of the lattice, \i.e., $p \in P$, for a total of $(N-1)^2$ single-qubit gates. The hopping Hamiltonian is defined as 
\begin{equation}
V_{\rm hop}= -J \sum_{p \in P}  V_{{\rm hop},p},
\label{VhopSQ}
\end{equation}
where $V_{{\rm hop}, p}$ is the hopping associated with the plaquette $p$ and defined as
\begin{equation}
\label{VhopPlaquette}
V_{{\rm hop},p}= \sum_{s=0}^1 \sum_{s \sp{\prime}=0}^1 \bra{s \sp{\prime}} U_p \ket{s}a_{s \sp{\prime}}^{\dagger}[i+1, j+1]a_{s}[i,j]+\mathrm{h.c.}
\end{equation}
The overall sign of the hopping Hamiltonian $V_{\rm hop}$ is taken to be negative, with $J > 0$, but identical dynamics is obtained when all $J \rightarrow -J$.
The effect of this hopping Hamiltonian can be exemplified by considering a single plaquette in Fig.~\ref{singleQGa} which executes the $X$ gate when hopping takes place
\begin{equation*}
H_{X}=-J V_{\rm hop, X}=-J \{ a_{0}^{\dagger}[R] a_{1}[L]+a_{1}^{\dagger}[R] a_{0}[L]+\mathrm{h.c.} \},
%\label{HsingleHopping}
\end{equation*}
where $a_{s}[S]$, $s=0,1$, $S=L, R$ annihilates the state $\ket{s}_S$, \i.e., $a_{s}[S]\ket{s}_S= \ket{\rm vac}_S$. 

Let us suppose that the system starts at $t=0$ in the state $\ket{\Psi_{\rm in}}= \ket{\varphi_{\rm in}}_L \otimes \ket{\rm vac}_R= (\alpha \ket{0}_L+\beta \ket{1}_L)\otimes \ket{\rm vac}_R$. One can observe that $H_X \ket{\Psi_{\rm in}} = -J \ket{\Psi_{\rm out}}$ with $\ket{\Psi_{\rm out}}=\ket{\rm vac}_L \otimes \ket{\varphi_{\rm out}}_R=\ket{\rm vac}_L \otimes X_R \ket{\varphi_{\rm in}}_R$ and $H_X \ket{\Psi_{\rm out}}=-J \ket{\Psi_{\rm in}}$, thus undoing the $X$ gate. 

As long as $X=U_p$ is a unitary matrix, we see that the dynamics is not influenced by what unitary is implemented. and it is equivalent to a continuous-time quantum walk on a line with $2$ sites, where $\ket{\Psi_{\rm in}}$ and $\ket{\Psi_{\rm out}}$ play the role of the discretized allowed positions. If we would consider a chain of $L-1$ gates applied like in Fig.~\ref{singleQG}, in series, it can be easily shown that the dynamics is equivalent to that of a continuous-time quantum walk on a line with $L$ sites (see also Appendix \ref{FKPqc}). \par

%Since the system is always in the single-particle subspace, the quantum states are linear superposition of $2 (N+1)$ states namely $\{\ket{0}_{0} \ket{\rm vac}_{1} \ket{\rm vac}_2 \dots \ket{\rm vac}_{N}$, $\ket{1}_{0} \ket{\rm vac}_{1} \ket{\rm vac}_2 \dots \ket{\rm vac}_{N}$, $\ket{\rm vac}_{0} \ket{0}_{1} \ket{\rm vac}_2 \dots \ket{\rm vac}_{N}$, $\dots$, $\ket{\rm vac}_0 \ket{\rm vac}_1 \dots \ket{\rm vac}_{N-1} \ket{1}_{N} \}$. We can then represent these quantum states in a $2(N+1)$-dimensional Hilbert space $\mathcal{H}_{Q} \otimes \mathcal{H}_{clock}$ space composed of a qubit and a $N+1$-dimensional quantum system that plays the role of the clock, via the unitary map $W$ defined as\begin{align*}
%W \ket{\rm vac}_0 \ket{\rm vac}_1 \dots \ket{0}_k \dots \ket{\rm vac}_N= & \ket{0} \ket{k}, \\
%W \ket{\rm vac}_0 \ket{\rm vac}_1 \dots \ket{1}_k \dots \ket{\rm vac}_N= & \ket{1} \ket{k}.
%\end{align*} 
%We also understand that this mapping can be done for every track in the lattice in Fig.~\ref{figLattice}.\par

Let us now examine the combined effect of $V_{\rm hop}$ and $H_{\rm valid}$.
The hopping Hamiltonian $V_{\rm hop}$ is not diagonal in the eigenbasis of $H_{\rm valid}$ and can cause transitions from valid strings to valid strings, which, in quantum optics language, are resonant. In addition, it can induce transitions from a valid connected string to an invalid, disconnected string which is a far off-resonant  (by $\Delta$), suppressed transition.

In \cite{terhalLloyd} it was thus argued that a perturbative treatment of this Hamiltonian $H$ gives rise to an effective Hamiltonian $H_{\rm eff}$ which only contains resonant controlled-hopping terms, \i.e. particles can only move forward when their move keeps the particles in the valid string subspace. Explicitly,

\begin{equation}
H_{\mathrm{eff}}= -J \sum_{p \in P} H_{{\rm cond.hop}, p}+\mathcal{O} ( \lVert V_{\rm hop} \rVert^2/\Delta ),
\label{HeffTL}
\end{equation}
where we defined the conditional hopping Hamiltonian in a plaquette $p$ bordered by top and bottom sites resp. $(i+1,j)$ and $(i,j+1)$ as
\begin{equation}
\label{condHopSQ}
H_{{\rm cond.hop}, p}= \bm{n}[i+1, j]\bm{n}[i, j+1] \otimes V_{{\rm hop},p}.
\end{equation}

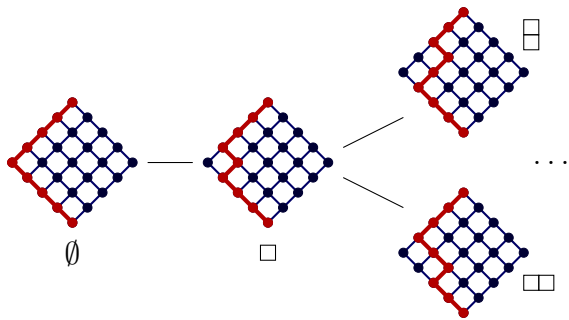
\begin{figure}[htb]
\centering
\begin{tikzpicture}[scale=0.2]
\node at (0,-6) {$\emptyset$};
\draw (12.5,-6.5) rectangle (13.5,-5.5);
\draw (30,-8.5) rectangle (31,-7.5);
\draw (31,-8.5) rectangle (32,-7.5);
\draw (30,8.5) rectangle (31,7.5);
\draw (30,9.5) rectangle (31,8.5);
\pic [scale=0.2] at (0,0) {lattice};
\pic [scale=0.2] at (13,0) {lattice};
\pic [scale=0.2] at (26,6) {lattice};
\pic [scale=0.2] at (26,-6) {lattice};
\node at (32, 0) {$\dots$};
\draw (5,0)--(8,0);
\draw (18,1)--(22,3);
\draw (18,-1)--(22,-3);
\draw [ultra thick, color=red!70!black] (0,4)--(-4,0)--(0,-4);
\foreach \x in {0,..., 4}
\filldraw [color=red!70!black]   (0-\x*1, 4-\x*1) circle (0.3cm);
\foreach \x in {0,..., 4}
\filldraw [color=red!70!black]   (-4+\x*1, 0-\x*1) circle (0.3cm);
\draw [ultra thick, color=red!70!black] (0+13,4)--(-3+13,1)--(-2+13,0)--(-3+13,-1)--(0+13,-4);
\foreach \x in {0,..., 3}
\filldraw [color=red!70!black]   (13-\x*1, 4-\x*1) circle (0.3cm);
\filldraw [color=red!70!black] (13-2,0) circle (0.3cm);
\foreach \x in {0,..., 3}
\filldraw [color=red!70!black]   (10+\x*1, -1-\x*1) circle (0.3cm);
\draw [ultra thick, color=red!70!black] (26,10)--(24,8)--(25,7)--(23,5)--(26,2);
\foreach \x in {0,1,2}
\filldraw [color=red!70!black] (26-\x*1,10-\x*1) circle (0.3cm);
\filldraw [color=red!70!black] (25, 7) circle (0.3cm);
\foreach \x in {0,1}
\filldraw [color=red!70!black] (24-\x*1,6-\x*1) circle (0.3cm);
\foreach \x in {0,1,2}
\filldraw [color=red!70!black] (24+\x*1,4-\x*1) circle (0.3cm);
\draw [ultra thick, color=red!70!black] (26,-2)--(23,-5)--(25,-7)--(24,-8)--(26,-10);
\foreach \x in {0,1,2,3}
\filldraw [color=red!70!black] (26-\x*1,-2-\x*1) circle (0.3cm);
\foreach \x in {0,1}
\filldraw [color=red!70!black] (24+\x*1,-6-\x*1) circle (0.3cm);
\foreach \x in {0,1,2}
\filldraw [color=red!70!black] (24+\x*1,-8-\x*1) circle (0.3cm);
\end{tikzpicture}
\caption{Coherent quantum walk of the connected string which can be viewed as a quantum walk on Young's lattice \cite{gossetTerhal,DOR:quantum-crys}.}
\label{QWstring}
\end{figure}
Assuming that the system starts in the configuration in which all particles are on the left (see Fig. \ref{QWstring}), the dynamics of the string under $H_{\mathrm{eff}}$ in Eq.~(\ref{HeffTL}) can be nicely exactly solved, see \cite{gossetTerhal, janzing2007}. In particular, the forward motion of the string in the open-wedge region, in the limit of a large lattice size, has a constant velocity \cite{gossetTerhal}.

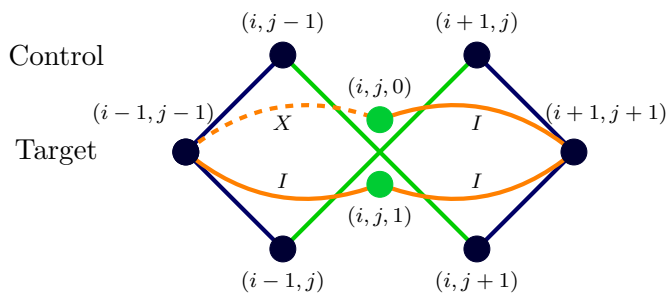
\begin{figure}
\centering
\begin{tikzpicture}[scale=0.43]
\draw [ultra thick, color=black!60!blue] (-6, 0)--(-3, 3);
\draw [ultra thick, color=black!60!blue] (-6, 0)--(-3, -3);
\draw [ultra thick, color=green!80!black] (-3, -3)--(0, 0);
\draw [ultra thick, color=green!80!black] (-3, 3)--(0, 0);
\draw [ultra thick, color=green!80!black] (0, 0)--(3, 3);
\draw [ultra thick, color=green!80!black] (0, 0)--(3, -3);
\draw [ultra thick, color=black!60!blue] (3,3)--(6,0);
\draw [ultra thick, color=black!60!blue] (3,-3)--(6,0);
\draw [ultra thick, color=orange,  dashed] (-6,0) to [bend left] (0,1);
\draw [ultra thick, color=orange] (0,1) to [bend left] (6,0);
\draw [ultra thick, color=orange] (-6,0) to [bend right] (0,-1);
\draw [ultra thick, color=orange] (0,-1) to [bend right] (6,0);
\filldraw [color=black!80!blue]   (-6, 0) circle (0.4cm);
\filldraw [color=black!80!blue]   (-3, 3) circle (0.4cm);
\filldraw [color=green!80!blue]   (0, 1) circle (0.4cm);
\filldraw [color=green!80!blue]   (0, -1) circle (0.4cm);
\filldraw [color=black!80!blue]   (-6, 0) circle (0.4cm);
\filldraw [color=black!80!blue]   (-3, -3) circle (0.4cm);
\filldraw [color=black!80!blue]   (3, -3) circle (0.4cm);
\filldraw [color=black!80!blue]   (3, 3) circle (0.4cm);
\filldraw [color=black!80!blue]   (6, 0) circle (0.4cm);
\node at (-3, 0.9) {\scriptsize{$X$}};
\node at (-3, -0.9) {\scriptsize{$I$}};
\node at (+3, 0.9) {\scriptsize{$I$}};
\node at (+3, -0.9) {\scriptsize{$I$}};
\node at (0, 2) {\scriptsize{$(i,j,0)$}};
\node at (0, -2) {\scriptsize{$(i,j,1)$}};
\node at (-10, 3) {\small{Control}};
\node at (-10, 0) {\small{Target}};
\node at (-3,4) {\scriptsize{$(i, j-1)$}};
\node at (3,4) {\scriptsize{$(i+1, j)$}};
\node at (-7,1.2) {\scriptsize{$(i-1, j-1)$}};
\node at (7,1.2) {\scriptsize{$(i+1, j+1)$}};
\node at (-3,-4) {\scriptsize{$(i-1, j)$}};
\node at (3,-4) {\scriptsize{$(i, j+1)$}};
\end{tikzpicture}
\caption{The CNOT gate. The control particle moves on the upper track and the target particle moves on the middle track, while a bystander particle not involved in the logic can move on the bottom track. In the middle track one has two split-sites which both connect via green edges to the other four sites. These four edges are modified from standard edges to implement the conditional logic, see Eq.~\ref{eq:cnot-delta}. The particle on the middle track can go through either of the split-sites, but only if it goes past one site it undergoes an $X$ gate.}
\label{cnotFig}
\end{figure}
\subsection{Multi-Qubit Logic}
\label{sec:mql}

To get computational universality, one could only focus on circuits involving single- and two-qubit gates, for instance Hadamard, single-qubit $T={\rm diag}(1, e^{i \pi/4})$ gate and the CNOT gate \cite{nielsenChuang}. The $T$ gate requires complex hopping parameters \cite{terhalLloyd}, but in our transmon implementation only real flip-flop interactions are available, see the arguments in Section \ref{sec:real-stoq}. However, we can explicitly construct a way of performing the Toffoli gate achieving universality with Hadamard and Toffoli \cite{ShiToffoli}.  In Appendix \ref{secControlledHadamard}, we also show how universality with real gates can be achieved using Hadamard, CNOT and controlled-Hadamard gates. The controlled-Hadamard can be implemented with similar resources as the CNOT gate.

In the following two sections we review the CNOT construction presented in \cite{terhalLloyd} and introduce the construction of the Toffoli gate. We point out that these constructions may have applications that go beyond the present model, as units that implement particular gates in a modular architecture approach to quantum computation as proposed in related work \cite{banchiToffoli}. 

\subsubsection{CNOT Gate}
\label{subsecCNOT}

The CNOT construction is depicted in Fig.~\ref{cnotFig}. The Hamiltonian of the region of the lattice where the CNOT is implemented has to be modified. In particular, the central site is replaced by two split-sites and the target particle is directed to one of the two split-sites depending on the state of the control particle. Accordingly, an $X$ gate is applied if the control is in the $\ket{1}$ state, while the identity gate is applied if it is in the $\ket{0}$ state. Finally, the CNOT is completed via hopping of the particle from the intermediate split-sites to a final site with the application of an identity gate. This way of implementing the CNOT resembles that of a railroad switch \cite{nagaj2010}. The working principle of the CNOT also resembles that of a quantum spin transistor that has been more recently proposed in Refs. \cite{zinnerSpinNature, zinnerSpinTrans}.\par

 How can we make sure that the logic is implemented correctly? The idea is to give energy penalties to configurations that perform incorrect logic. This means that if the control is in internal state $\ket{0}$ and the target particle is at the intermediate site $(i, j, 1)$, this invalid configuration should have a penalty $\Delta$ compared to the valid configuration in which the target particle is at $(i, j,0)$. Analogously, a penalty is given to configurations in which the control particle has internal state $\ket{1}$ and the target particle is at site $(i,j,0)$. This is done by letting the attractive interactions between the control particle at sites $(i,j-1)$ and $(i+1,j)$ and the intermediate sites $(i,j,0)$ and $(i,j,1)$ depend on the spin state of the control particle. The attractive interactions between the sites $(i-1,j)$ and $(i,j+1)$ and the split-sites are the same as before. More precisely, the green edges in Fig.~\ref{cnotFig} are chosen as
\begin{multline}
-\Delta \sum_{s=0,1} \left(n_{s}[i,j-1] {\bf n}[i,j,s]+n_{s}[i+1,j] {\bf n}[i,j,s]\right)+ \\
-\Delta \sum_{s=0,1} \left({\bf n}[i-1,j] {\bf n}[i,j,s]+{\bf n}[i,j+1] {\bf n}[i,j,s])\right).
\label{eq:cnot-delta}
\end{multline}
In addition, the orange hopping edges $V_{\rm hop}$ in Fig.~\ref{cnotFig} to the split-sites are of the form in Eq.~(\ref{VhopPlaquette}) where the $X$ and $I$ labels indicate whether the hopping affects the internal state.  \par

Using the CNOT idea one can implement any controlled unitary, and in particular the controlled Hadamard, which can be used together with the CNOT to construct a Toffoli gate as shown in Appendix \ref{secControlledHadamard}.

\subsubsection{Direct Toffoli Gate}
\label{subsec:Tof}

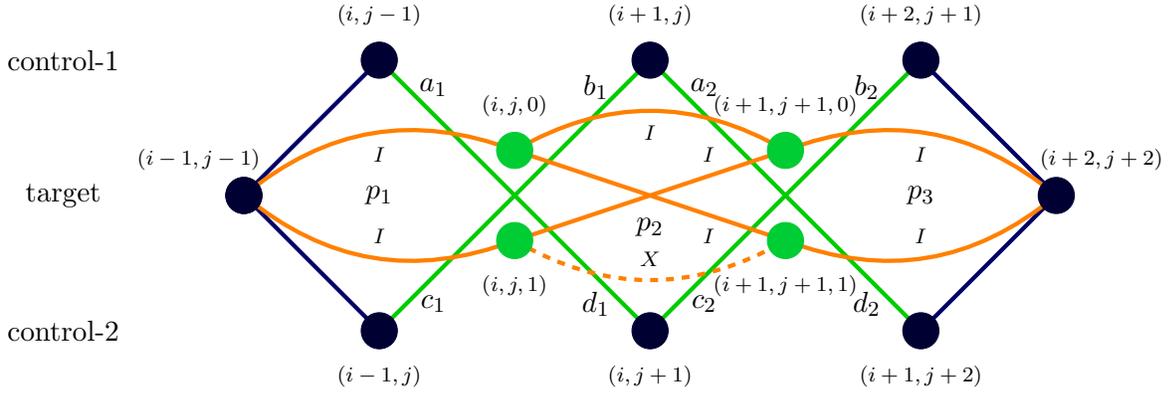
\begin{figure*}[htb]
\centering
\begin{tikzpicture}[scale=0.6]
\draw [ultra thick, color=black!60!blue] (-6, 0)--(-3, 3);
\draw [ultra thick, color=black!60!blue] (-6, 0)--(-3, -3);
\draw [ultra thick, color=green!80!black] (-3, -3)--(0, 0);
\draw [ultra thick, color=green!80!black] (-3, 3)--(0, 0);
\draw [ultra thick, color=green!80!black] (0, 0)--(3, 3);
\draw [ultra thick, color=green!80!black] (0, 0)--(3, -3);
\draw [ultra thick, color=green!80!black] (3,3)--(6,0);
\draw [ultra thick, color=green!80!black] (3,-3)--(6,0);
\draw [ultra thick, color=green!80!black] (6,0)--(9, 3);
\draw [ultra thick, color=black!60!blue] (9,3)--(12, 0);
\draw [ultra thick, color=green!80!black] (6,0)--(9, -3);
\draw [ultra thick, color=black!60!blue] (9,-3)--(12, 0);
\draw [ultra thick, color=orange] (-6,0) to [bend left] (0,1);
\draw [ultra thick, color=orange] (-6,0) to [bend right] (0,-1);
\draw [ultra thick, color=orange] (0,1) to [bend left] (6,1);
\draw [ultra thick, color=orange, dashed] (0,-1) to [bend right] (6,-1);
\draw [ultra thick, color=orange] (0,1) to (6,-1);
\draw [ultra thick, color=orange] (0,-1) to (6,1);
\draw [ultra thick, color=orange] (6,1) to [bend left] (12,0);
\draw [ultra thick, color=orange] (6,-1) to [bend right] (12,0);
\filldraw [color=black!80!blue]   (-6, 0) circle (0.4cm);
\filldraw [color=black!80!blue]   (-3, 3) circle (0.4cm);
\filldraw [color=green!80!blue]   (0, 1) circle (0.4cm);
\filldraw [color=green!80!blue]   (0, -1) circle (0.4cm);
\filldraw [color=black!80!blue]   (-6, 0) circle (0.4cm);
\filldraw [color=black!80!blue]   (-3, -3) circle (0.4cm);
\filldraw [color=black!80!blue]   (3, -3) circle (0.4cm);
\filldraw [color=black!80!blue]   (3, 3) circle (0.4cm);
\filldraw [color=green!80!blue]   (6, 1) circle (0.4cm);
\filldraw [color=green!80!blue]   (6, -1) circle (0.4cm);
\filldraw [color=black!80!blue]   (9, 3) circle (0.4cm);
\filldraw [color=black!80!blue]   (9, -3) circle (0.4cm);
\filldraw [color=black!80!blue]   (12, 0) circle (0.4cm);
\node at (-3, 0.9) {\scriptsize{$I$}};
\node at (-3, -0.9) {\scriptsize{$I$}};
\node at (3,1.4) {\scriptsize{$I$}};
\node at (3, -1.4) {\scriptsize{$X$}};
\node at (4.3, -0.9) {\scriptsize{$I$}};
\node at (4.3, 0.9) {\scriptsize{$I$}};
\node at (9, 0.9) {\scriptsize{$I$}};
\node at (9, -0.9) {\scriptsize{$I$}};
\node at (-10, 3) {\small{control-1}};
\node at (-10, 0) {\small{target}};
\node at (-10, -3) {\small{control-2}};
\node at (-3,4) {\scriptsize{$(i, j-1)$}};
\node at (3,4) {\scriptsize{$(i+1, j)$}};
\node at (9,4) {\scriptsize{$(i+2, j+1)$}};
\node at (9,-4) {\scriptsize{$(i+1, j+2)$}};
\node at (-7,0.8) {\scriptsize{$(i-1, j-1)$}};
\node at (13,0.8) {\scriptsize{$(i+2, j+2)$}};
\node at (-3,-4) {\scriptsize{$(i-1, j)$}};
\node at (3,-4) {\scriptsize{$(i, j+1)$}};
\node at (0, 2) {\scriptsize{$(i,j,0)$}};
\node at (0, -2) {\scriptsize{$(i,j,1)$}};
\node at (6, 2) {\scriptsize{$(i+1,j+1,0)$}};
\node at (6, -2) {\scriptsize{$(i+1,j+1,1)$}};
\node at (-1.8,2.4) {\small{$a_1$}};
\node at (1.8,2.4) {\small{$b_1$}};
\node at (-1.8,-2.4) {\small{$c_1$}};
\node at (1.8,-2.4) {\small{$d_1$}};
\node at (6-1.8,2.4) {\small{$a_2$}};
\node at (6+1.8,2.4) {\small{$b_2$}};
\node at (6-1.8,-2.4) {\small{$c_2$}};
\node at (6+1.8,-2.4) {\small{$d_2$}};
\node at (-3,0) {\small{$p_1$}};
\node at (3,-0.7) {\small{$p_2$}};
\node at (9,0) {\small{$p_3$}};
\end{tikzpicture}
\caption{Direct Toffoli gate.}
\label{toffoliFig}
\end{figure*}

We now show that a Toffoli gate can be constructed in close analogy to the CNOT discussed in the previous subsection. The main idea is conveyed in Fig.~\ref{toffoliFig}. We consider the realization of a Toffoli gate in which the target qubit is \emph{sandwiched} between the two control qubits.  As seen in Fig.~\ref{toffoliFig} in order to do the Toffoli gate we need to modify a larger region compared to the CNOT. Specifically, we now split two sites in two. In the first plaquette $p_1$ we apply the first step of the CNOT choosing the control-$1$ particle as control. There is one difference, namely we do not immediately apply the $X$-gate, but we just conditionally direct the target particle to site $(i,j,0)$ or site $(i,j,1)$ depending on the state of the control-$1$ particle. After this first intermediate step, we continue to do conditional operations depending on the state of the second control particle, which, in the meantime, needs to have completed a hopping to site $(i,j+1)$ in order for the target particle to move forward. As an example, if the target particle is at $(i,j,1)$ and the second control is the $s=1$ state, the target particle will be allowed to only hop to the site $(i+1, j+1, 1)$ with the application of an $X$-gate, thus correctly realizing the logic of the Toffoli gate. The system works similarly in the remaining three cases in which the identity gate is applied. From the second split-sites $(i+1, j+1,0)$ and $(i+1, j+1,1)$ the target hops to the final single site at $(i+2, j+2)$ with the application of an identity gate. This construction of the Toffoli can be viewed as a double, sequential railroad switch. \par 
In Appendix \ref{sec:tof} we work these ideas out mathematically, showing that the effective Hamiltonian, obtained in lowest-order perturbation theory, applies the correct Toffoli gate logic.
It is clear that with the same idea we can implement any controlled-controlled-$U$ gate, just by substituting the hopping term that implements the $X$-gate with a hopping term that implements a generic single-qubit unitary as described in Sec.~\ref{reviewSec}. 

\subsection{Dual Rail Representation}
\label{dualRailSec}

We represent a particle with spin $s=0,1$ by a pair of transmon qubits (qubit $s=0$ or qubit $s=1$) with an excitation in either one of the qubits. In this dual-rail representation, the hopping terms across a plaquette $p$ implementing $U_p$, Eq.~(\ref{VhopPlaquette}), become simple flip-flop terms moving the excitation
\begin{equation*}
%\label{VhopPlaquetteQubits}
\sum_{s,s'} \bra{s \sp{\prime}} U_p \ket{s}\sigma_{s \sp{\prime}}^{+}[i+1, j+1]\sigma_{s}^{-}[i,j]+\mathrm{h.c.}
\end{equation*}
This mapping leads to the couplers shown in Fig.~\ref{circuitGates} where $U_p$ can be the $X$, $I$ or $H$ gate. 

In the dual-rail representation, the total number operator ${\bf n}[\nu]$ for a particle at site $\nu$ is the sum of the number operators for the two transmon qubits at the site, counting whether a single excitation is present for the pair or none. The internal-state dependent number operator $n_{s}[\nu]$ for $s=0,1$ just counts whether transmon qubit $s=0,1$ has an excitation or not. This implies that the standard attractive edge terms can be written as doubled cross-Kerr couplers as in Fig.~\ref{strongZZcoup}. For the CNOT, due to the split-site, one then requires several more cross-Kerr interactions as shown in Fig.~\ref{fig:CNOT-CKerr}. Realizing this connectivity is clearly a challenging aspect of our proposal.

If we use transmon qubits, we additionally have local $- \Omega \frac{\sigma^z}{2}$ terms in the Hamiltonian (see Eq.~(\ref{eq:transmon}) and Eq.~(\ref{HQ})) besides engineered flip-flop and cross-Kerr interactions. If all transmon qubits have identical frequencies, we can move to the rotating frame associated with that frequency without changing the Hamiltonian. This follows because any such rotating frame change leaves the cross-Kerr terms invariant while the on-track flip-flop interactions only depend on difference in frequencies between qubits on a track. Thus, in fact, it suffices to demand that all qubits on one track have the same frequency, in this case these extra single-qubit Z terms do not affect the dynamics.

As suggested in Table \ref{tab:numbers} and indicated with colors in Fig.~\ref{layoutConcept} we imagine keeping qubits on adjacent tracks quite far detuned to avoid spurious intertrack excitation hopping. 

Even the qubits on a single track can have small differences in frequencies: in a chosen rotating frame its implies that the hopping is slightly off-resonant and thus less effective. We numerically study this effect in Sec.~\ref{sec:dis}.
\section{Numerical Studies of Small Lattices}
\label{sec:error}

In this subsection we analyze errors in the model in Section \ref{reviewSec}, which originate from the fact that our effective Hamiltonian is arrived at in lowest-order perturbation theory.  We start by looking at the dynamics in the rotated grid without the application of any gates. We study the probability that the string becomes disconnected since we intuively expect it to relate to the probability of incorrect logic if we were to apply gates like the CNOT or Toffoli.

\begin{figure}[htb]
\centering
\includegraphics[scale=0.42]{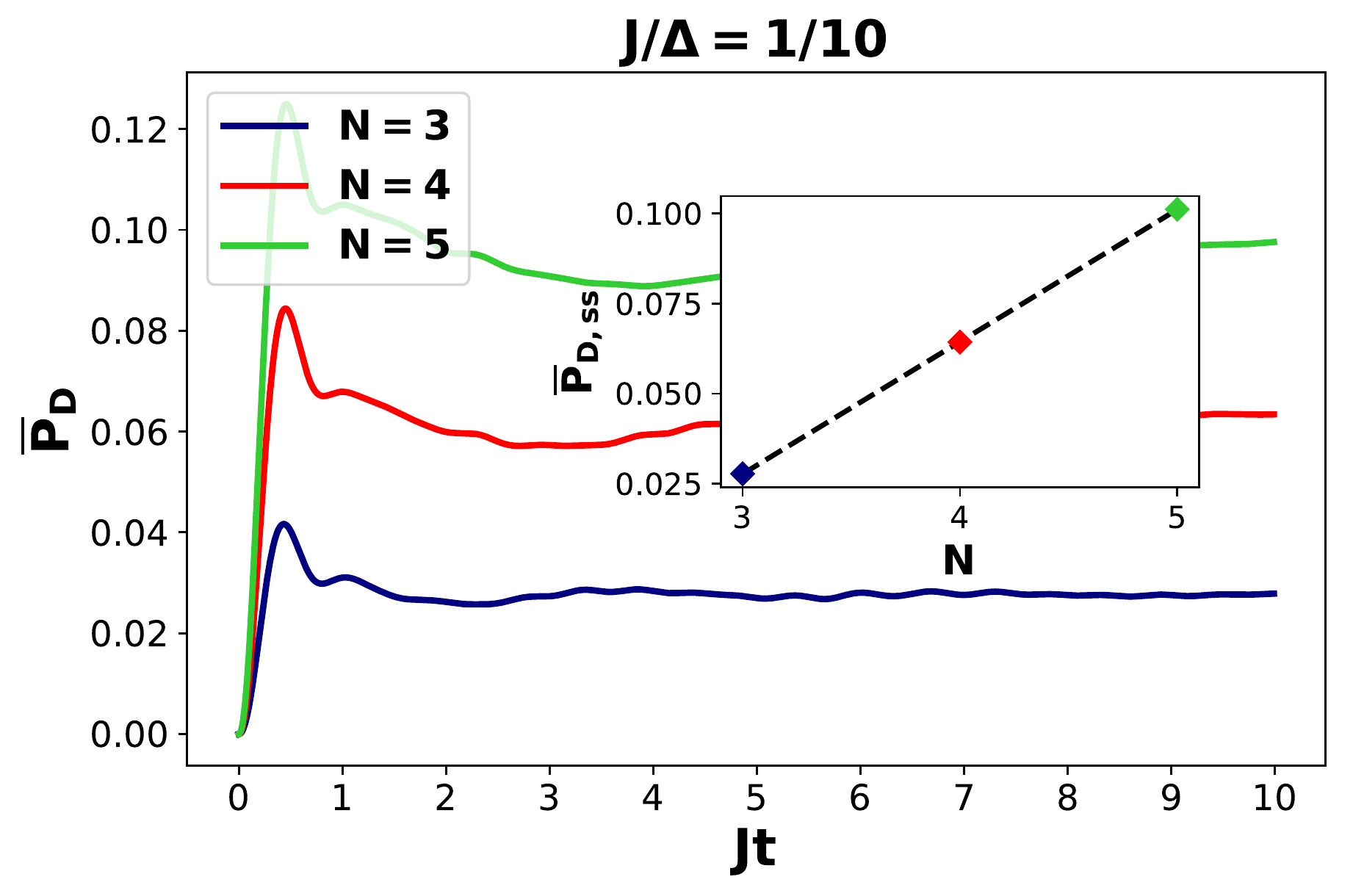}
\caption{Time-averaged probability $\overline{P}_D$ for the string to be disconnected as a function of time for different lattice sizes. The inset shows the scaling of the steady state probability $\overline{P}_{D, ss}$, evaluated at a final time, with $N$.}
\label{fig::Pdisc}
\end{figure} 

For smalll lattice sizes we have numerically studied the time-averaged probability for the string to be disconnected, defined as 
\begin{equation}
\label{eq::timeAVG}
\overline{P}_D(t)= \frac{1}{t}\int_0^t dt \sp{\prime} P_D(t \sp{\prime}).
\end{equation}
We consider the time-averaged probability since $P_D(t)$ itself is an oscillatory function in time, demonstrating that these string-disconnect errors are coherent errors not accumulating in time. 
Fig.~\ref{fig::Pdisc} shows this time-averaged probability $\overline{P}_D(t)$ as a function of time $t$ for different lattice sizes, fixing $J/\Delta=1/10$. We see that, unlike the instantaneous probability $P_D(t)$, the time-averaged probability reaches a steady value after a time of few $1/J$. In the inset we see that this steady-state average probability $\overline{P}_{D,ss}$ scales linearly with the lattice size for these lattice sizes.  \par 
\begin{figure}[htb]
\centering
\includegraphics[scale=0.43]{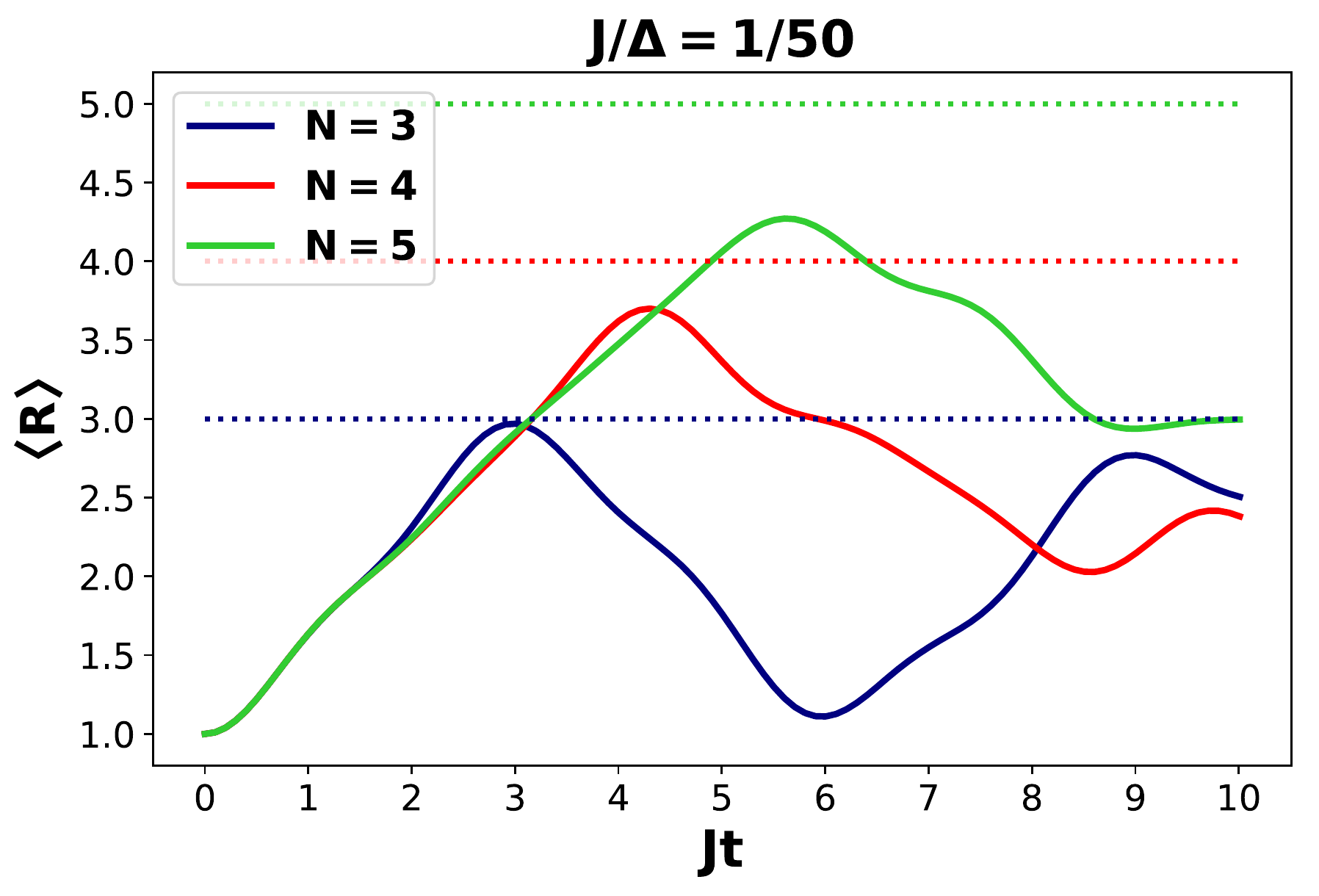}
\caption{Average position of the particle on the central track as a function of time for different lattice sizes $N$. The dashed horizontal lines identify the maximum position that can be reached for each lattice size. $\langle R \rangle=1$ is the initial position.}
\label{fig::xCT}
\end{figure}
\begin{figure}
\centering
\begin{subfigure}[t]{0.4 \textwidth}
\centering
\includegraphics[scale=0.28]{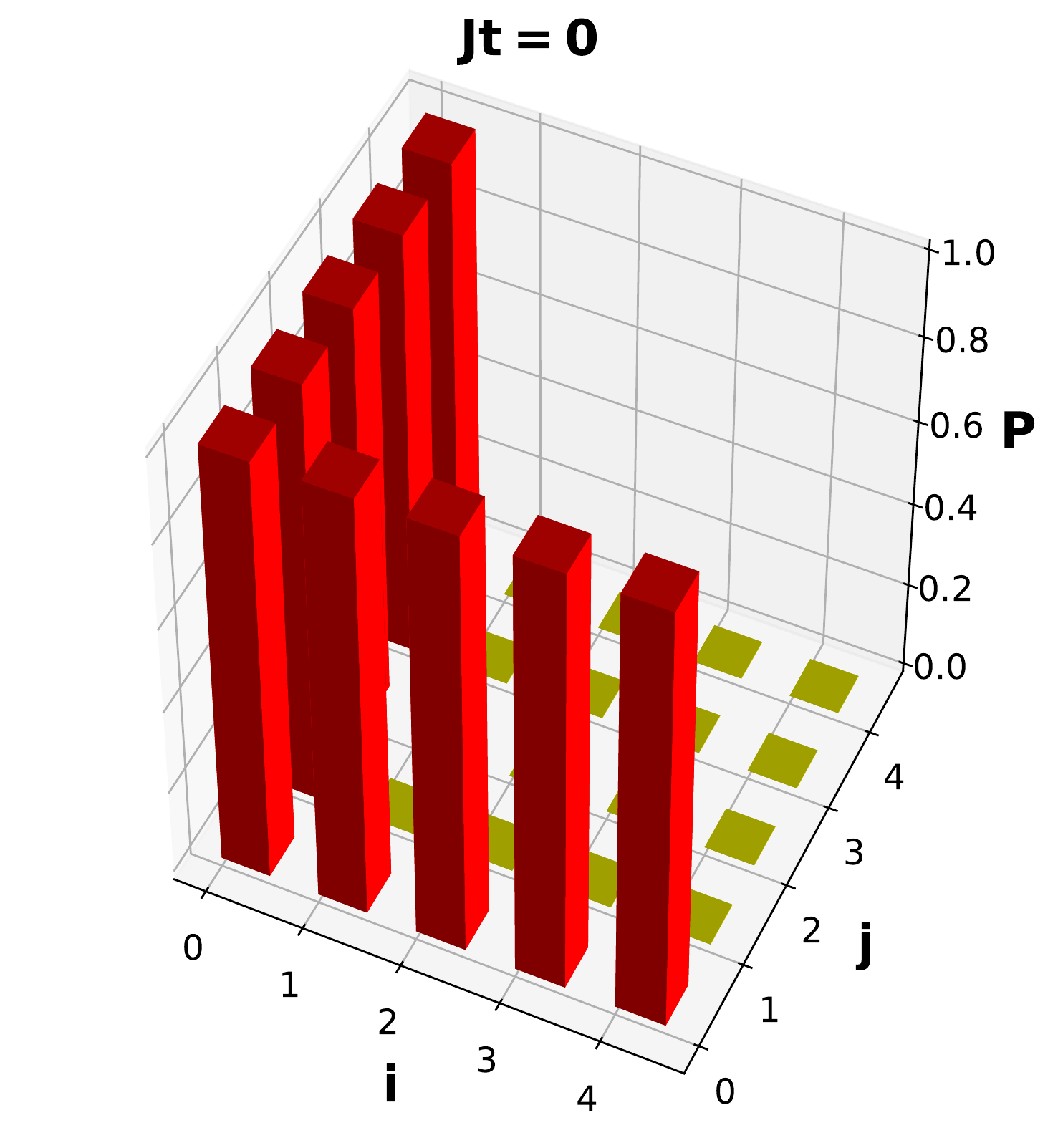}
\caption{}
\end{subfigure}
\begin{subfigure}[t]{0.4 \textwidth}
\centering
\includegraphics[scale=0.28]{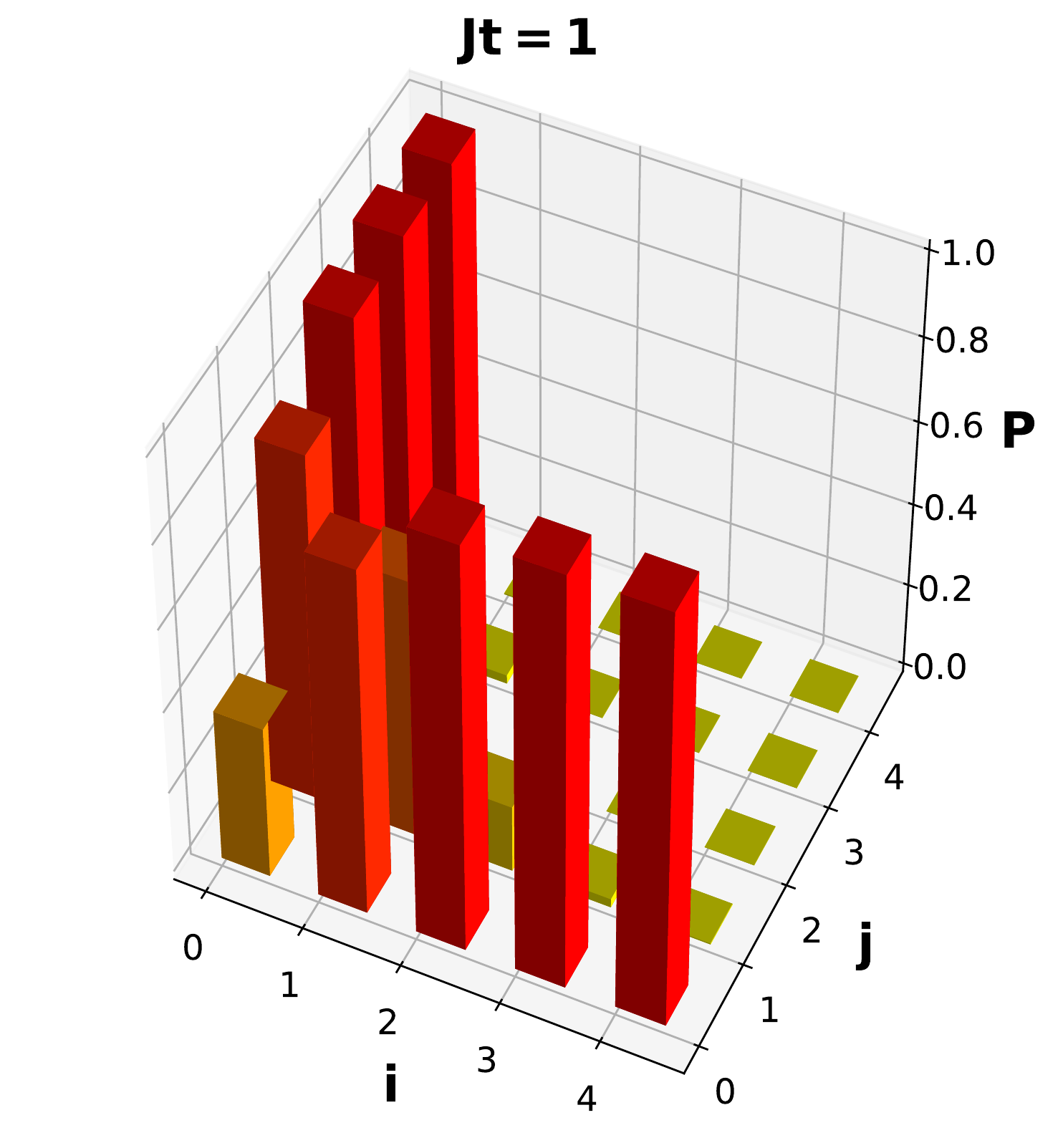}
\caption{}
\end{subfigure}
\begin{subfigure}[t]{0.4 \textwidth}
\centering
\includegraphics[scale=0.28]{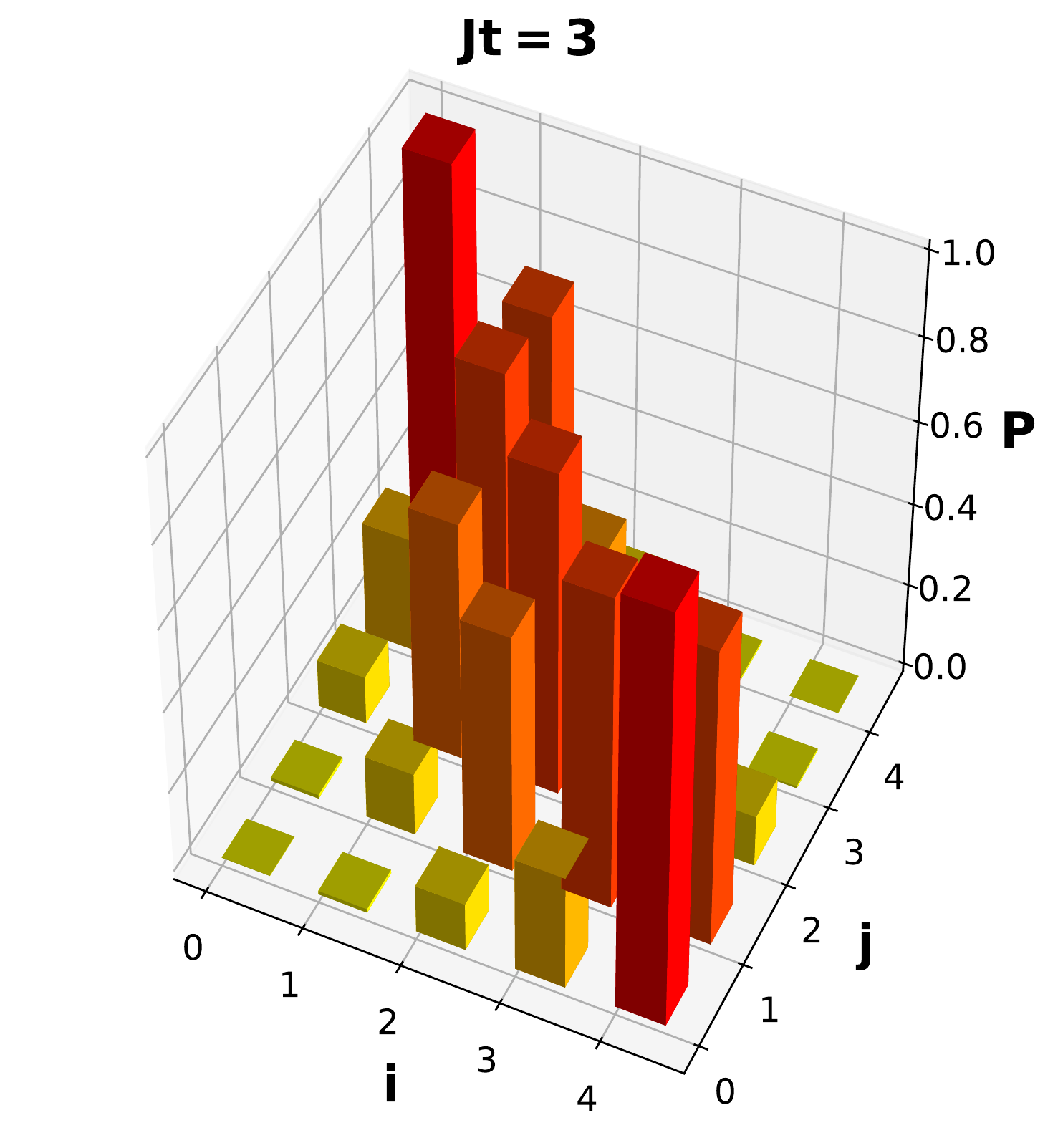}
\caption{}
\end{subfigure}
\begin{subfigure}[t]{0.4 \textwidth}
\centering
\includegraphics[scale=0.28]{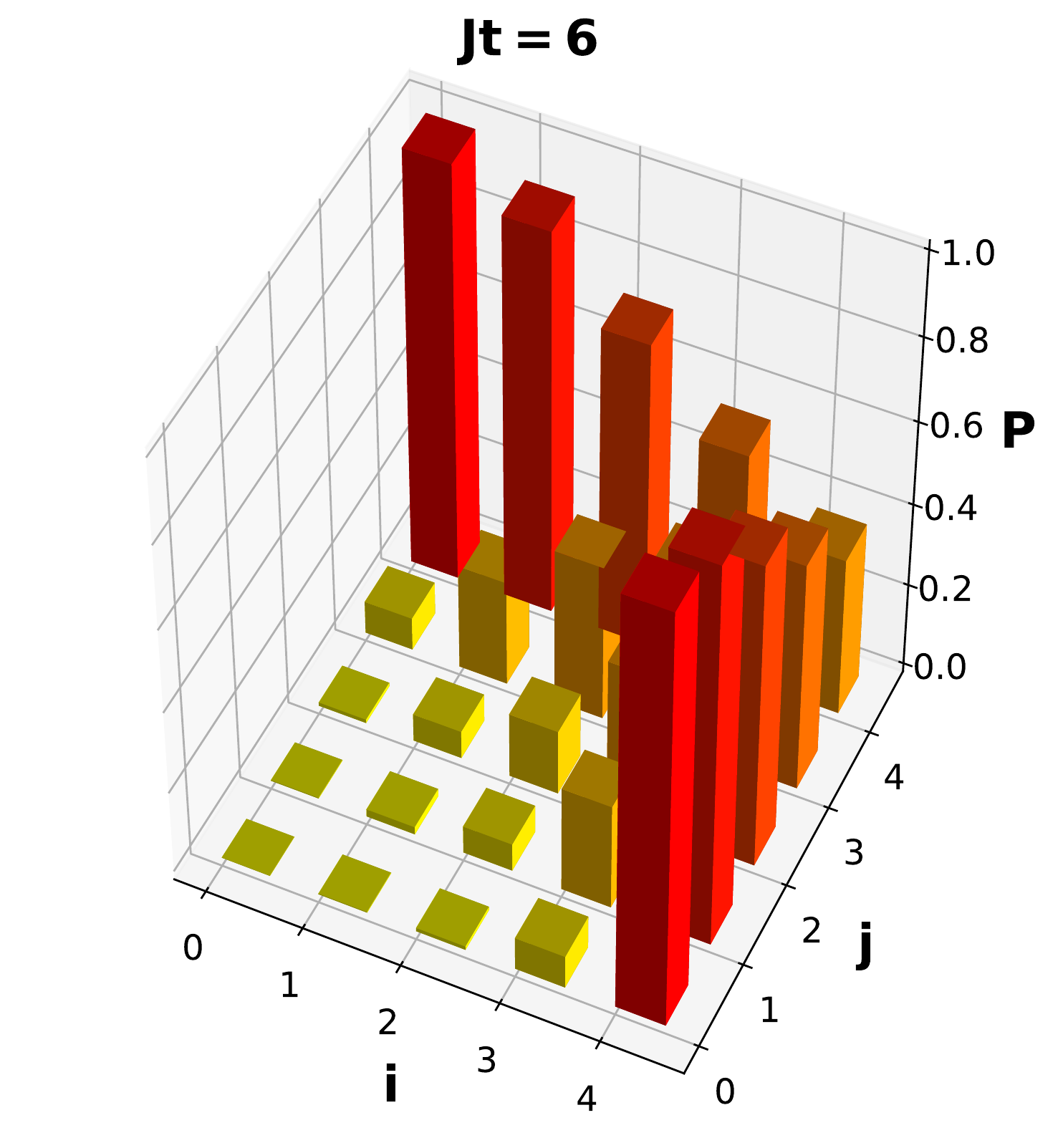}
\caption{}
\end{subfigure}
\caption{Plots of wavefront at different times. The bars represent the probability to find a particle in that position.}
\label{fig:screenWF}
\end{figure}
We further analyze the dynamics of the string of particles in Fig.~\ref{fig::xCT}. We plot the instantaneous average position of the particle on the central track, which is representative of the speed of the string. Formally $R=\sum_{i=1}^N \bm{n}[i,i]/N$. In this plot we take $J/\Delta$ to be quite small, \i.e. $J/\Delta=1/50$, so that the string is unlikely to be disconnected. We see that at short times of order $\sim 1/J$ the particle moves at a constant velocity, independent of lattice size, approximately equal to $0.6 /(J t)$ (lattice sites per unit of dimensionless time $Jt$). This agrees with the analysis of Ref.~\cite{gossetTerhal} in which, based on the solution for $N \rightarrow + \infty$, it is argued that the string should move at constant velocity. In Fig.~\ref{fig::longTimeDis} we see how the average position of the particle on the middle track oscillates over time for much longer times (in red). 

In Fig.~\ref{fig:screenWF} we further show some screenshots of the wavefront for lattice size $N=5$ at different short times. We see that the particles tend to stay together as expected and move forward in a correlated way. \par

We now numerically examine the effect of perturbative errors on the execution of a CNOT gate as in Fig.~\ref{cnotFig}. In our simulations we remove the spectator track at the bottom of Fig.~\ref{cnotFig}. We then initialize the control and the target particles on the left sites with theirs spins in one fixed basis state. Our main goal is to evaluate the probability that the CNOT is implemented succesfully or not. To this end we are interested in the probability that the particles are found on the right with the correct CNOT logic implemented on their internal states. 
\begin{figure}
\centering
\includegraphics[scale=0.42]{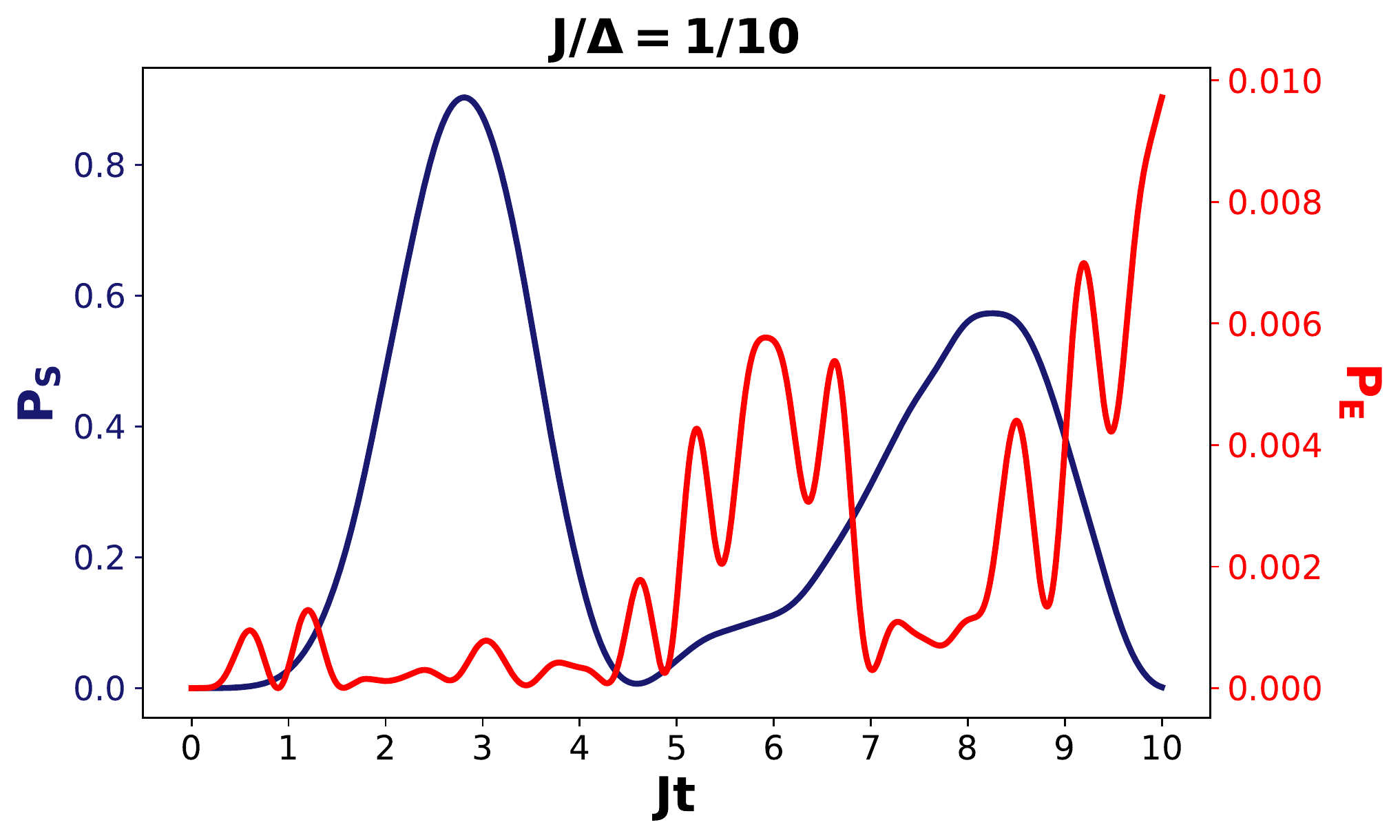}
\caption{Example of probability of success (blue line, left axis) and error (red line, right axis) as a function of time for $J/\Delta=1/10$.}
\label{fig::exPSPE}
\end{figure} 
We are interested in relatively short `first-passage' times $\sim 1/J$ since we know that the string moves at a speed $\sim 1/J$. 

We show a typical time evolution in Fig.~\ref{fig::exPSPE}, where we plot the probability of success $P_S$ and the probability of error $P_E$ defined as follows. $P_E$ is the probability of finding both particles on the very right side, at the end of the CNOT region, but with wrong internal states according to the CNOT logic, while the success probability $P_S$ is the probability for finding them on the right but in the correct state instead. Needless to say, the probability for arrival on the right $P_S+P_E \leq 1$. We see that the probability of success quickly increases, reaching a high maximum above at approximately $Jt=3$, while the error stays relatively low, although we start to see an increase at the end of the simulation, reaching $\approx 1 \%$. \par 

\begin{figure}[htb]
\centering
\includegraphics[scale=0.45]{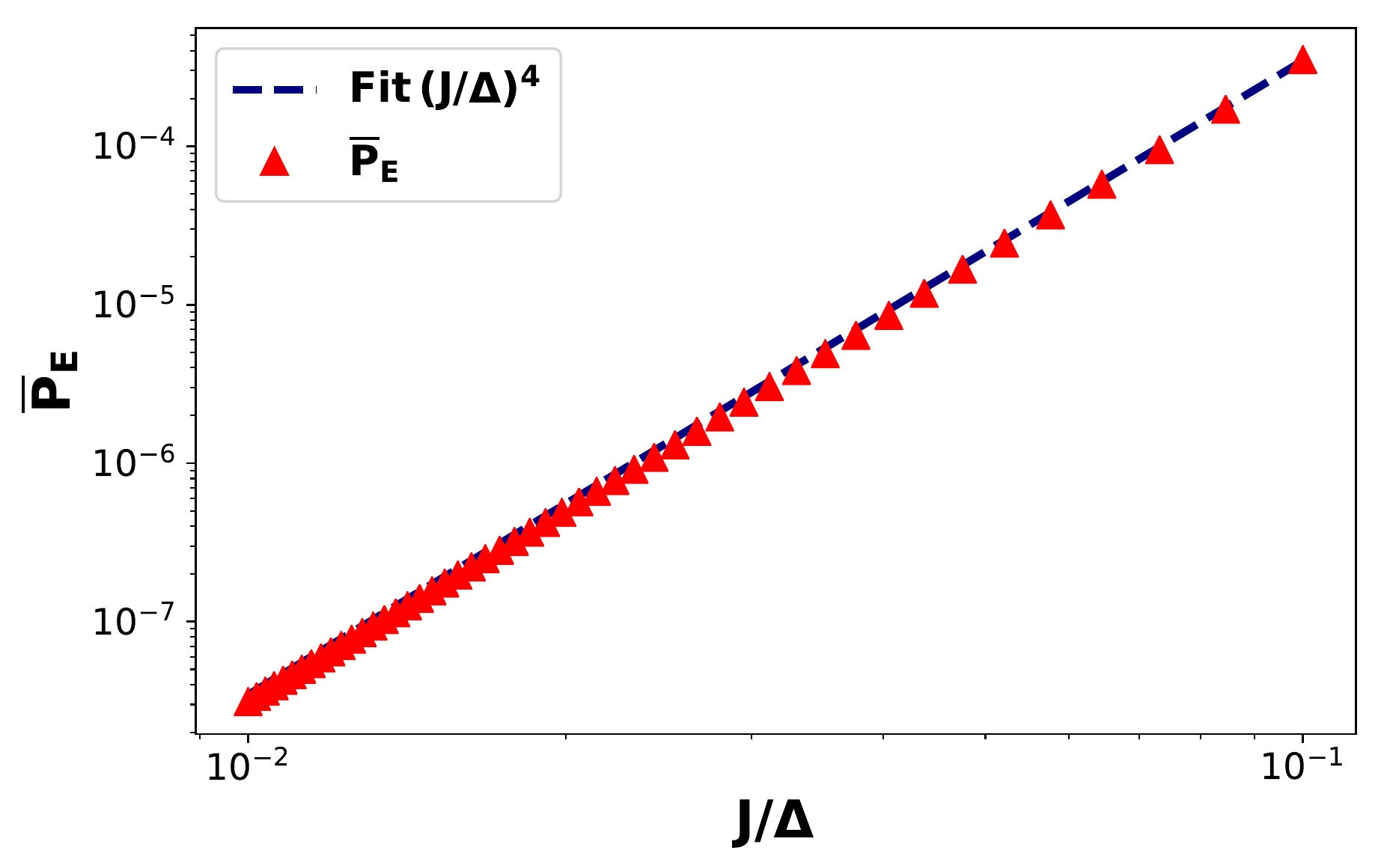}
\caption{Time-averaged probability of error, as a function of $J/\Delta$ for CNOT at $Jt=3$. We use a logarithmic scale on both axes.}
\label{errorScaling}
\end{figure}

In Fig.~\ref{errorScaling} we examine the scaling of the time-averaged error $\overline{P}_E(t)$ as a function of the ratio $J/\Delta$. Naturally, we expect this error to decrease when we decrease the ratio $J/\Delta$ while looking at the same $Jt$.  Fig.~\ref{errorScaling} shows that this probability scales as $(J/\Delta)^{4}$, and we find that this scaling is insensitive to our choice of $Jt$.

\par
\begin{figure}[htb]
\centering
\includegraphics[scale=0.45]{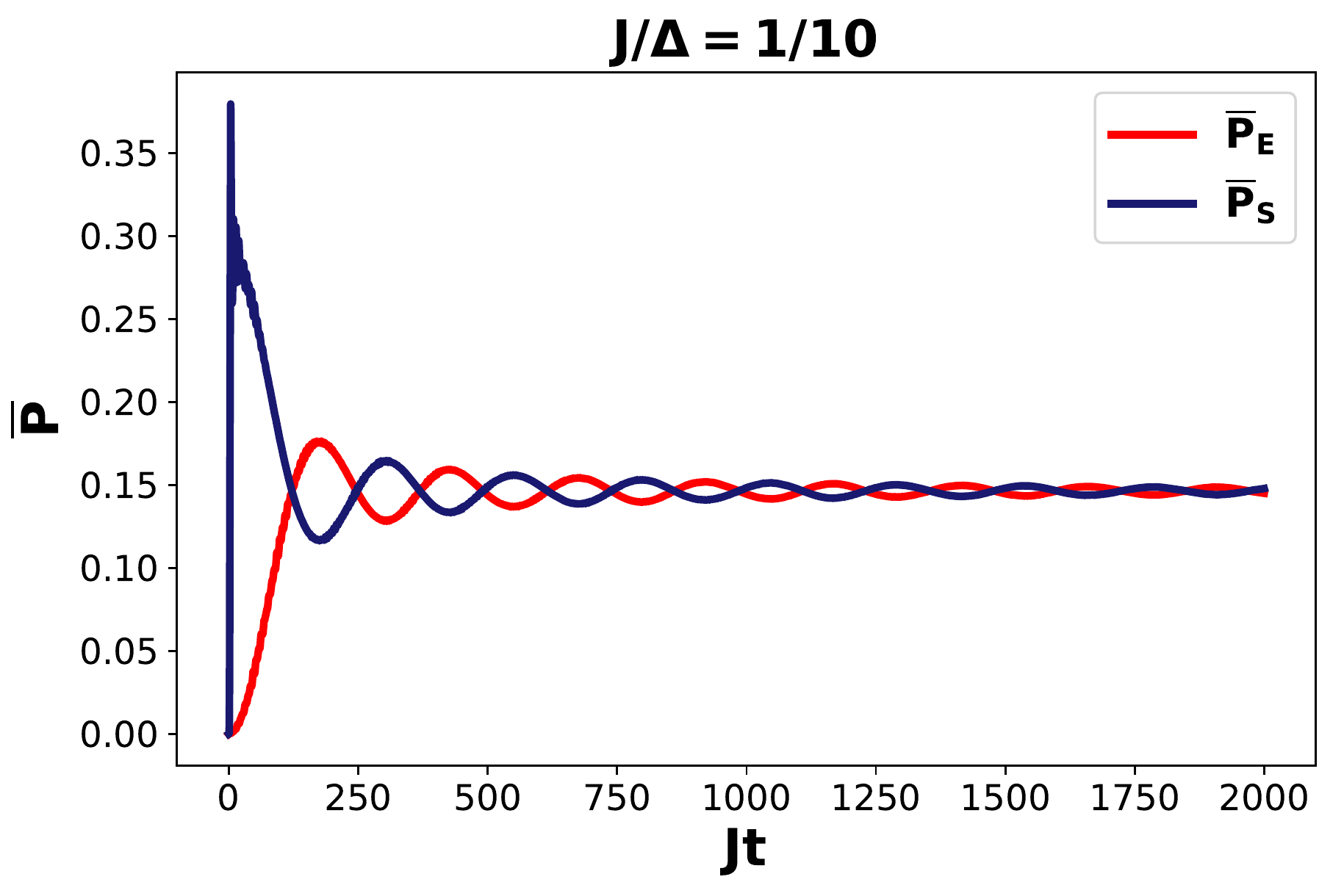}
\caption{Long-time behaviour for the time-averaged probabilities of error and success.}
\label{fig::longTimeError}
\end{figure}

 
It is important to understand that the CNOT construction works under the assumption that the forward motion of the computation is guaranteed, so that the gate is not done and then undone many times. Since we only simulate a small CNOT region of a few sites, such accumulation of error by doing and undoing the gate incorrectly cannot be prevented, and is thus not representative of the expected behavior of the CNOT embedded in a large lattice. For example, Fig.~\ref{fig::longTimeError} shows that when we average over very long times in our simulation, $\overline{P}_E$ and $\overline{P}_S$ become the same. This is a reflection of the fact that incorrect spin states and correct spin states at the sites on the right of the CNOT region have the same energy. It also shows that the probability that the CNOT gate fails becomes 1 in this long-time limit. 

\subsection{Static Disorder}
\label{sec:dis}
\begin{figure}[htb]
\centering
\begin{subfigure}[t]{0.4 \textwidth}
\centering
\includegraphics[scale=0.4]{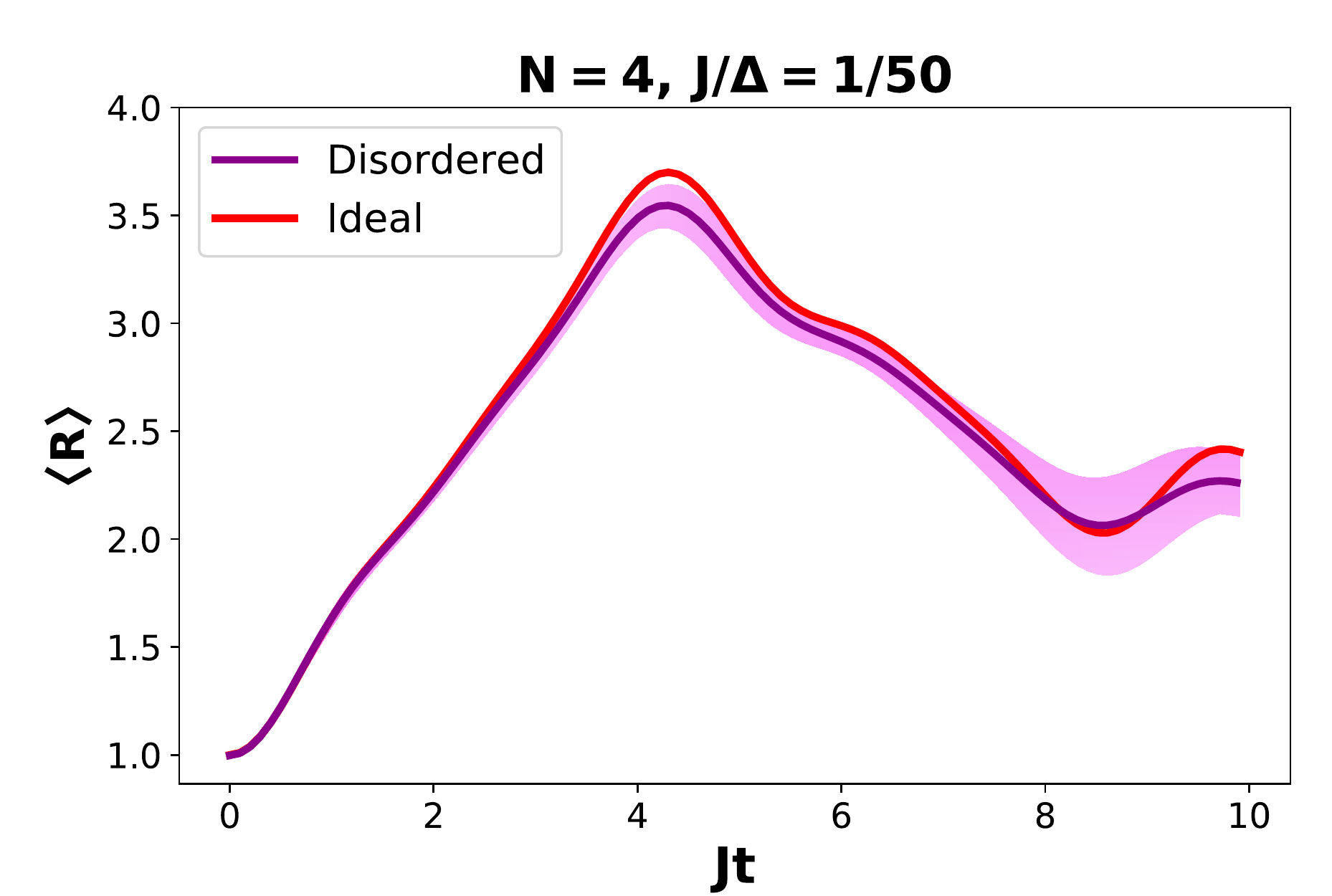}
\caption{Ideal time evolution vs. average time evolution with $10 \%$ Gaussian disorder on the hopping parameter $J$ (meaning that the standard deviation of the Gaussian is taken to be $\sigma=J \times 0.1$).}
\label{fig::disorderedEvA}
\end{subfigure}

\begin{subfigure}[t]{0.4 \textwidth}
\centering
\includegraphics[scale=0.4]{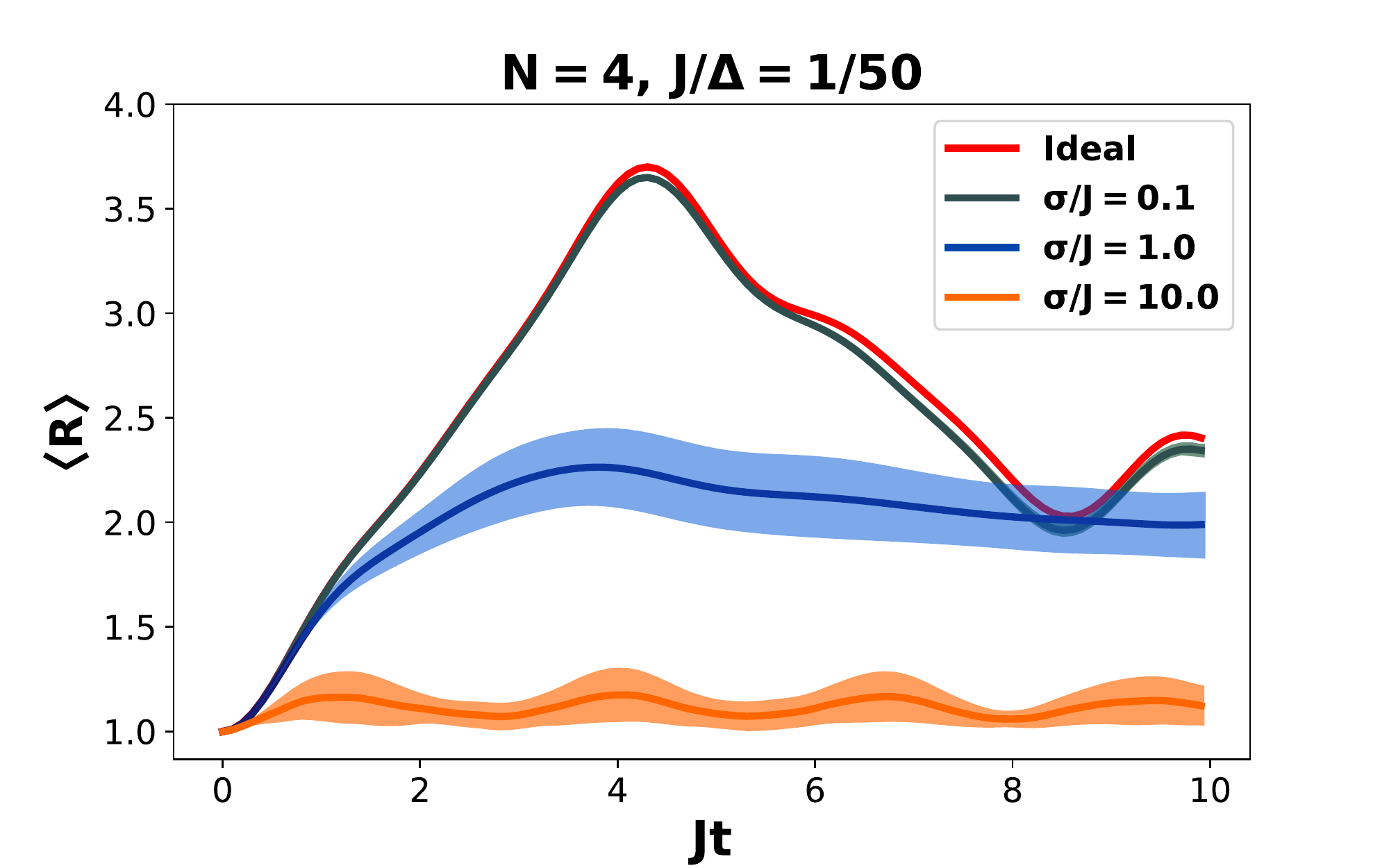}
\caption{Ideal time evolution vs. time evolution with Gaussian disorder on the on-site energy. For $\sigma/J=0.1$ the light shaded area is not visible by eye.} 
\label{fig::disorderedEvB}
\end{subfigure}

\caption{Ideal vs. disordered time evolution. The disordered data are averaged over $50$ simulation runs. The light shaded areas represent the standard deviation of the disordered data at each time, which shall not be confused with the standard deviation $\sigma$ of the distribution of $J$ . $\langle R \rangle=1$ is the initial position.}
\label{fig::disorderedEv}
\end{figure}

\begin{figure}
\centering
\includegraphics[scale=0.4]{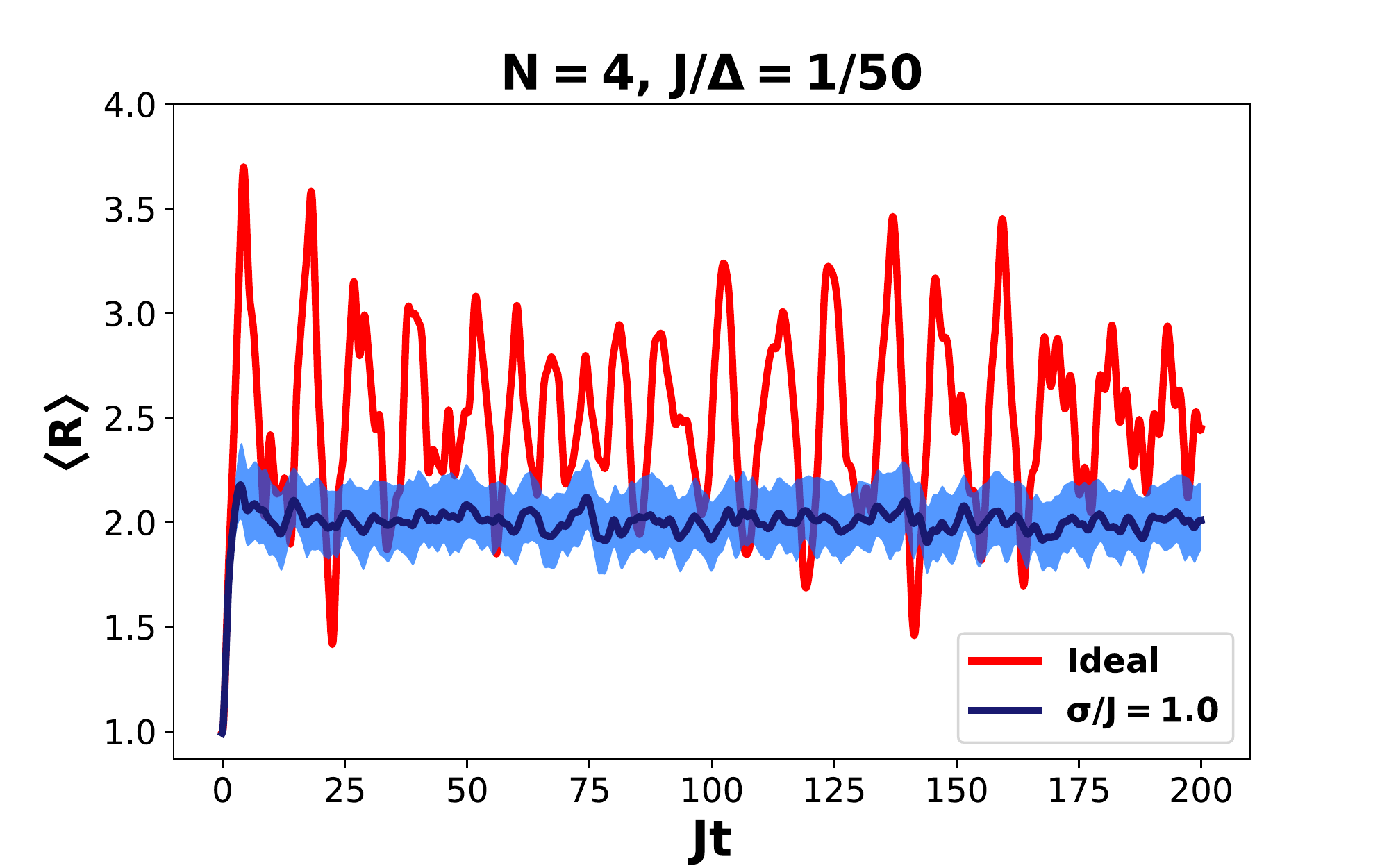}
\caption{Ideal long-time evolution vs. on-site disordered long-time evolution. As in Fig. \ref{fig::disorderedEv}, the light shaded area represents the standard deviation of the disordered data at each time.}
\label{fig::longTimeDis}
\end{figure}

We numerically examine how static disorder modifies the string dynamics. We focus on two kinds of disorder, namely variations in the hopping parameter $J$ and variations of the on-site energy. In the dual-rail representation, the latter corresponds to variations of the transmon qubit frequency between qubits on different sites, possibly on one track. This model does not consider disorder in the qubit frequencies of a pair of transmon qubits, which should in principle be equal. Said in the language of the model in Sec.~\ref{reviewSec}, we do not examine spin-dependent disorder.
It is known that on-site or hopping disorder of the dynamics of a single particle on a line leads to Anderson localization with a localization length which depends on the disorder strength. Here the question is to understand what localization of the string occurs at what disorder strength and when this localization would prevent the execution of the computation.

In Fig.~\ref{fig::disorderedEv} we show results for the case $N=4$: we compare the ideal time evolution of the average position of the particle on the central track with the disordered one. For disorder in the hopping parameter, we take $J$ to vary by roughly $10\%$, which is reasonable in our transmon implementation. In Fig~\ref{fig::disorderedEvA} we see that this amount of disorder slightly influences the dynamics of the string, but it still reaches the center of the grid with approximately the same velocity as in the ideal case. 
%Localization is also not observed for larger (unphysical) disorder of order $\sigma \sim J \times 10$. 

The situation is different in the case of disorder on the on-site energy as shown in Fig.~ \ref{fig::disorderedEvB}. As argued previously, when all on-site energies, translating into transmon qubit frequencies, are identical, one can rotate away this frame. When frequencies differ, the flip-flop coupling will be off-resonant and thus less effective in bringing about forward motion.
In Fig.~\ref{fig::disorderedEvB} we see that increasing the ratio between the standard deviation and the hopping parameter, we pass from a configuration in which there is essentially no localization ($\sigma/J=0.1$) to a localized configuration ($\sigma/J=10.0$). This tells us that in order for the string to propagate we need an on-track variation of the on-site energy that is $\le J$. In our transmon implementation and parameters chosen as in Table \ref{tab:numbers}, this means a variation of no more than $15 \, \mathrm{MHz}$ on the qubit frequencies, which can be achieved by careful design. \par 

Finally, Fig.~\ref{fig::longTimeDis} shows the long-time behaviour of the position of the central particle for the ideal case versus the case with disorder of the on-site energy. In the ideal case, while on average the particle is in the middle, it still presents oscillations inherent to the unitary dynamics. The averaged disorder evolution instead presents little oscillations and we also notice that average position is not exactly in the middle (the middle position is $\langle R \rangle=2.5$), but slightly closer to the initial position. 

From these numerics we cannot readily extrapolate what happens for larger $N$. The interesting effect of small disorder is to localize the string where indeed more localization occurs with more disorder. An open question is how the localized $\langle R \rangle$ depends on $N$, i.e. does the stationary value for $\langle R \rangle$ at fixed disorder strength decrease as a function of $N$ (and if so, as what function).

We have not included numerical studies of disorder on the cross-Kerr coupling. As long as $\Delta/J$ is sufficiently large to warrant the perturbative picture, we do not expect such disorder to play a large role in the string dynamics.

%BMT AC did you write something on disorder in cross-Kerr elsewhere, if so, then this has to be in line with what is above..
% AC: no I didn't but I would say that it has the same effect as disorder on the on-site energy

\section{Cross-Kerr and flip-flop couplings between superconducting transmon qubits}
\label{transmonImpl}

In this section we describe details of the required couplers to implement the Hamiltonian quantum computing scheme using superconducting transmon qubits. 

\subsection{Cross-Kerr interaction}
\label{zzIntSec}
% BMT this section needs  a bit more physics/intuition description
% it also needs/has (?) a discussion of why this element is more scalable than the Walraff/Steele version
% and what limits practical implementation, requiring transmons to be grounded, no capacitive 
%coupling to ground..

% hbar=1 from the very start, 2 pi factors..

%  off-resonant hopping...
%AC I slightly modified the text

The most challenging interaction to engineer in our problem is the strong cross-Kerr interaction between transmon qubits. There are several proposals for the implementation of this kind of interaction with transmon qubits in the literature \cite{braumuller2016, reiner2016,neumeierLeib, jinRossini, MARCOS2014634, Kounalakis2018, leibZoller}. In particular, we will build on ideas similar to Refs. \cite{neumeierLeib, jinRossini, MARCOS2014634, Kounalakis2018}, in which a simple junction in parallel with a capacitance is used to implement the desired cross-Kerr interaction. This element also induces a linear flip-flop interaction which is undesired. However, the capacitance and the Josephson junction give rise to flip-flop terms of opposite signs and consequently we can find a value of the capacitance that exactly cancels the linear term. This is basically the same idea of an LC filter, which is based on the fact that the impedance of a capacitor and an inductor have different signs \cite{pozar}.  \par

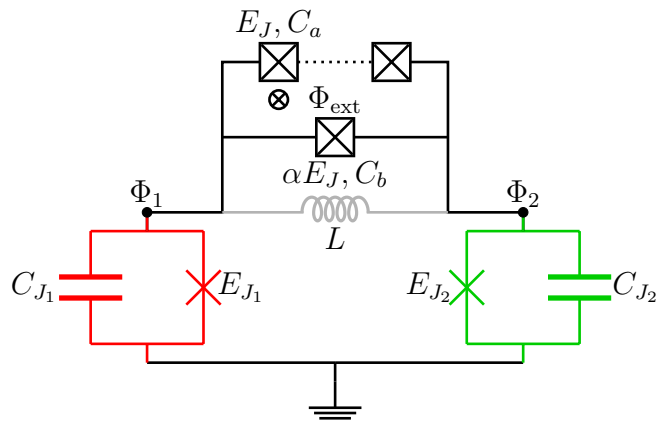
\begin{figure}
\centering
\begin{circuitikz}[scale=0.5, american voltages]
\draw[line width=1.0] (0,0)
to [short] (-5,0);
\draw[line width=1.0, color=red] (-5,0)
to [short] (-5, 0.5)
to [short] (-6.5, 0.5)
to [C] (-6.5, 3.5)
to [short] (-3.5, 3.5)
to [short] (-3.5, 0.5)
to [short] (-5, 0.5);
\draw[line width=3.0, black] (-3.5, 2) node[cross=7.5,color=red,rotate=0, line width=1.0] {};
\draw[line width=1.0] (0,0)
to [short] (5,0);
\draw[line width=1.0, color=green!80!black] (5,0)
to [short] (5, 0.5)
to [short] (6.5, 0.5)
to [C] (6.5, 3.5)
to [short] (3.5, 3.5)
to [short] (3.5, 0.5)
to [short] (5, 0.5);
\draw[line width=3.0, black] (3.5, 2) node[cross=7.5,color=green!80!black,rotate=0, line width=1.0] {};
%array
\draw [line width=1.0] (-5,3.5)--(-5, 4)--(-3, 4)--(-3, 8)--(-2, 8);
\draw [line width=1.0, color=red] (-5,3.5)--(-5, 4);
\draw [line width=1.0] (-2, 7.5) rectangle (-1, 8.5);
\draw[line width=3.0, black] (-1.5, 8) node[cross=7.5,color=black,rotate=0, line width=1.0] {};
\draw [line width=1.0] (5,3.5)--(5, 4)--(3, 4)--(3, 8)--(2, 8);
\draw [line width=1.0, color=green!80!black] (5,3.5)--(5, 4);
\draw [line width=1.0] (2, 7.5) rectangle (1, 8.5);
\draw[line width=3.0, black] (1.5, 8) node[cross=7.5,color=black,rotate=0, line width=1.0] {};
\draw [line width=1.0, dotted] (-1, 8)--(1,8);
%junction
\draw [line width=1.0] (-3, 6)--(-0.5, 6);
\draw [line width=1.0] (-0.5, 5.5) rectangle (0.5, 6.5);
\draw[line width=3.0, black] (0, 6) node[cross=7.5,color=black,rotate=0, line width=1.0] {};
\draw [line width=1.0] (3, 6)--(0.5, 6);
%inductance
\draw [line width=1.0, color=gray!60!white] (-3, 4) 
to [short] (-2, 4)
to [L, line width=0.6, color=gray!60!white] (2,4)
to [short] (3,4);
\node at (-8, 2) {$C_{J_1}$}; 
\node at (8, 2) {$C_{J_2}$}; 
\node at (-2.5, 2) {$E_{J_1}$};
\node at (2.5, 2) {$E_{J_2}$};
\node at (0, 5) {$\alpha E_J, C_{b}$};
\node at (0, 3.3) {$L$};
\node at (-1.5, 9) {$E_J, C_{a}$};
\node at (0, 7) {$\Phi_{\rm ext}$};
\draw[black,fill=black] (-5,4) circle (0.7ex);
\draw[black,fill=black] (5,4) circle (0.7ex);
\node at (5, 4.5) {$\Phi_{2}$};
\node at (-5, 4.5) {$\Phi_{1}$};
\draw[line width=1.0] (0,0)
to (0,-0.5) node [ground]{};
\draw[line width=1.0] (-1.5, 7) circle (7pt);
\draw[line width=3.0, black] (-1.5, 7) node[cross=3.5,color=black,rotate=0, line width=1.0] {};
\end{circuitikz}
\caption{Attractive cross-Kerr coupler between two grounded transmon qubits. The parameter $\alpha < 1$ is the ratio of the (small) black-sheep junction Josephson energy versus the Josephson energy of a junction in the array. The different colors of the transmons denote that they should have different frequencies as shown in the color scheme in Fig. ~\ref{layoutConcept}. Additionally, the total coupling capacitance is $C_c= C_a/N_J+ C_b$. The inductance is shown in grey as it is not an essential element of the coupler and could be omitted. $\Phi_1$ and $\Phi_2$ represent the node flux variables, while $\Phi_{\rm ext}$ the external flux in the loop formed by the junction and the array of junctions.}
\label{capFig}
\end{figure} 

In order to limit the value of the capacitance needed to achieve this cancellation, and thus the cross-talk between the qubits on the same track (see discussion in Subsec. \ref{subsec::CKstrength}), we consider, instead of a simple junction, a modified direct coupler shown in Fig.~\ref{capFig}. It represents two transmons coupled by a junction in parallel with an array of $N_J$ junctions and an inductance $L$.
% BMT AC DPDV but doesn't the inductance put some constraints on some space/size. 
% AC: I wouldn't say fundamental ones, for the range of inductances we are talking here \sim nanoHenry. For instance, they are used in the "gmon" architecture, although there they are smaller there. To make them more compact one could in turn use an array of junctions (many junctions).

%BMT2 I don't know what is modified about this coupler
%AC: I am referring to the simple junction case, I added few words.
The transmons have different frequencies since they sit on adjacent tracks according to our scheme in Fig.~\ref{layoutConcept}.  In the following paragraphs we show how to obtain the Hamiltonian of two coupled transmon qubits from the circuit in Fig.~\ref{capFig} through various approximations: the final Hamiltonian is in Eq.~(\ref{HQ}). 

We assume that the array of junctions is operated in the limit of $E_J/E_C  \gg1$, with $E_{Ca}= e^2/(2 C_a)$, as well as the limit in which the resonant frequencies of the array are larger than any other frequency of the problem. In this limit, the internal degrees of freedom of the array can be eliminated and the array can be treated with an effective potential
\begin{equation}
\label{uArray}
U_{{\rm array}, m}(\varphi)= -N_J E_J \cos \biggl(\frac{\varphi+2 \pi m}{N_J}\biggr),
\end{equation}
with $\varphi$ the superconducting phase difference across the element and $m \in \{0, 1, \dots, N_J-1\}$ \cite{ManucharyanPhd, manucharyanPhaseSlip, rastelliPop, nazarov_blanter_2009}. 
% BMT AC do we have an implementation of this already, how big? Fluxonia have up to 80 junctions.
% BMT Ok
The effective Hamiltonian depends on the parameter $m$, which labels the different classical metastable minima of the total potential of the array in the limit of $E_J/E_{Ca} \rightarrow + \infty$. In the phase-slip model of the array of junctions, first discussed in Ref.~\cite{matveev2002}, we associate a quantum state $\ket{m}$ with energy given by Eq.~(\ref{uArray}) to each metastable minimum of the array. The index $m$ of a minimum represents the number of vortices, \i.e., the number of $2 \pi$-turns that the phase along the array undergoes. 

In the limit of $E_J/E_{Ca} \gg 1$ (classical limit), we expect that given a certain state $\ket{m}$ the amplitude of tunneling to a different $\ket{m \sp{\prime}}$ is small and we can effectively assume that the potential is given by Eq.~(\ref{uArray}). This is exactly the same working regime of the superinductances used for the fluxonium qubit \cite{Manucharyan113}. In what follows we will assume that the array of junctions is initially set in the state $\ket{m=0}$ \footnote{Note that we could also start with a different $\ket{m}$ and obtain the same results, but with external fluxes shifted by an integer of $2 \pi$.}, so that our effective array potential is (see also Ref. \cite{richerPop})
\begin{equation*}
U_{{\rm array},0}(\varphi)=-N_J E_J \cos \frac{\varphi}{N_J} 
% \approx -N_J E_J+\frac{E_J \varphi^2}{2 N_J}
\end{equation*}

The Josephson junction array will also add an additional capacitance given by $C_a/N_J$, with $C_a$ the capacitance of a single junction in the array. The total coupling capacitance is then $C_c= C_a/N_J +C_b$, with $C_b$ the capacitance of the small (black sheep) junction in parallel. It is worth mentioning that a system composed of an array of three large junctions in parallel with a black sheep junction has been analyzed in Refs.~\cite{snail, voolSnail}, and nicknamed the SNAIL, with the goal of obtaining a potential that gives rise to a three-wave mixing term (third-order in annihilation/creation operators), but without cross-Kerr (quartic). Here instead we would like to limit the quadratic term, while keeping the cross-Kerr interaction. \par 

We obtain the Lagrangian of the circuit in Fig.~\ref{capFig} as
\begin{equation*}
\mathcal{L}= \frac{C_{J_1}}{2}\dot{\Phi}_1^2+\frac{C_{J_2}}{2}\dot{\Phi}_2^2 +\frac{C_c}{2}(\dot{\Phi}_1-\dot{\Phi}_2)^2- U_{\rm tot}(\Phi_1,\Phi_2),
\end{equation*}
with the total potential 
\begin{multline}
U_{\rm tot}(\Phi_1,\Phi_2)= -E_{J_1} \cos \biggl[ \frac{2 \pi}{\Phi_0} \Phi_1 \biggr] -E_{J_2} \cos \biggl[ \frac{2 \pi}{\Phi_0} \Phi_2 \biggr] +  \frac{1}{2 L} \biggl(\Phi_1-\Phi_2 \biggr)^2- \\ \alpha E_J \cos \biggl[\frac{2 \pi}{\Phi_0}(\Phi_1-\Phi_2)\biggr] - N_J E_J \cos \biggl[\frac{2 \pi}{\Phi_0} \frac{\Phi_1-\Phi_2+\Phi_{\rm ext}}{N_J} \biggr].
\label{eq:tot-pot}
\end{multline}
We define the conjugate variables $Q_{1,2}= \frac{\partial \mathcal{L}}{\partial \dot{\Phi}_{1,2}}$, in terms of which the Hamiltonian (obtained via a Legendre transform) reads
\begin{equation*}
H= \frac{Q_1^2}{2 \tilde{C}_{J_1}}+\frac{Q_2^2}{2 \tilde{C}_{J_2}}+ \frac{Q_1 Q_2}{\tilde{C}_{c}}+U_{\rm tot}(\Phi_1,\Phi_2),
\end{equation*}
where we defined the capacitances 
\begin{eqnarray*}
\frac{1}{\tilde{C}_{J_1}}= \frac{C_{J_2}+C_c}{{\rm det}({\bf C})} ,\;& \dfrac{1}{\tilde{C}_{J_2}} = \dfrac{C_{J_1}+C_c}{{\rm det}({\bf C})} ,\;& \frac{1}{\tilde{C}_c}= \frac{C_c}{{\rm det}({\bf C})},
\end{eqnarray*}
with the determinant of the capacitance matrix ${\bf C}$ as
\begin{equation*}
{\rm det}({\bf C})=C_{J_1} C_{J_2}+(C_{J_1}+C_{J_2})C_c.
\end{equation*}

We rewrite $Q_{1,2}= 2 e \mathfrak{n}_{1,2}$ with $\mathfrak{n}_m$ the number of Cooper pairs and the superconducting phases $\varphi_{1,2}= 2 \pi \Phi_{1,2}/\Phi_{0}$, In order to quantize the problem we promote $\mathfrak{n}_{m}$ and $\varphi_{m}$ to operators with commutation relation $[\hat{\varphi}_m, \hat{\mathfrak{n}}_m]=iI $, but we will continue to write $\hat{\mathfrak{n}}$ as $\mathfrak{n}$ and $\hat{\varphi}$ as $\varphi$.

\begin{figure}[htb]
\centering
\includegraphics[scale=0.4]{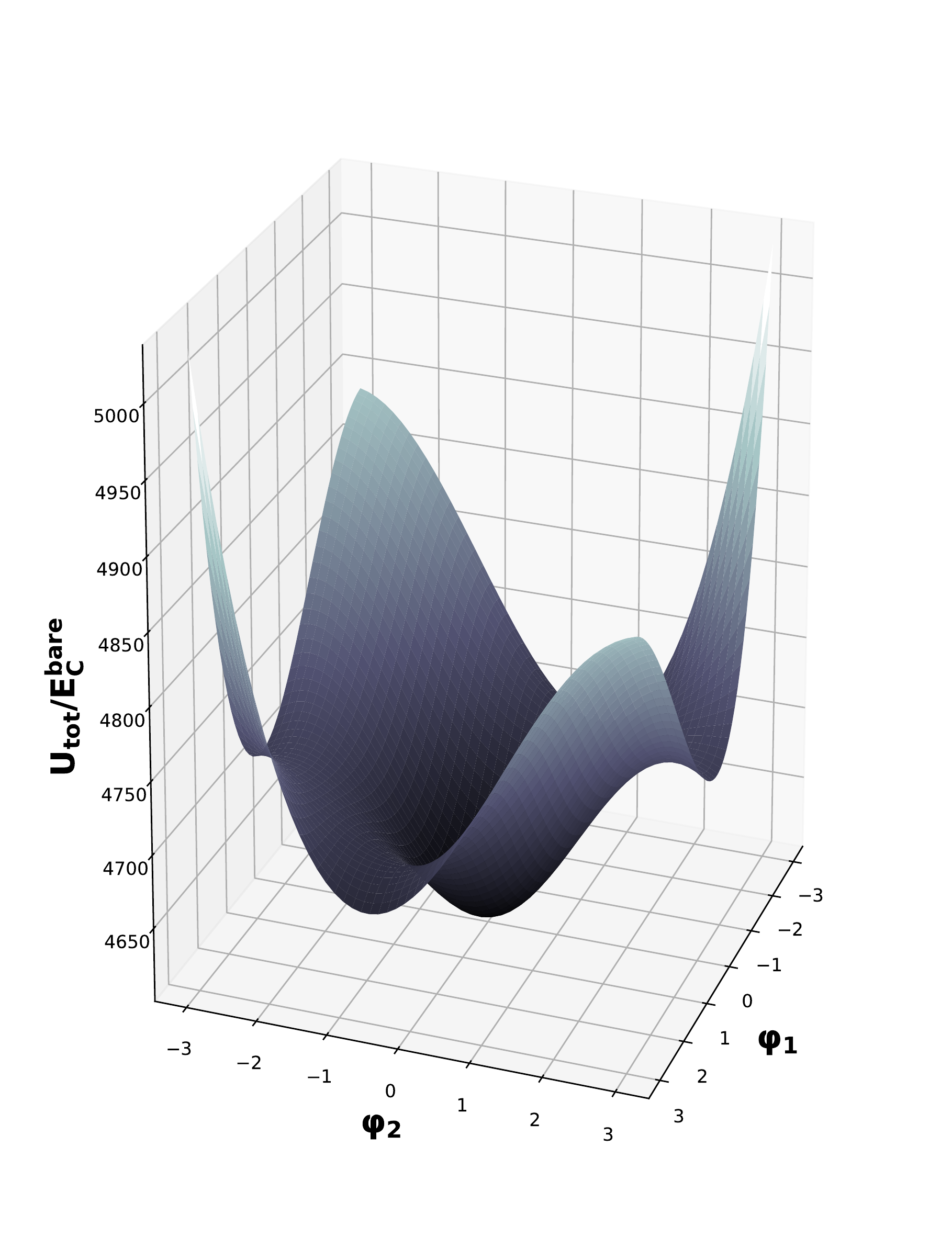}
\caption{Total potential $U_{\rm tot}(\varphi_1, \varphi_2)$ for the parameters in Table \ref{tableParameters} where $\alpha$ is set to $0.043$, showing that the global minimum occurs at the origin.}
\label{totPot}
\end{figure}

Using this notation, we rewrite our Hamiltonian as
\begin{equation}
\label{totHam}
H=4 E_{C_1} \mathfrak{n}_1^2+4 E_{C_2} \mathfrak{n}_2^2+8 E_{\rm cap}^{\rm coup} \mathfrak{n}_1 \mathfrak{n}_2 +U_{\rm tot}(\varphi_1,\varphi_2),
\end{equation}
with charging energy $E_{C_m}= e^2/(2 \tilde{C}_{J_m})$ and a capacative coupling energy between the two transmons equal to 
\begin{equation}
E_{\rm cap}^{\rm coup}=e^2/(2 \tilde{C}_c).
\label{def:Ecap}
\end{equation}
Let us now focus on the coupling part of the potential $U_{\rm tot}$ in terms of the phase difference $\varphi= \varphi_1 - \varphi_2$ at the nodes in Fig.~\ref{capFig}:
\begin{equation*}
U_{c}(\varphi)= \frac{E_L}{2} \varphi^2-\alpha E_{J} \cos \varphi-N_J E_J \cos \biggl( \frac{\varphi+\varphi_{\rm ext}}{N_J} \biggr),
\end{equation*}
Here we introduced the inductive energy $E_L= \Phi_0^2/(4 \pi^2 L)$.  Additionally, we fix the external flux to the value $\varphi_{\rm ext}= 2 \pi \Phi_{\rm ext}/\Phi_0=N_J \pi$. We will now assume that it is possible to Taylor-expand the coupling potential up to fourth order, as it is usually done for transmon qubits. This is a good approximation as long as we work in the transmon regime $E_{J_m}/E_{C_m} \gg 1$. In addition, we should also guarantee that the total potential has a global minimum at the origin. Expanding the coupling potential around $\varphi=0$ gives 
\begin{equation*}
U_{c}(\varphi)/E_J= -\alpha+N_J+\frac{1}{2} \biggl[\alpha+\frac{E_L}{E_J}-\frac{1}{N_J} \biggr]\varphi^2- \frac{1}{24} \biggl[\alpha-\frac{1}{N_J^3} \biggr]\varphi^4+\mathcal{O}(\varphi^6).
\end{equation*} 
We see that by setting 
\begin{equation}
\alpha+\frac{E_L}{E_J}-\frac{1}{N_J}=0.
\label{cancel}
\end{equation}
the quadratic term vanishes completely, while the quartic term would be approximately the same as the case with a simple junction, since the contribution due to the array scales as $1/N_J^3$. One should keep in mind that this point is a maximum of the coupling potential, but always (at least) a relative minimum of the total potential. 
In fact, the reason for including the inductor $L$ in the circuit of Fig.~\ref{capFig} is to make sure that the origin is a global minimum of the total potential, and not only a metastable minimum. In Fig.~\ref{totPot} we plot an example of this total potential showing that it has a global minimum in the origin and this holds true for all the parameters that we have considered in the next subsection. However, the inductance is not necessary if we accept to work in a metastable minimum.

As we will see, we shall aim to meet the condition in Eq.~(\ref{cancel}) only approximately as the nonlinearity $\varphi^4$ renormalizes the hopping strength in the transmon qubit subspace. 
% BMT2 change
It is convenient to define the following energies
\begin{subequations}
\begin{equation}
E_{\rm ind}^{\rm coup}= E_J \biggl(\alpha+\frac{E_L}{E_J}-\frac{1}{N_J} \biggr),
\label{eq:ind-coup}
\end{equation}
\begin{equation}
E_{\rm Kerr}^{\rm coup}= E_J \biggl(\alpha-\frac{1}{N_J^3} \biggr).
\end{equation}
\end{subequations}
We now perform a Taylor expansion for the total Hamiltonian in Eq.~\ref{totHam} and rewrite it as
\begin{equation*}
H= \sum_{m=1}^2 H_m +H_{\rm lin}+H_{CK} +H_{\rm non.lin},
\end{equation*}
where we defined the Hamiltonian of the single transmons as
\begin{equation*}
H_m= 4 E_{C_m} \mathfrak{n}_m +\frac{E_{L_m}}{2} \varphi_m^2-\frac{E_{K_m}}{24} \varphi_m^4,
\end{equation*}
with for $m=1,2$
\begin{equation}
E_{L_m}=E_{J_m}+E_{\rm ind}^{\rm coup}, E_{K_m}= E_{J_m}+E_{\rm Kerr}^{\rm coup}.
\label{eq:defLK}
\end{equation}
The linear coupling Hamiltonian $H_{\rm lin}$ is given by 
\begin{equation*}
H_{\rm lin}= 8 E_{\rm cap}^{\rm coup} \mathfrak{n}_{1}\mathfrak{n}_{2}-E_{\rm ind}^{\rm coup} \varphi_{1} \varphi_{2},
\end{equation*}
while the term responsible for the cross-Kerr interaction is
\begin{equation*}
H_{\rm CK}= -\frac{E_{\rm Kerr}^{\rm coup}}{4} \varphi_1^2 \varphi_2^2.
\end{equation*}
We also introduced what we called a nonlinear Hamiltonian
\begin{equation*}
H_{\rm non.lin}= \frac{E_{\rm Kerr}^{\rm coup}}{6} (\varphi_1^3 \varphi_2+\varphi_1 \varphi_2^3).
\end{equation*}
We introduce annihilation and creation operators for the transmon modes 
\begin{subequations}
\begin{equation*}
\varphi_m= \biggl(\frac{2 E_{C_m}}{E_{L_m}} \biggr)^{1/4}\bigl(a_m+a_{m}^{\dagger} \bigr),
\end{equation*}
\begin{equation*}
\mathfrak{n}_m= \frac{i}{2} \biggl(\frac{E_{L_m}}{2E_{C_m}} \biggr)^{1/4}\bigl(a_m^{\dagger}-a_m \bigr).
\end{equation*}
\end{subequations}

We now rewrite the previously defined Hamiltonian using annihilation and creation operators, performing additionally several rotating wave approximations (RWA):
\begin{itemize}
\item Transmon Hamiltonian:
\begin{equation}
\frac{H_m}{\hbar} \overset{\mathrm{RWA}}{=} (\omega_m+\delta_m) a_{m}^{\dagger} a_m+ \frac{\delta_m}{2}a_{m}^{\dagger} a_{m}^{\dagger}a_m a_m,
\label{eq:transmon}
\end{equation}
with the transmon frequency $\omega_m= \sqrt{8 E_{C_m} E_{L_m}}/\hbar$ and the anharmonicity $\delta_m= -E_{K_m} E_{C_m}/(\hbar E_{L_m})$. 
\item Linear coupling Hamiltonian:
\begin{equation*}
\frac{H_{\rm lin}}{\hbar} \overset{\mathrm{RWA}}{=}  J_{\rm lin} (a_1 a_2^{\dagger}+a_{1}^{\dagger}a_2),
\end{equation*}
with the linear hopping parameter $J_{\rm lin}$ given by
\begin{equation}
\label{gl}
J_{\rm lin}= J_{\rm cap}+J_{\rm ind},
\end{equation}
where we have a capacitive and an inductive $J_{\rm ind}$ contribution, \i.e.
\begin{subequations}
\label{gcapIndEq}
\begin{equation}
\label{gcapEq}
J_{\rm cap}=  \frac{1}{\hbar} 2 E_{\rm cap}^{\rm coup} \biggl( \frac{E_{L_1}}{2 E_{C_1}} \biggr)^{1/4}\biggl( \frac{E_{L_2}}{2 E_{C_2}} \biggr)^{1/4},
\end{equation}
\begin{equation}
\label{gindEq}
J_{\rm ind}=  -\frac{1}{\hbar} E_{\rm ind}^{\rm coup} \biggl(\frac{2 E_{C_1}}{E_{L_1}} \biggr)^{1/4}\biggl(\frac{2 E_{C_2}}{E_{L_2}} \biggr)^{1/4}.
\end{equation}
\end{subequations}
In these equations we clearly see that capacitive and inductive coupling give a flip-flop interaction of opposite sign. 
\item Cross-Kerr Hamiltonian
\begin{equation}
\frac{H_{\rm CK}}{\hbar}\overset{\mathrm{RWA}}{=} J_{\rm CK} \biggl(a_{1}^{\dagger} a_1 a_{2}^{\dagger}a_2+ \frac{1}{2} a_{1}^{\dagger}a_1+ \frac{1}{2} a_{2}^{\dagger}a_2+ \frac{1}{4} a_1 a_1 a_{2}^{\dagger} a_{2}^{\dagger}+\frac{1}{4} a_1^{\dagger} a_1^{\dagger} a_{2} a_{2} \biggr),
\label{CKapprox}
\end{equation}
with 
\begin{equation}
J_{\rm CK}= -\frac{1}{\hbar} E_{\rm Kerr}^{\rm coup} \biggl(\frac{2 E_{C_1}}{E_{L_1}} \biggr)^{1/2}\biggl(\frac{2 E_{C_2}}{E_{L_2}} \biggr)^{1/2} \equiv -\Delta.
\label{eq:CK}
\end{equation}
In Eq.~(\ref{CKapprox}) the RWA means that we have omitted all fast rotating terms which are products of unequal numbers of creation and annihilation operators. The last two terms in Eq.~(\ref{CKapprox}) which represent off-resonant two-photon processes contribute little and disappear when projecting onto the transmon qubit subspaces. We see that when the transmon qubits become more harmonic, i.e. $E_J/E_{C_m}$ increases for $m=1,2$, $J_{\rm CK}$ becomes progressively smaller as $E_{L_m}/E_{C_m}$ defined through Eq.~(\ref{eq:defLK}) increases. 

\item Nonlinear Hamiltonian
\begin{multline*}
\frac{H_{\rm non.lin}}{\hbar} \overset{\mathrm{RWA}}{=} J_{\rm non.lin}^{(1)} [(a_1 a_2^{\dagger}+a_1^{\dagger} a_2)+(a_1 a_2^{\dagger}+a_1^{\dagger} a_2)a_1^{\dagger}a_1+ \\ a_1^{\dagger} a_1 (a_1 a_2^{\dagger}+a_1^{\dagger} a_2)+ (a_1 a_2^{\dagger} a_1^{\dagger}a_1+a_1^{\dagger}a_1 a_1^{\dagger} a_2)]+ \\
  J_{\rm non.lin.}^{(2)} [(a_1 a_2^{\dagger}+a_1^{\dagger} a_2)+(a_1 a_2^{\dagger}+a_1^{\dagger} a_2)a_2^{\dagger}a_2+ \\ a_2^{\dagger} a_2 (a_1 a_2^{\dagger}+a_1^{\dagger} a_2)+ (a_1 a_2^{\dagger} a_2^{\dagger}a_2+a_2^{\dagger}a_2 a_1^{\dagger} a_2)],
\end{multline*}
with for $m=1,2$
\begin{subequations}
\begin{equation*}
J_{\rm non.lin}^{(1)}= \frac{1}{\hbar} \frac{E_{CK}}{6} \biggl(\frac{2 E_{C_1}}{E_{L_1}} \biggr)^{3/4} \biggl(\frac{2 E_{C_2}}{E_{L_2}} \biggr)^{1/4}
\end{equation*}
\begin{equation*}
J_{{\rm non.lin}}^{(2)}= \frac{1}{\hbar} \frac{E_{CK}}{6} \biggl(\frac{2 E_{C_1}}{E_{L_1}} \biggr)^{1/4} \biggl(\frac{2 E_{C_2}}{E_{L_2}} \biggr)^{3/4}.
\end{equation*}
\end{subequations}
Again RWA means that fast-rotating terms with unequal numbers of creation and annihilation operators are omitted. In the transmon qubit space all these nonlinear terms act as flip-flop interactions.
\end{itemize}
When we project the various terms of the full Hamiltonian onto the first two levels of each transmon qubit, we obtain the final Hamiltonian
\begin{equation}
\label{HQ}
\frac{H_Q}{\hbar}= -\frac{\Omega_1}{2} \sigma_1^z -\frac{\Omega_2}{2} \sigma_2^z +J_{\rm hop}(\sigma_1^+ \sigma_2^-+\sigma_1^- \sigma_2^+) -\Delta n_1 n_2,
\end{equation} 
Here the fully-dressed transmon frequency is $\Omega_m= \omega_m+\delta_m-\Delta/2$, while the total {\em forbidden} hopping strength is 
\begin{equation}
\label{ghEq}
J_{\rm hop}= J_{\rm cap}+J_{\rm ind}+J_{\rm non.lin.},
\end{equation}
with 
\begin{equation}
\label{Jnonlin}
J_{\rm non.lin}=3 \sum_m J_{{\rm non.lin}}^{(m)}.
\end{equation}
Since we never drive the transmon qubits, except for creating the initial excitations, the qubit approximation is justified. 
%BMT2 small change

 We observe that the dressed transmon frequency will depend on the number of cross-Kerr interactions that a transmon qubit participates in. This means that if want transmon qubits to have the same dressed frequencies, then the bare transmon frequency $\omega_m+\delta_m$ in Eq.~(\ref{eq:transmon}) needs to be different if the qubit participates in a different number of cross-Kerr couplers. Qubits which differ in this way occur at the boundaries of the computational lattice in Fig.~\ref{layoutConcept} as well as on the split-sites in the CNOT (or Toffoli) region shown in Fig.~\ref{fig:CNOT-CKerr}.
\par 

\subsection{Cross-Kerr Coupling Strength and Cross-Talk}
\label{subsec::CKstrength}
When we consider our coupler in a larger system we have to analyze the problem of unwanted long-range coupling (cross-talk), which is always potentially present in superconducting qubit architectures. In order to mitigate this problem one requires that the coupling capacitances $C_c$ are much smaller than the transmon capacitances $C_J$, so that the inversion of the capacitance matrix remains local in a first-order approximation in $C_c/C_J$.  \par 

\begin{figure}
\centering
\includegraphics[scale=0.06]{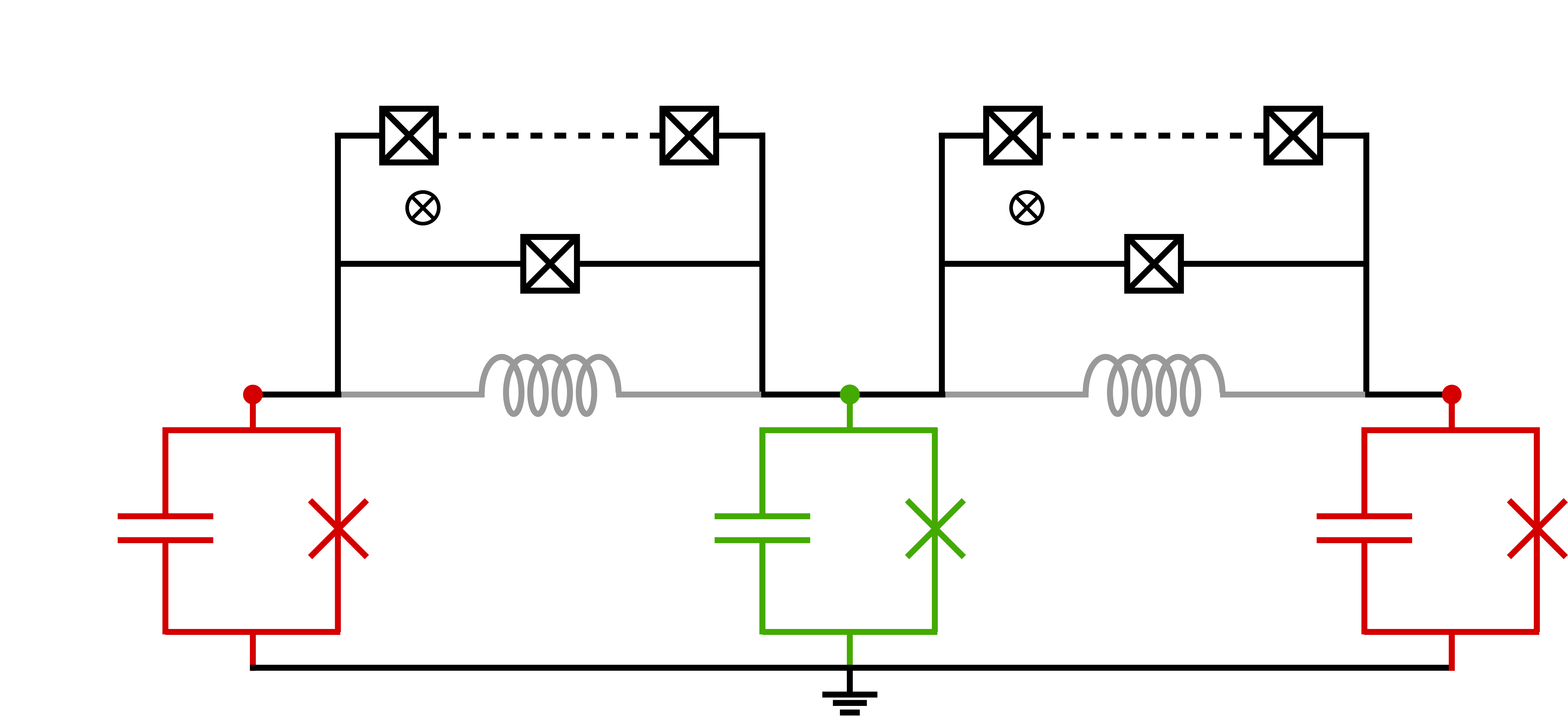}
\caption{Two resonant transmons (red) coupled to a common detuned transmon (green). Parameters are not shown for simplicity. The resonant transmons can either be two transmons forming a pair or two transmon qubits on adjacent sites on a track.}
\label{fig::NNcircuit}
\end{figure}

Cross-talk via two cross-Kerr couplers would be particular undesirable for qubits on the same track as these have the same frequency, see e.g. Fig.~\ref{fig::NNcircuit}. In order to limit this problem, our idea is to limit the inductive contribution to the linear coupling while keeping the cross-Kerr approximately the same, which is what the coupler analyzed in the previous subsection achieves. This should limit the capacitance needed to \emph{filter} the hopping interaction between different tracks and thus moderate the problem of cross-talk.

%BMT2 I made the hopping of the cross-Kerr into forbidden hopping as it is wholly unclear what hopping we want and what not, I use cross-talk hopping for the issue in Fig. 21

There are however several trade-offs that one should consider in choosing parameter strengths. First of all, we need to require the ratio $E_J/E_{C_a}$ of each junction of the array to be much larger than one, in order to prevent phase slips. We will always fix $E_J/E_{C_a} = 100$. This is in contrast with our desire to have a small effective coupling capacitance. To reduce this capacitance we could also increase the number of junctions, but from a practical point of view we would like to limit $N_J$. 
% BMT2 why is this a space consideration?
% AC a bit but not only. In general, fewer junctions -> easier, more reliable fabrication (the SNAIL itself as proposed in the Yale paper has three junctions because three is the minimum number to achieve pure three-wave mixing)
In what follows, we just assume a set of typical and reasonable experimentally achievable parameters, and compute the typical forbidden hopping strength, the cross-Kerr strength as well as the cross-talk hopping induced by two couplers. In particular, in order to evaluate the order of magnitude of this cross-talk hopping, we consider a circuit as in Fig.~\ref{fig::NNcircuit} where two transmons on the same track (red) are both coupled to a transmon on a different track (green). 

We will focus on the case $N_J=4$ and take the bare charging energy of the bare transmons as in Fig.~\ref{capFig} to be equal, \i.e. $E_{C}^{\rm bare}\equiv  e^2/(2 C_J)$ as our unit of energy and also set $\hbar=1$. We neglect the intrinsic capacitance of the coupling junction $C_b=0$.

\begin{center}
    \begin{tabular}{| p{2cm}  | p{2cm} | }
    \hline
    $E_{J_1}$ & $80$  \\ \hline 
    $E_{J_2}$ & $60$  \\ \hline
    $E_{C_a}$ & $12$  \\ \hline 
    $E_{J}$  & $1200$  \\ \hline 
    $E_{L}$  & $255$ \\
    \hline
    \end{tabular}
    \captionof{table}{Parameters in units of $E_C^{\rm bare}$.}
    \label{tableParameters}
\end{center}

\begin{figure}
\centering
\begin{subfigure}[t]{0.45 \textwidth}
\centering
\includegraphics[scale=0.45]{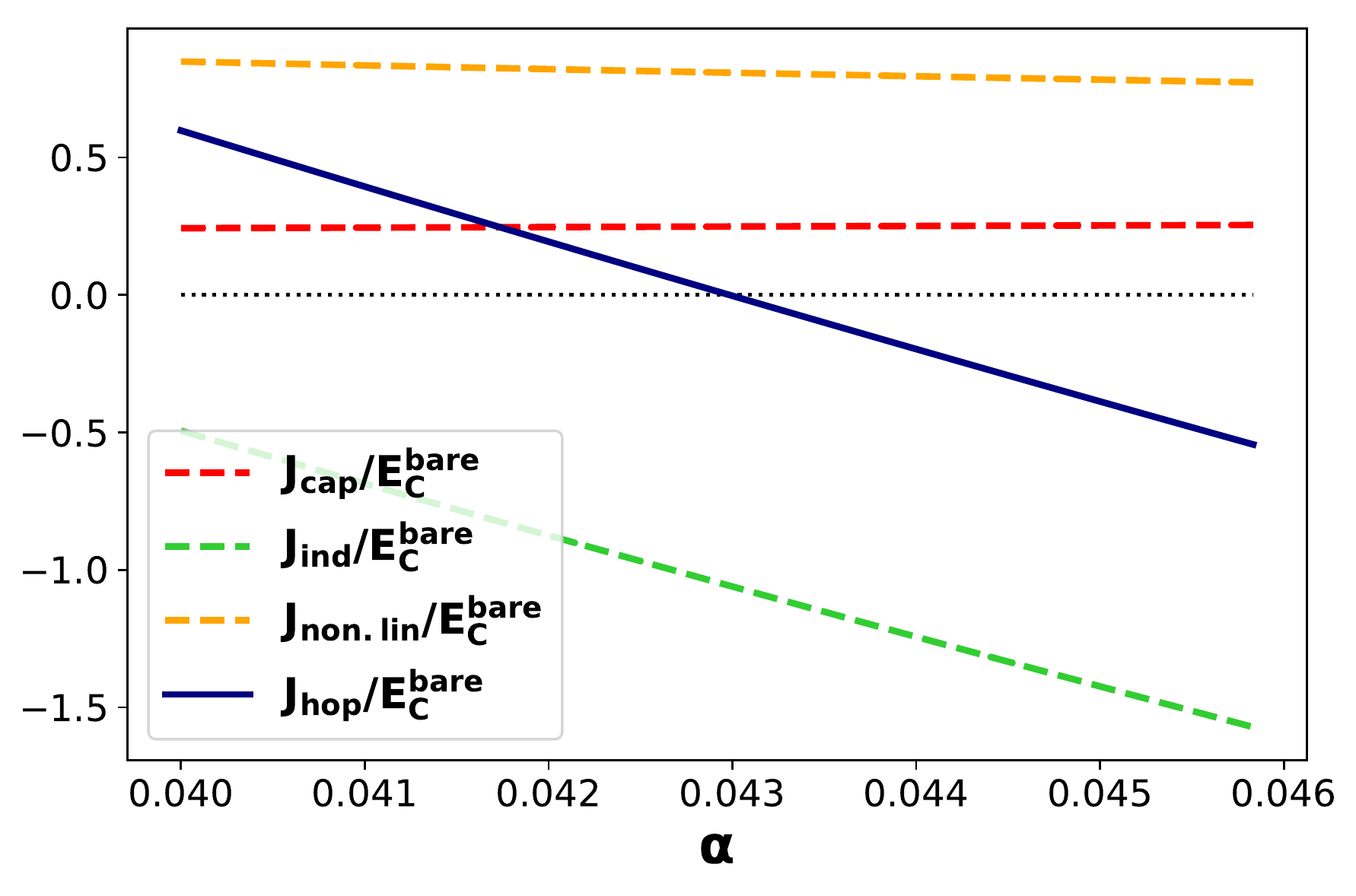}
\subcaption{Different contributions to the forbidden hopping parameter with their strengths as defined in Eqs.~\ref{gcapIndEq} and Eq.~\ref{Jnonlin}  (dashed lines) and the total forbidden hopping strength (solid line) as defined in Eq.~(\ref{ghEq}).}
\label{hopFigP}
\end{subfigure}

\begin{subfigure}[t]{0.45 \textwidth}
\centering
\includegraphics[scale=0.45]{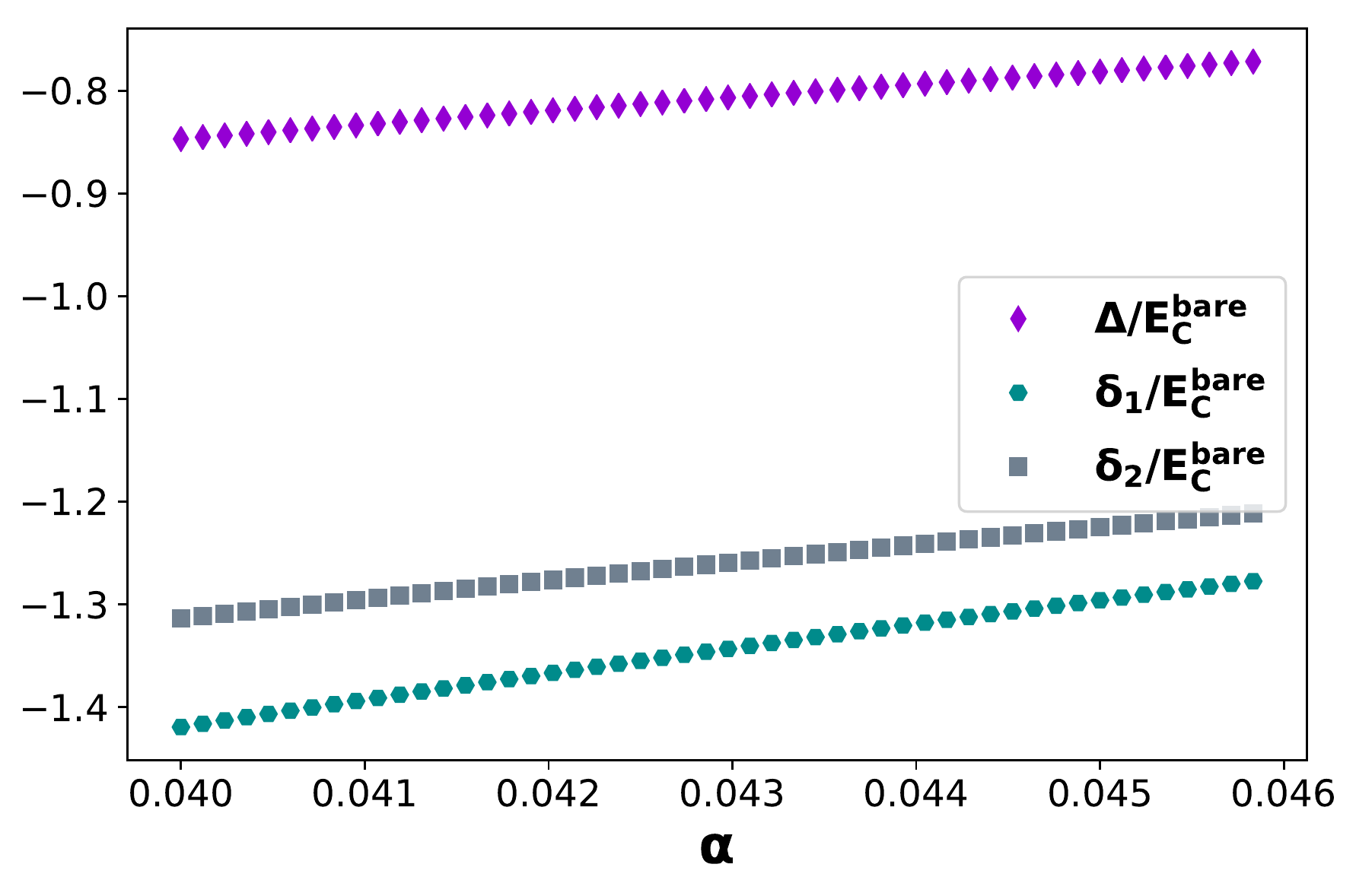}
\subcaption{Cross-Kerr coupling and anharmonicities of the two transmons (see Eqs. \ref{eq:transmon} and Eq. \ref{eq:CK} for the definitions).}

\label{ckFig}
\end{subfigure}
\caption{Forbidden hopping (a) and cross-Kerr (b) coupling strengths for parameters $\alpha$ close to the condition of canceling hopping in Eq.~(\ref{cancel}). In Fig.~\ref{hopFigP} we see that the hopping coupling is zero at $\alpha \approx 0.043$, corresponding to a Josephson energy of the small junction equal to $\alpha E_J/E_C^{\rm bare}= 51.72$.  At this point, $\Delta/E_C^{\rm bare} \approx 0.8$.}
\end{figure}

The chosen parameters are listed in Table \ref{tableParameters}. In Fig.~\ref{hopFigP} we see how it is possible to achieve a hopping term that is exactly equal to zero. This is possible with an effective capacitance that is much smaller than the transmon capacitance, in particular $C_c \approx 1/(48) C_J$. 
With these parameters we can get a high cross-Kerr interaction $\Delta/E_C^{\rm bare} \approx 0.8$ as we can see from Fig.~\ref{ckFig}. Assuming a typical bare charging energy $E_C^{\rm bare}/h \approx 200 \mathrm{MHz}$, we then get $\Delta \approx 2 \pi \times 160 \mathrm{MHz}$.

%BMT2 AC one sentence on how you get from this to Delta=100 MHz, THIS IS THE IMPORTANT POINT
% need to choose a E_C^bare
% AC I added text. Actually assuming Ec^bare = 200 MHz one would get \Delta = 2\pi 160 MHz. It should be clear that in the initial table with 100 MHz we mean \mathcal{O}. I would leave it in this way.

Considering the circuit in Fig.~\ref{fig::NNcircuit}, and using again the parameters listed in Table \ref{tableParameters}, we estimate a cross-talk hopping between the red transmons of approximately $0.004 E_C^{\rm bare}$, which corresponds approximately to $2 \pi \times 0.8 \mathrm{MHz}$. If these two red transmons form a pair, then this hopping should ideally be of zero strength. If these two red transmons are two transmons on the same track at nearest sites, then this term will weakly contribute to the total flip-flop interaction. 

%BMT2 AC how many MHz is this cross-talk then, when you choose E_C^bare.
% AC added text

More precisely, if we choose the perturbative regime $J/\Delta=1/10$, where $J$ is now the desired on-track hopping strength of Table \ref{tab:numbers}, the weak on-track flip-flop interaction that we discuss in the next section can be chosen as $J \approx 0.08 E_C^{\rm bare}$. Thus the contribution from the cross-talk `cross-Kerr mediated' hopping is negligliby small.

\subsection{Weak On-Track Flip-Flop Interactions}
As we have seen in the previous section the hopping interaction is naturally obtained with transmon qubits, by simply using capacitances or inductances. As we know, we need hopping interactions between qubits on a track, which are much weaker than the cross-Kerr interactions on the edges. On the other hand, these interactions should be larger than the forbidden hopping and cross-talk hopping strengths.

The general form of the capacitive coupling between two equal transmons can be deduced from the analysis done in Sec.~\ref{zzIntSec} and is of the general form (see Eq.~(\ref{gl})):
\begin{equation*}
J_{\rm cap}= \frac{1}{\hbar} 2 E_{\rm cap}^{\rm coup} \biggl( \frac{E_{L}}{2 E_C} \biggr)^{1/2}.
\end{equation*} 
with $E_{\rm cap}^{\rm coup}$ in Eq.~(\ref{def:Ecap}). In this expression we have assumed that the transmons are coupled via the cross-Kerr coupler to other parts of the circuits so that the inductive energy $E_L$ includes the inductance due to this coupling as in Eq.~(\ref{eq:defLK}). Similarly, one can imagine that $E_C$ is a dressed charging energy which includes the capacitive energy contributions from different couplings. Assuming $E_C/h \approx 200 \mathrm{MHz}$, we obtain the magnitude of a typical flip-flop interaction to be $J \approx 2\pi \times 10-15 \mathrm{MHz}$, which can be easily obtained with a small capacitance. 

%BMT2 AC I moved a sentence, ok
%AC ok

% BMT AC here J is clearly positive, why should it be?
% AC: actually it is just a consequence of how we define the annihilation and creation operators. By giving a phase to them it becomes negative. 
% BMT right

We see that a capacitive coupling always gives rise to a coupling of the same sign. We do hovever need a sign change to implement the Hadamard gate. An immediate solution would be to use inductances to implement couplings of negative signs. However, in order to keep the coupling small, one would need to work in the regime of large inductances and thus use arrays of junctions. 

There is a much simpler way to achieve this sign change which is to couple both qubits capacitively to a common resonator, and require the qubits to be far detuned from the resonators' frequencies, working in the so-called dispersive regime. Let $g_{\rm res}$ be the coupling to the resonator of the two qubits, let $\delta < 0$ be their anharmonicity, let $\Omega$ be their frequency (which should be the same for qubits on one track) and let $\omega_{\rm res}$ be the frequency of the resonator. One has an effective flip-flop coupling between the two qubits \cite{blais2004,koch2007} of strength
\begin{equation*}
J_{\rm eff}= \frac{g_{\rm res}^2}{\Omega-\omega_{\rm res}}-\frac{1}{2} \frac{(\sqrt{2} g_{\rm res})^2}{\Omega+\delta-\omega_{\rm res}},
\end{equation*}
It is clear that we can obtain couplings of different signs by properly selecting the frequency of the resonator $\omega_{\rm res}$. For instance by taking $\omega_{\rm res}> \Omega$, we would get a negative effective hopping as expected. The circuit that implements the Hadamard gate is represented in Fig.~\ref{circuitGates}.
It is clear, that besides a Hadamard gate, any real rotation 
\begin{equation*}
U(\theta)=\left(\begin{array}{cc} \cos(\theta) & \sin(\theta) \\ -\sin(\theta) & \cos(\theta) \end{array} \right)
\end{equation*}
could be engineered using (resonator-mediated) flip-flop couplings. The two pairs of transmon qubits need to be four-way coupled as in Fig.~\ref{Hgate} where the strength of the flip-flop interactions should be set to $J_{00}=J \cos(\theta)$, $J_{10}=J \sin(\theta)$ etc. where $J$ is the global hopping strength. \par

% BMT deal/check \hbar and 2pi's.

\subsection{Real Interactions and Time-Reversal}
\label{sec:real-stoq}

Since we obtain the flip-flop interaction from a passive capacitive coupling, \i.e., we do not use any external drives, and we work in the transmon regime, we are giving up the possibility to have complex-valued couplings in our chosen transmon basis. This is a consequence of the fact that the Hamiltonians are invariant under a time-reversal transformation, and additionally, the basis we have chosen is the eigenbasis of an operator invariant under time-reversal symmetry, namely the basis of the uncoupled transmon Hamiltonians. More precisely, the capacitive coupling $Q_1 Q_2$ and the inductive coupling $\Phi_1 \Phi_2$ are invariant under the time-reversal transformations $Q_{m} \rightarrow Q_m$, $\Phi_{m} \rightarrow -\Phi_m$ \cite{brito2015}. We may point out that passivity alone, meaning no external time-dependent drive, does not automatically imply reciprocity in circuit-QED systems as shown in Refs.~\cite{kochTimeRev, vogtCirculator}. 
\section{Challenges}
\label{transmon:noise}

As the present computational model does not specify a means for error correction, the effect of coherent errors and stochastic noise has immediate consequences for its viability. In the next paragraphs we discuss two specific challenges for using the proposed architecture for some form of quantum computation or quantum learning.

\subsection{Loss of Excitations}
As mentioned at the beginning of the paper and in \cite{terhalLloyd}, the loss of transmon excitations puts a severe limitation on the computation time. Suppose that we have a single hopping particle on a line with $L$ sites. We associate with each site a decay rate $\gamma$ that destroys the particle, \i.e., it brings the system from one of the sites $\ket{k}$ to a vacuum state $\ket{\rm vac}$. Formally, we can model such dissipation using a Lindblad master equation 
\begin{equation*}
\frac{d \rho}{dt}= -i [H_{L}, \rho]+ \sum_{k=0}^{L-1} \gamma D [\ket{\rm vac}\bra{k}]\;\rho,
\end{equation*}
with the Lindblad dissipator $D[A]\rho= A^{\dagger} \rho A-1/2 \{A^{\dagger} A, \rho \}$ and $H_L$ the Hamiltonian of a quantum walk on a line, see e.g. Eq.~(\ref{HQW}). 
It is not hard to show that this implies that the decay rate of the excitation is $\gamma$ (since the single particle can only be annihilated at one of the sites), so that the decay rate does not depend on the length of the line $L$ or the depth of the computation. Of course, the total probability of particle loss does still increase with the depth of the computation as the computation time linearly increases with the computational depth.

\subsection{Arrival Time and Measurement}
\label{sec:timing-issues}

% BMT add ancillas during comp add and remove...

One question in the Hamiltonian computing models, both the single-clock as well as the multi-clock model, is to determine which qubits to measure to read out the computation. Ideally, the computation would finish in a known amount of time and only the relevant computational qubits are measured. For example, for the Feynman model as realized on the 2D lattice in Appendix \ref{FKPqc}, these could be the qubits in the final column of the computation. However, its is known that, at a fixed time, the probability to get to the last site of the walk on a line with $L$ sites decreases with $1/L$ and various ideas have been formulated to address this problem, see e.g. \cite{CLN:clock}. 

One simple method is to pad the quantum circuit with I gates, \i.e. make the walk, say, twice as long, and measure at a random, sufficiently long, time to determine where the excitations reside. In this way, the probability for determining the total output of the computation at a random time can be made arbitrarily close to 1. In the model in Appendix \ref{FKPqc} it would correspond to measuring a whole 2D rectangular region of qubits at the end of the lattice. We call such region of qubits to be measured at a specific (random) time the measurement region.

%AC: the notation now should agree with that in Gosset-Terhal

Similarly, in \cite{gossetTerhal} it was envisioned that the actual quantum circuit with non-trivial gates is encoded in a $K \times K$ region of the lattice located in the left corner with $K=N/4$, while in the rest of the lattice only trivial identities are applied.  If all particles are found somewhere in a measurement region which is beyond the region where the computation takes place, then the quantum computation has been completed. In this model the measurement region is again a fully-2D region. 

This disadvantage of not knowing where the computation is at a point in time thus implies some qubit overhead, but perhaps worse, it requires being able to read-out 2D regions of qubits simultaneously. Such read-out is a standard requirement in the usual stationary-qubit circuit model and it necessitates using the third dimension to accomodate the hardware of read-out resonators, measurement feedlines \cite{mariantoni2016}, which adds unwanted complexity.

\begin{figure}[htb]
\centering
\includegraphics[scale=0.07]{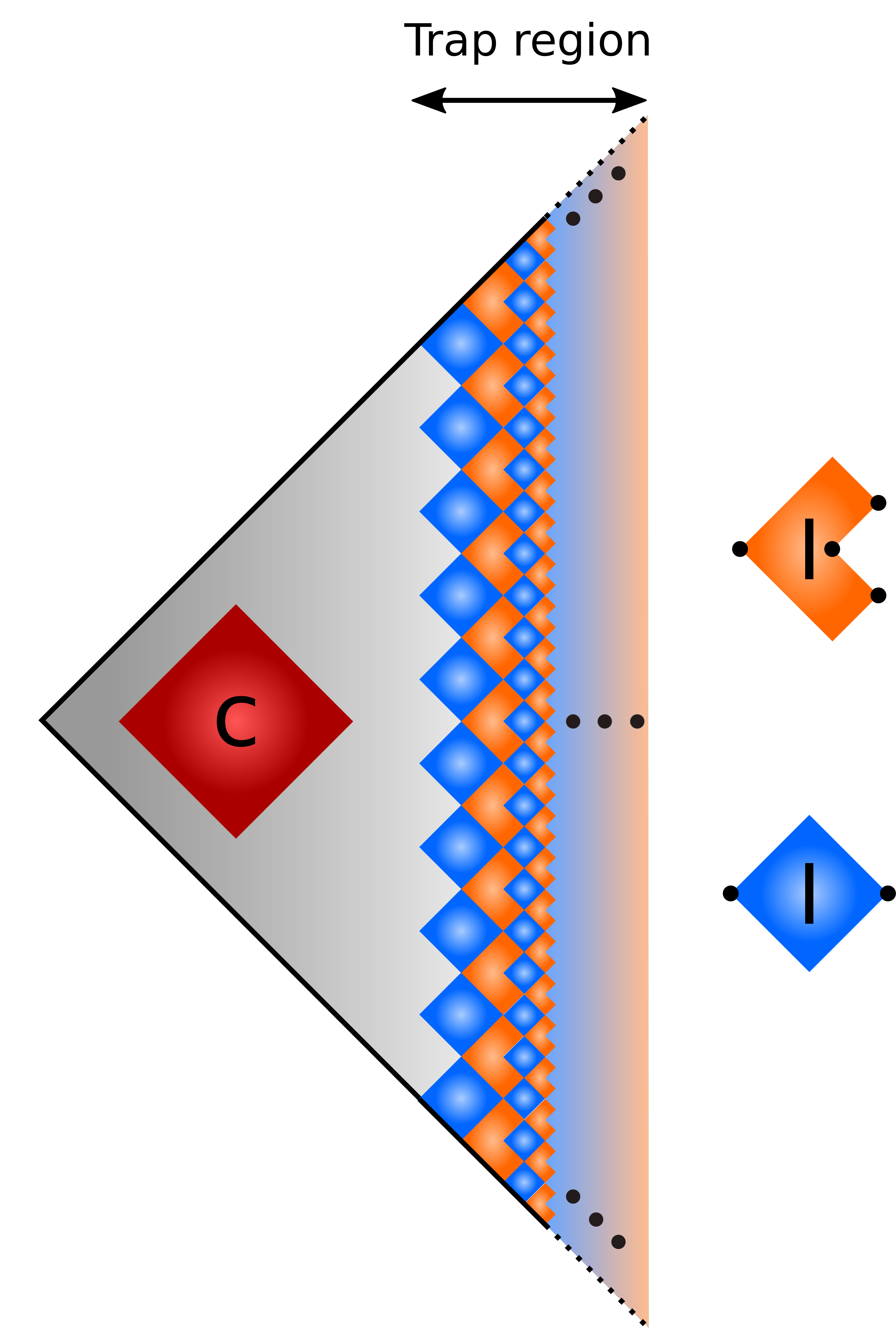}
\caption{Sketch of a lattice terminated with an entropic trap region. The red square region is where the computation takes place, meaning that non-trivial gates are applied. All blue plaquettes beyond this region execute $I$ gates. In the trap region, new tracks with particles get added in each layer, e.g. an orange `isometry' unit maps a single input track with one particle onto 3 tracks each with a particle. The isometry could otherwise execute the I gate on the input particle. Translated to transmon qubits, we add two pairs of transmon qubits, each with a single excitation, for an orange plaquette. These excitations can only move forward in the right direction and spawn new excitations until they hit the trap end.}
\label{dif::entropicTrap}
\end{figure}

A modification of the rotated lattice geometry of \cite{gossetTerhal} would be to terminate the lattice region in which the computation is followed by an MERA-like expanding {\em trap} region of $O(1)$ or at most $r=O(\log N)$ sites wide, see Fig.~\ref{dif::entropicTrap}. In the trap region new dummy pairs of transmon qubits are placed with their excitations, as shown in Fig.~\ref{dif::entropicTrap}. These new dummy excitations can only move forward deeper into the trap region and spawn the forward motion of a next layer of dummy qubits. In the trap region the string is rapidly increasing in length and one expects that this expansion will lead to an irreversible effect on the dynamics, \i.e. entropically trap the string, while the overall dynamics remains unitary. It is an interesting open question how to analyze the dynamics of the space-time circuit-to-Hamiltonian construction applied to a MERA circuit (see \cite{metz:bsc}) or a regular circuit appended by a MERA-like trap region.

One imagines that in such scheme, only the transmon qubits in the trap region whose causal history involves the computational region $C$ need to measured \footnote{One could even imagine applying CNOTs between the computational qubits and the dummy qubits so that the output of the computation is redundantly copied onto the dummy qubits and all qubits are measured.}. Since the trap region is, say, $O(1)$ in width, it would imply that space may be available to let all read-out lines come in from the right side of the planar lattice. 

For the Feynman single-clock model, we discuss a different way of making sure that the computation arrives in a certain time due to Peres \cite{peres1985} (and used in \cite{christandl2004} in the context of perfect quantum state transfer) in Appendix \ref{sec:peres}. Since this solution is not scalable to large system sizes and we don't know how to apply it to the multi-clock model, we do not advocate it as the preferred solution.

\section{Discussion}
\label{conclSec}

We have observed that adding disorder in the on-site or hopping terms can lead to a localization of time, \i.e. the average position of the particle on the middle track can become frozen. It would be interesting to obtain more theoretical or experimental evidence for this behavior for larger lattice sizes $N$. A difficulty is that disorder breaks the elegant map from the 2D rotated lattice model onto that of $N-1$ non-interacting fermions hopping on a line, see e.g. \cite{gossetTerhal}. In the non-interacting fermion model the on-site energy of one fermion will now depend on how many fermions are to the left of the site where the fermion sits. Hence a new theoretical analysis, going beyond an Anderson localization analysis \cite{DKS:localization}, may be required to address this question.

% BMT AC N site lattice is N-1 11's in the string so N-1 hopping fermions 
% AC yes N-1 is correct!

We did not explicitly examine the effect of disorder in the execution of logical gates such as the Hadamard or $X$ gate. Since the hopping strengths $J$ determine the unitary gate which is executed, it is clear that such disorder leads to inaccurate computation. Fascinating is that static disorder alters the application of unitary gates $U$ and $U^{\dagger}$ to possibly irreversible gates $M$ and $M^{\dagger}$ which could lead to inadvertently measuring the computational state.

An interesting open question is whether the wavefront dynamics can be viewed as a non-classical `bunching' effect of photons in the following sense. Instead of a pair of transmon qubits we could use two bosonic modes and encode $\ket{0} \equiv \ket{\alpha}_0 \ket{\rm vac}_1$ and $\ket{1} \equiv \ket{\rm vac}_0 \ket{\alpha}_1$ where $\ket{\alpha}$ is a coherent state. 
%Let us assume for simplicity that the modes are entirely harmonic but the cross-Kerr couplers are as before \footnote{This assumption is not warranted in the circuit-QED setting.}. 
We can view all flip-flop interactions in our scheme as bosonic beam-splitter interactions, evolving the coherent amplitudes at two adjacent sites on a single track into new coherent amplitudes. 
Since we have non-linear bosonic interactions, we cannot efficiently solve the system's dynamics, 
although a mean-field approximation of the non-linearity would enable an efficient simulation. 
It will be interesting to understand how this is different from our single-excitation encoding, since the dual-rail coherent state encoding would suffer less from excitation loss. We expect that the coherent state encoding would not lead to wavefront dynamics, nor to the correct application of logical gates.
 
% DPDV say something about the use of dissipation?

One could envision incorporating quantum error correction, include the initialization of ancilla qubits during the computation, but this may require going to a 3D version of the model. A mathematical version of such 3D model is discussed in \cite{DOR:quantum-crys}: in this model the computation would move forward as a surface of a crystal which is growing from a corner. Going to 3D means a higher-connectivity per (transmon) degree of freedom and facing practical challenges related to 3D superconducting qubit hardware \cite{mariantoni2016}. In addition, error correction requires the inclusion of ancilla qubits during the computation which are measured and reset to remove entropy build-up, hindering the overall passivity of the scheme.  \par

BMT acknowledges support from the European Research Council (EQEC, ERC Consolidator Grant No. 682726). DD was supported by Intelligence Advanced Research Projects Activity (IARPA) under contract
W911NF-16-0114. AC acknowledges support from the Excellence Initiative of the Deutsche Forschungsgemeinschaft. We thank Fabian Hassler, Martin Rymarz and Stefano Bosco for useful discussions.

\appendix

\section{Hamiltonian for the direct Toffoli Gate}
\label{sec:tof}

We describe the ideas behind the direct Toffoli gate mathematically.  We consider how the Hamiltonian $H_{\rm valid}$ is modified as compared to the standard case, Eq.~(\ref{HstringTL}). We need to modify the edge Hamiltonian corresponding to the green edges in Fig.~\ref{toffoliFig}. We call the set of these edges $E_{\rm Tof}$. $H_{\rm valid}$ can be written as

\begin{equation*}
H_{\rm valid}= -\Delta \sum_{(\mu, \nu)\in E \backslash E_{\rm Tof}} \bm{n}[\mu] \bm{n}[\nu \sp{\prime}]-\Delta \sum_{e \in E_{\rm Tof}}H_{e}.
\end{equation*}
The terms for the edges $c_1, d_1$ and $a_2, b_2$ do not change as compared to the standard case, hence we write
\begin{subequations}
\begin{alignat}{4}
H_{c_1}= & \bm{n}[i-1,j]\bm{n}[i, j],  \\[1ex]
H_{d_1}= & \bm{n}[i,j+1]\bm{n}[i, j],\\[1ex]
H_{a_2}=& \bm{n}[i+1,j]\bm{n}[i+1, j+1], \\[1ex]
H_{b_2}=& \bm{n}[i+2,j+1]\bm{n}[i+1, j+1].
\end{alignat}
\label{standEdge}
\end{subequations}
The remaining edges in $E_{\rm Tof}$ are chosen as
\begin{subequations}
\begin{equation}
H_{a_1}=  n_{s_1=0}[i,j-1] \bm{n}[i,j,0]+ n_{s_1=1}[i,j-1] \bm{n}[i,j,1] ,
\end{equation}
\begin{equation}
H_{b_1}= n_{s_1=0}[i+1,j] \bm{n}[i,j,0]+ n_{s_1=1}[i+1,j] \bm{n}[i,j,1] ,
\end{equation}
\begin{equation}
H_{c_2}=  n_{s_2=0}[i,j+1] \bm{n}[i+1,j+1,0]+ \\ n_{s_2=1}[i,j+1] \bm{n}[i+1,j+1,1] ,
\end{equation}
and
\begin{equation}
H_{d_2}=  n_{s_2=0}[i+1,j+2] \bm{n}[i+1,j+1,0]+ \\ n_{s_2=1}[i+1,j+2] \bm{n}[i+1,j+1,1] .
\end{equation}
\label{toffoliEdge}
\end{subequations}
As for the CNOT one can check that valid strings have energy $E_0=-(N_{\rm track}-1) \Delta$, while the gap still remains $\Delta$. In particular, the first excited subspace of the new $H_{\rm valid}$ is formed by the subspace of strings that are broken at one position and strings that are connected but incorrect.

The hopping Hamiltonian $V_{\rm hop}$ needs to be modified at plaquettes $p_1, p_2, p_3 \in P_{\rm Tof}$ in Fig. \ref{toffoliFig}. $V_{\rm hop}$ is written as
\begin{equation*}
V_{\rm hop}= -J \sum_{p \in P \backslash P_{\rm Tof}} V_{\rm hop,p} -J \sum_{p \in P_{\rm Tof}} \tilde{V}_{\rm hop, p},
\end{equation*}
where the three terms in the last sum are given by
\begin{subequations}
\begin{equation}
\tilde{V}_{\rm hop,p_1}=  \sum_{s_1=0}^{1} \sum_{k=0}^{1} (a_{s_1}^{\dagger}[i,j,k] a_{s_1}[i-1, j-1]+\mathrm{h.c.}),
\end{equation}
\begin{multline}
\tilde{V}_{\rm hop, p_2}= \sum_{s_2=0}^{1} (a_{s_2}^{\dagger}[i+1, j+1,0] a_{s_2}[i,j,0]+\mathrm{h.c.})+ \sum_{s_2=0}^{1} (a_{s_2}^{\dagger}[i+1, j+1,0] a_{s_2}[i,j,1]+\mathrm{h.c.}) + \\
 \sum_{s_2=0}^{1} (a_{s_2}^{\dagger}[i+1, j+1,1] a_{s_2}[i,j,0]+\mathrm{h.c.})+ \sum_{s_2=0}^{1} (a_{\bar{s}_2}^{\dagger}[i+1, j+1,1] a_{s_2}[i,j,1]+\mathrm{h.c.}),
\end{multline}
and
\begin{equation}
\tilde{V}_{\rm hop,p_3}= \sum_{s_2=0}^{1} \sum_{k=0}^{1}(a_{s_2}^{\dagger}[i+2, j+2] a_{s_1}[i+1,j+1,k] +\mathrm{h.c.}).
\end{equation}
\label{eq:hops-all}
\end{subequations}

To obtain the effective Hamiltonian, we can invoke Schrieffer-Wolff perturbation theory (as in \cite{terhalLloyd}), although in our lowest-order application its effect is rather immediate. For SW theory, we can use Sec. 3 of \cite{bravyiDiVincenzo} and Sec. 4 of \cite{bravyiHastings}. Let us denote by $P_-$ the projector onto the groundspace of $H_{\rm valid}$, i.e., the subspace of valid strings. We also define $P_+=I-P_-$ the projector onto the subspace of invalid (either disconnected and/or incorrect) strings. Given a generic linear operator $O$ we define the operators
\begin{alignat*}{2}
& O_{--}= P_- O P_- \quad , \quad O_{++}= P_+ O P_+ \\ &O_{+-}= P_+ O P_- \quad , \quad O_{-+}= P_- O P_+.
\end{alignat*} 
An operator $O$ is said to be block-diagonal if $O_{+-}=O_{-+}=0$, while block-off-diagonal if $O_{++}=O_{--}=0$. Assuming $2 \lVert V \rVert/\Delta < 1$, the effective low energy Hamiltonian is defined in general as
\begin{equation*}
H_{\mathrm{eff}}= (e^{S}H e^{-S})_{--},
\end{equation*}
with $S$ an anti-hermitian and block-off-diagonal operator such that $e^{S}H e^{-S}$ is also block-diagonal and $\lVert S \rVert <\pi/2$. The operator $S$ can be Taylor-expanded and a recursive relation can be obtained for each power, but here we only use the lowest-order term $S \approx 0$ so the SW transformation $e^S\approx I$. In lowest-order one simply has 
\begin{equation*}
H_{\mathrm{eff}}= H_{--}+\mathcal{O} ( \lVert V_{\rm hop} \rVert^2/\Delta )= (V_{\rm hop})_{--}+\mathcal{O} ( \lVert V_{\rm hop} \rVert^2/\Delta ).
\end{equation*}
For the Toffoli region, we can write the effective Hamiltonian as
\begin{equation*}
H_{\mathrm{eff}}= -J \sum_{p \in P \backslash P_{\rm Tof}} H_{{\rm cond.hop},p}-J \sum_{p \in P_{\rm Tof}} \tilde{H}_{{\rm cond.hop}, p}+\mathcal{O} ( \lVert V_{\rm hop} \rVert^2/\Delta ),
\end{equation*}
where the second sum involves the conditional hopping terms related to plaquettes $p_1, p_2, p_3 \in P_{\rm Tof}$. We have

\begin{subequations}
\begin{equation*}
\tilde{H}_{{\rm cond.hop},p_1}= \sum_{s_1=0}^1 \biggl(\sum_{s=0}^1 n_{s_1}[i, j-1] \bm{n}[i-1, j] a_{s}^{\dagger}[i,j,s_1]a_{s}[i-1, j-1]+ \mathrm{h.c.} \biggr)
\end{equation*}
The term corresponding to the second plaquette is more complicated: on the valid subspace, it acts as
\begin{equation*}
\tilde{H}_{{\rm cond.hop}, p_2}=\sum_{s_1=0}^1 \sum_{s_2=0}^1  \sum_{s=0}^1 n_{s_2}[i, j+1] n_{s_1}[i+1, j] \biggl(a_{s\oplus s_1 s_2}^{\dagger}[i+1, j+1, s_2]a_{s}[i,j,s_1]+ \mathrm{h.c.}\biggr),
\end{equation*}
exhibiting the Toffoli gate logic on the internal spin $s$. 
Analogously to the term for the plaquette $p_1$, we have
\begin{equation*}
\tilde{H}_{{\rm cond.hop}, p_3}= \sum_{s_2=0,1} \sum_{s=0}^1 n_{s_2}[i+1, j+2] \bm{n}[i+2, j+1] \biggl( a_{s}^{\dagger}[i+2, j+2]a_{s}[i+1, j+1, s_2]+\mathrm{h.c.} \biggr).
\end{equation*}
\end{subequations}

To probe our understanding of the Toffoli construction, we can ask if all the hopping terms described above are really necessary. Let us consider again Fig. \ref{toffoliFig} and suppose that the first control qubit is in the $s_1=1$ state. The target qubit will then be directed to the site $(i,j,0)$. At this point we are tempted to say that no matter what the state of the second control is, the identity has to be applied and so the hopping from $(i,j,0)$ to $(i+1,j+1,1)$ looks useless. This is however not the case, and the reason why this is incorrect is that we selected a particular direction of the computation, \i.e., from left to right, while terms that hop in the opposite direction are always present by the Hermiticity of the Hamiltonian. 

If we remove the identity connecting $(i,j,0)$ and $(i+1,j+1,1)$ and try to run the circuit backwards one can show that it cannot implement a Toffoli gate. Suppose that the target particle is in $(i+2, j+2)$ and the second control is in the state $s_2=1$. If the target particle runs backwards, it would be directed to site $(i+1, j+1, 1)$. However, at this point even if the first control qubit is in $s_1=0$, the target can never go further backwards with the application of the identity since the hopping from $(i+1, j+1,1)$ to $(i,j,0)$ is missing. We conclude that if the particles hop backwards the (inverse) Toffoli gate is not applied. For this reason, the interactions depicted in Fig.~\ref{toffoliFig} and written in Eqs.~(\ref{standEdge}), (\ref{toffoliEdge}) and (\ref{eq:hops-all}) are all necessary. As a final remark, and in analogy with the CNOT, the control particles $1$ and $2$ must obviously undergo the identity in the Toffoli region.

\section{Feynman model with hopping particles on a 2D lattice}
\label{FKPqc}

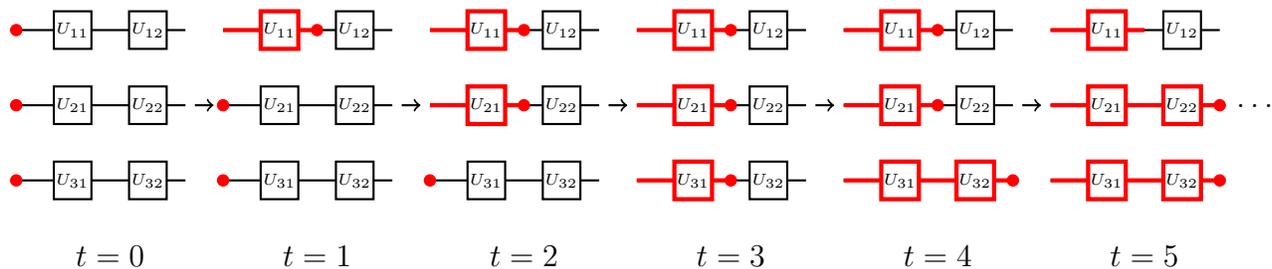
\begin{figure*}
\centering
\begin{tikzpicture}[scale=0.5]
\pic [scale=0.5] at (0,0) {quantumCircuit};
\foreach \x in {0,...,4}
\draw [thick, ->] (4.75+\x*5.5, -2)--(5.25+\x*5.5, -2);
\pic [scale=0.5] at (5.5,0) {quantumCircuit};
\pic [scale=0.5] at (11,0) {quantumCircuit};
\pic [scale=0.5] at (16.5,0) {quantumCircuit};
\pic [scale=0.5] at (22,0) {quantumCircuit};
\pic [scale=0.5] at (27.5,0) {quantumCircuit};
\foreach \x in {0,1,2}
\filldraw [color=red]   (0, -2*\x) circle (0.15cm);
%Second block
\draw[ultra thick, red] (5.5,0)--(6.5,0);
\draw[ultra thick, red] (6.5, -0.5) rectangle (7.5, 0.5);
\draw[ultra thick, red] (7.5,0)--(8,0);
\filldraw [color=red]   (8, 0) circle (0.15cm);
\foreach \x in {1,2}
\filldraw [color=red]   (5.5, -2*\x) circle (0.15cm);
%Third block
\foreach \y in {0,1}{
\draw[ultra thick, red] (11,0-2*\y)--(12,0-2*\y);
\draw[ultra thick, red] (12, -0.5-2*\y) rectangle (13, 0.5-2*\y);
\draw[ultra thick, red] (13,0-2*\y)--(13.5,0-2*\y);
\filldraw [color=red]   (13.5, 0-2*\y) circle (0.15cm);
}
\filldraw [color=red]   (11, 0-4) circle (0.15cm);
%4th block
\foreach \y in {0,1,2}{
\draw[ultra thick, red] (16.5,0-2*\y)--(17.5,0-2*\y);
\draw[ultra thick, red] (17.5, -0.5-2*\y) rectangle (18.5, 0.5-2*\y);
\draw[ultra thick, red] (18.5,0-2*\y)--(19,0-2*\y);
\filldraw [color=red]   (19, 0-2*\y) circle (0.15cm);
}
%5th block
\foreach \y in {0,1}{
\draw[ultra thick, red] (22,0-2*\y)--(23,0-2*\y);
\draw[ultra thick, red] (23, -0.5-2*\y) rectangle (24, 0.5-2*\y);
\draw[ultra thick, red] (24,0-2*\y)--(24.5,0-2*\y);
\filldraw [color=red]   (24.5, 0-2*\y) circle (0.15cm);
}
\draw[ultra thick, red] (22, -4)--(23, -4);
\draw[ultra thick, red] (23, -4.5) rectangle (24, -3.5);
\draw[ultra thick, red] (24,-4)--(25, -4);
\draw[ultra thick, red] (25, -4.5) rectangle (26, -3.5);
\draw[ultra thick, red] (26,-4)--(26.5, -4);
\filldraw [color=red]   (26.5, -4) circle (0.15cm);
%6th block
\foreach \y in {0,1,2}{
\draw[ultra thick, red] (27.5,0-2*\y)--(28.5,0-2*\y);
\draw[ultra thick, red] (28.5, -0.5-2*\y) rectangle (29.5, 0.5-2*\y);
\draw[ultra thick, red] (29.5,0-2*\y)--(30,0-2*\y);
}
\foreach \y in {0,1}{
\draw[ultra thick, red] (30,-4+2*\y)--(30.5,-4+2*\y);
\draw[ultra thick, red] (30.5, -4.5+2*\y) rectangle (31.5,-3.5+2*\y);
\draw[ultra thick, red] (31.5,-4+2*\y)--(32,-4+2*\y);
\filldraw [color=red]   (32, -4+2*\y) circle (0.15cm);
}
\node at (2.5, -6) {$t=0$};
\node at (2.5+5.5, -6) {$t=1$};
\node at (2.5+2*5.5, -6) {$t=2$};
\node at (2.5+3*5.5, -6) {$t=3$};
\node at (2.5+4*5.5, -6) {$t=4$};
\node at (2.5+5*5.5, -6) {$t=5$};
\node at (33, -2) {$\dots$};
\end{tikzpicture}
\caption{Example of depth-2 quantum circuit with 3 qubits in which gates are executed in a `snake-like' order.}
\label{snakeQuantumCircuit}
\end{figure*}
%\twocolumngrid 

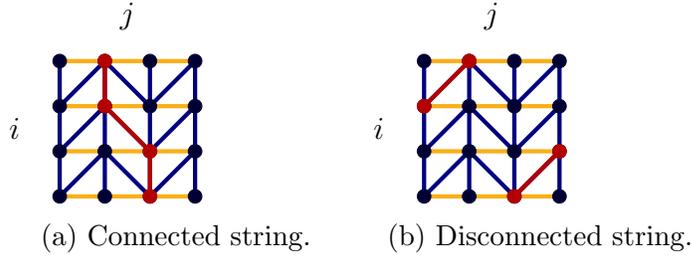
\begin{figure}[htb]
\centering
\begin{subfigure}[t]{0.3\textwidth}
\begin{tikzpicture}[scale=0.3]
\foreach \y in {0,...,3}
\draw [ultra thick, yellow!30!orange] (0,0-\y*2)--(6, 0-\y*2);
\foreach \x in {0,...,3}
\draw [ultra thick, black!50!blue] (0+2*\x,0)--(0+2*\x, -6);
\foreach \y in {0,1,2}
\draw [ultra thick, black!50!blue] (0,-2-\y*2)--(2,0-\y*2)--(4,-2-\y*2)--(6,0-\y*2);
\foreach \x in {0,...,3}
\foreach \y in {0,...,3}
\filldraw [color=black!80!blue]   (0+\x*2, 0-\y*2) circle (0.3cm);
%connected string
\filldraw [color=red!70!black]   (2, 0) circle (0.3cm);
\filldraw [color=red!70!black]   (2, -2) circle (0.3cm);
\filldraw [color=red!70!black]   (4, -4) circle (0.3cm);
\filldraw [color=red!70!black]   (4, -6) circle (0.3cm);
\draw [ultra thick, red!70!black] (2,0)--(2,-2)--(4,-4)--(4,-6);
\node at (3,2) {$j$};
\node at (-2,-3) {$i$};
\end{tikzpicture}
\subcaption{Connected string.}
\end{subfigure}
\begin{subfigure}[t]{0.3\textwidth}
\begin{tikzpicture}[scale=0.3]
\foreach \y in {0,...,3}
\draw [ultra thick,yellow!30!orange] (0,0-\y*2)--(6, 0-\y*2);
\foreach \x in {0,...,3}
\draw [ultra thick, black!50!blue] (0+2*\x,0)--(0+2*\x, -6);
\foreach \y in {0,1,2}
\draw [ultra thick, black!50!blue] (0,-2-\y*2)--(2,0-\y*2)--(4,-2-\y*2)--(6,0-\y*2);
\foreach \x in {0,...,3}
\foreach \y in {0,...,3}
\filldraw [color=black!80!blue]   (0+\x*2, 0-\y*2) circle (0.3cm);
%disconnected string
\filldraw [color=red!70!black]   (2, 0) circle (0.3cm);
\filldraw [color=red!70!black]   (0, -2) circle (0.3cm);
\filldraw [color=red!70!black]   (6, -4) circle (0.3cm);
\filldraw [color=red!70!black]   (4, -6) circle (0.3cm);
\draw [ultra thick, red!70!black] (2,0)--(0,-2);
\draw [ultra thick, red!70!black] (4,-6)--(6,-4);
\node at (3,2) {$j$};
\node at (-2,-3) {$i$};
\end{tikzpicture}
\subcaption{Disconnected string.}
\end{subfigure}
\caption{Examples of connected and disconnected strings of particles in the lattice model with $N \times M$ sites (shown is $N=M=4$). The red dots denote the position of the particles at sites $(i,j)$.}
\label{newLatticeFig}
\end{figure}

We construct a lattice model similar to that in Fig.~\ref{figLattice}, but now the dynamics of the valid strings maps to that of a quantum walk on a line. This is the dynamics of the clock Hamiltonian proposed by Feynman \cite{feynman1986}. A known difference between the multi-clock (as exemplified in the Lloyd-Terhal model) versus this single-clock Feynman construction is that the former allows for a partially parallel, and hence faster, execution of gates. \par
Before introducing the new model, we describe the type of quantum circuit that we are going to implement. As in Sec.~\ref{reviewSec} we first focus for simplicity on the case in which only single-qubit gates are present and then extend the analysis in Section \ref{sec:mqubit} to circuits in which controlled two- and three-qubit gates need to be executed. 

Consider a quantum circuit with single-qubit gates shown in Fig.~\ref{snakeQuantumCircuit}. In practice, odd columns in the circuit are executed from top to bottom, while even columns are executed from bottom to top. The motion of the cursor reminds one of a snake that runs over the quantum circuit: this construction has previously been used in e.g. \cite{oliveiraTerhal, nagaj2012}. The reason why we restrict the analysis to this kind of ordering in the execution of the gates is to keep the interactions local on a 2D lattice. 
%\onecolumngrid 
\begin{figure*}
\centering
\begin{circuitikz}[scale=0.25]
\pic [scale=0.25] at (0,0) {latticeFeynman};
\pic [scale=0.25] at (9,0) {latticeFeynman};
\pic [scale=0.25] at (18,0) {latticeFeynman};
\pic [scale=0.25] at (27,0) {latticeFeynman};
\pic [scale=0.25] at (36,0) {latticeFeynman};
\pic [scale=0.25] at (45,0) {latticeFeynman};
\pic [scale=0.25] at (54,0) {latticeFeynman};
%1st block
\foreach \y in {0, ..., 3}
\filldraw [color=red!70!black]   (0, 0-\y*2) circle (0.3cm);
\draw [ultra thick, color=red!70!black] (0,0)--(0,-6);
%2nd block
\foreach \y in {1, ..., 3}
\filldraw [color=red!70!black]   (9, 0-\y*2) circle (0.3cm);
\filldraw [color=red!70!black]   (9+2, 0) circle (0.3cm);
\draw [ultra thick, color=red!70!black] (9,-2)--(9,-6);
\draw [ultra thick, color=red!70!black] (9+2,0)--(9,-2);
%3rd block
\foreach \y in {0,1}
\filldraw [color=red!70!black]   (18+2, 0-\y*2) circle (0.3cm);
\foreach \y in {2, 3}
\filldraw [color=red!70!black]   (18, 0-\y*2) circle (0.3cm);
\draw [ultra thick, color=red!70!black] (18+2,0)--(18+2,-2)--(18,-4)--(18,-6);
%4th block
\foreach \y in {0,1,2}
\filldraw [color=red!70!black]   (27+2, 0-\y*2) circle (0.3cm);
\filldraw [color=red!70!black]   (27, -6) circle (0.3cm);
\draw [ultra thick, color=red!70!black] (27+2,0)--(27+2,-4)--(27,-6);
%5th block
\foreach \y in {0,1,2,3}
\filldraw [color=red!70!black]   (36+2, 0-\y*2) circle (0.3cm);
\draw [ultra thick, color=red!70!black] (36+2,0)--(36+2,-6);
%6th block
\foreach \y in {0,1,2}
\filldraw [color=red!70!black]   (45+2, 0-\y*2) circle (0.3cm);
\filldraw [color=red!70!black]   (45+4,-6) circle (0.3cm);
\draw [ultra thick, color=red!70!black] (45+2,0)--(45+2,-4)--(45+4,-6);
%7th block
\foreach \y in {0,1}
\filldraw [color=red!70!black]   (54+2, 0-\y*2) circle (0.3cm);
\foreach \y in {0,1}
\filldraw [color=red!70!black]   (54+4,-6+2*\y) circle (0.3cm);
\draw [ultra thick, color=red!70!black] (54+2,0)--(54+2,-2)--(54+4,-4)--(54+4,-6);
\node at (64,-3) {$\dots$};
\foreach \x in {0,..., 6}
\node at (3+\x*9,-8) {$\ket{\x}$};
\foreach \x in {0,...,6}
\draw [thick, color=black, <->] (6.5+\x*9, -3) to [bend left] (8.5+\x*9,-3);
\end{circuitikz}
\caption{Motion of the string of particles which keeps the string connected and executes the gates in the circuit in the correct order.}
\label{stringMotionFig}
\end{figure*}
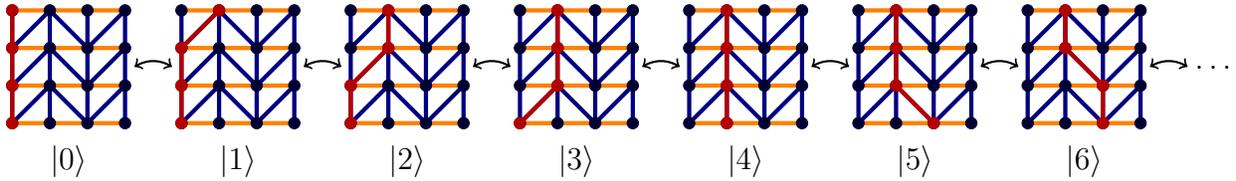

We consider the lattice in Fig. \ref{newLatticeFig}, consisting of a grid of sites $(i,j)$ with $i=1,\ldots N,j=1, \ldots M$, where particles can reside. Like for the model described in Sec. \ref{reviewSec}, we assume that there is one particle per horizontal line or track. The model encodes a quantum circuit with $N$ qubits, where each qubit undergoes $M-1$ unitary gates. 
The blue edges on the lattice represent \emph{attractive} interactions whose purpose is to keep the string of particles together. Note that, as in the Lloyd-Terhal model, each site participates in 4 attractive edges. The orange edges are associated with hopping terms which apply gates to the qubit that lives on each track as in Eq.~(\ref{VhopPlaquette}) in the main text. We denote the set of blue edges as $E_{b}$ and the set of orange edges as $E_{y}$. A generic unitary of our quantum circuit applied at an orange edge $e$ is denoted as $U_{ij}$ as shown in Fig.~\ref{snakeQuantumCircuit}. 

%BMT2 AC i changed yellow to orange, to me the edges look orange
%AC yes they are orange indeed. Yellow was a leftover of the previous versions.

Annihilation, creation and number operators are used in the same way as in Sec. \ref{reviewSec}.  We define the Hamiltonian $H=H_{\rm valid}+V_{\rm hop}$ as in Eq.~(\ref{basicH}) with 
\begin{equation*}
H_{\rm valid}=-\Delta \sum_{(\mu, \nu) \in E_b} \bm{n}[\mu]\bm{n}[\nu].
\end{equation*}
The groundspace of this Hamiltonian has eigenvalue $E_0=-\Delta (N-1)$ and consists of all possible strings of particles connected by the blue edges. The number of connected strings in this subspace is $L\equiv N \times (M-1)+1$. 
The first excited space is the space of strings that fail to be connected by blue edges at one position, having energy $E_0+\Delta$. 

The hopping Hamiltonian equals
\begin{equation*}
V_{\rm hop}=  -J\sum_{e\in E_y} V_{{\rm hop},e},
\end{equation*}
with the term for edge $e=(\mu, \nu)$ between sites $\mu=(i,j+1)$ and $\nu=(i,j)$ equal to 
\begin{equation*}
V_{{\rm hop}, e}= \sum_{s=0}^1 \sum_{s \sp{\prime}=0}^1 \bra{s \sp{\prime}} U_{ij} \ket{s} a_{s \sp{\prime}}^{\dagger} [i, j+1]a_{s}[i, j]+\mathrm{h.c.}
\end{equation*}
\par 

The effective conditional-hopping Hamiltonian $H_{\rm eff}$ induces a quantum walk on a line with $L$ sites, where the quantum computation follows the order depicted in Fig.~\ref{snakeQuantumCircuit}. An example of this effective motion is depicted in Fig.~\ref{stringMotionFig}. The system is initialized in a configuration in which all particles are on the left, denoted as $\ket{0}$. From this configuration there is only one \emph{move} that keeps the string connected which brings the string to configuration $\ket{1}$. From here there are two moves that keep the string connected. Either the string hops back to $\ket{0}$ or it hops forward to $\ket{2}$. The conditional hopping proceeds in this way until we reach the bottom, configuration $\ket{4}$, and then we can move backward or forward until we reach the top again etc. We understand that this is exactly the order in which the gates should be implemented \par 

Mathematically, the effective Hamiltonian $H_{\rm eff}$ is given by the projection of $H$ onto the groundspace of connected strings. Each connected string can be labeled as $\ket{k} \in \mathcal{H}_C$ with $k \in \{0,1,2, \dots, L-1\}$ as in Fig.~\ref{stringMotionFig} in tensor product with the internal state space of the particles. We can thus write 
\begin{equation*}
H_{\rm eff}=H_{F}= -J \sum_{k=0}^{L-1}  U_{k} \otimes \ket{k+1} \bra{k} +\mathrm{h.c.},
\end{equation*} 
where $U_{k}$ is the $k$-th single-qubit gate in our snake-like ordering. Properties of this effective Feynman Hamiltonian $H_F$ are well-known: it performs a unitary evolution in a computational space spanned by the orthogonal vectors
\begin{equation*}
\ket{\Psi_k}= (U_{k}U_{k-1} \dots U_1 \ket{\psi_0}) \otimes \ket{k}= \ket{\psi_k} \otimes \ket{k}, 
\end{equation*}
where $\ket{\psi_0}$ is some arbitary initial state. In the clock subspace $\mathcal{H}_C$ it acts as
\begin{equation*}
H_{F} \lvert_{\mathcal{H}_C}=-J \sum_{k=0}^{L-1} \ket{\Psi_{k+1}} \bra{\Psi_{k}} + \mathrm{h.c.},
\end{equation*}
which up to a trivial relabelling $\ket{\Psi_k} \equiv \ket{k+1}$ is just the Hamiltonian of a continuous-time quantum walk on a line with $L$ sites, i.e. 
\begin{equation}
\label{HQW}
H_{F} \lvert_{\mathcal{H}_C} = H_{L}=-J \sum_{k=1}^{L} \ket{k+1}\bra{k}+\mathrm{h.c.}
\end{equation}
Diagonalizing this Hamiltonian $H_L$ given eigenstates \cite{christandl2004, nagaj2010} 
\begin{equation*}
\ket{\tilde{l}}= \sqrt{\frac{2}{L+1}} \sum_{k=1}^{L} \sin \biggl(\frac{\pi l k}{L+1} \biggr)\ket{k}
\end{equation*}
and corresponding eigenvalues
\begin{equation*}
\label{eigvQW}
\tilde{\omega}_l= -2 J \cos \biggl(\frac{\pi l}{L+1}\biggr),
\end{equation*}
for $l=0,\dots, L-1$.
 \begin{figure}[htb]
\centering
\includegraphics[scale=0.45]{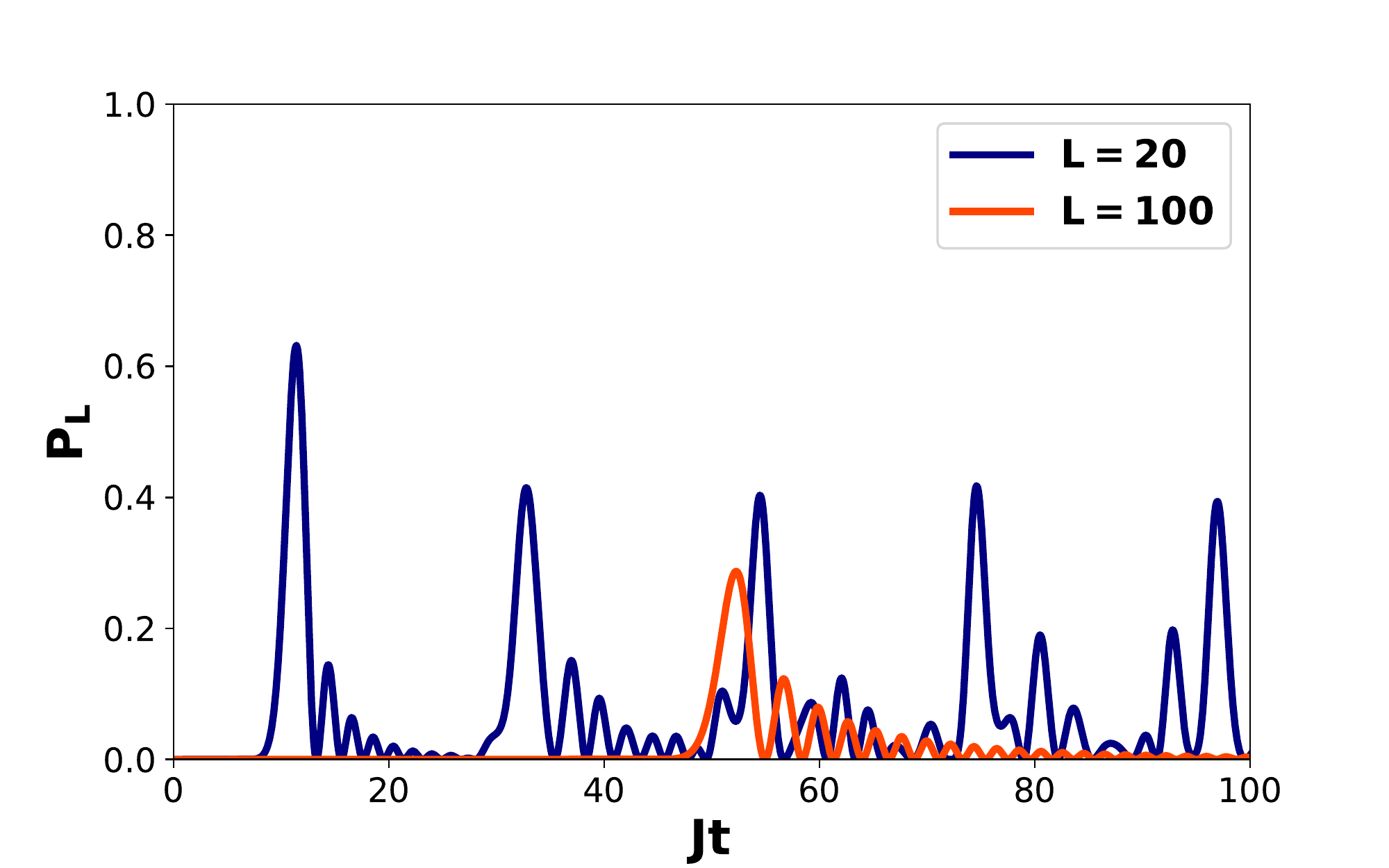}
\caption{Success probability $P_L(t)$ for $L=20$ and $L=100$ as a function of $Jt$.}
\label{successP}
\end{figure}

% BMT AC could you add two labels in Fast CNOT in correspondence of the Eq.() below, namely add (i,j) to top site on the right and (i+2,j-1) to bottom site, the rest can be inferred
% AC: done

The probability that starting from the initial computational state $\ket{1}$ the system is found in the final computational state $\ket{L}$ at some fixed time $t$ equals
\begin{equation*}
P_{L}(t)= \lvert \bra{L} e^{-i H_{L} t} \ket{1} \rvert^2.
\end{equation*} 
with
\begin{equation*}
\bra{L} e^{-i H_{L}t} \ket{1}= \sum_{l=1}^{L} \braket{L | \tilde{l}} \bra{\tilde{l}} e^{-i H_{L}t} \ket{1} = 
\frac{2}{L+1} \sum_{l=1}^L \sin \biggl(\frac{\pi l}{L+1} \biggr)\sin \biggl(\frac{\pi  L l}{L+1} \biggr) e^{-i \tilde{\omega}_l t}.
\end{equation*}

As a concrete example, we can plot in Fig. \ref{successP} the success probability for the case $L=20$ and $L=100$. As expected, we see how the probability decreases as a function of $L$. For instance, the maximum probability with $L=1000$ is approximately $0.1$.

\subsection{Peres' trick}
\label{sec:peres}

One can use a trick to reach the final stage of computation exactly. 
In 1985 Peres formulated a modification of the Feynman Hamiltonian which ensures that the computation is executed in a specific time \cite{peres1985}. This trick was rediscovered in \cite{christandl2004}, in the context of perfect quantum state transfer. It has also been argued that this way of achieving perfect state transfer is optimal from several points of view \cite{kayStateTransfer}.  

The idea is to choose different hopping strengths $J_{k}$ so that the Hamiltonian is proportional to the angular momentum operator $\mathcal{J}_{x}$ of a fictitious spin-$(L-1)/2$ particle. The basis states on the line $\ket{k}$ will be interpreted as the eigenkets of the operator $\mathcal{J}^2$ and $\mathcal{J}_{z}$. To this end, we relabel the basis states (again) as
\begin{equation*}
\ket{k} \rightarrow \ket{m=-(L-1)/2+k-1}, \quad k \in \{1, \dots, L\}.
\end{equation*}
which are eigenstates of $\mathcal{J}_z$ with eigenvalue $m$ ranging from $-\frac{L-1}{2}$ to $\frac{L-1}{2}$.
The Peres-Feynman Hamiltonian in this basis is then chosen as 
\begin{equation*}
H_{L,PF} = -\sum_{m=-(L-1)/2}^{(L-1)/2-1} J_{m} \ket{m+1}\bra{m}+\mathrm{h.c.} \equiv -J \mathcal{J}_x
\end{equation*}
with the choice
\begin{equation*}
J_m= \frac{J}{2} \sqrt{m(L-m)}
\end{equation*}
% BMT AC i changed some text to shorten so I just wanted to give J_m directly, could you fill this in?
% BMT still to do 
% AC: done!
Using this Hamiltonian it can be shown that at time $t=T=\pi/J$ we have the desired condition $\ket{\psi(T)}=\ket{L}$ with probability $1$.

Note that the coupling parameters $J_m$ scale with the number of gates. For this reason we cannot envision using this trick for a quantum computation of very long length.

\subsection{Multi-Qubit Logic}
\label{sec:mqubit}
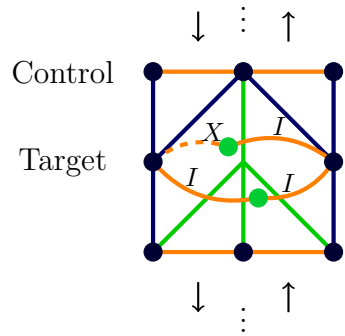
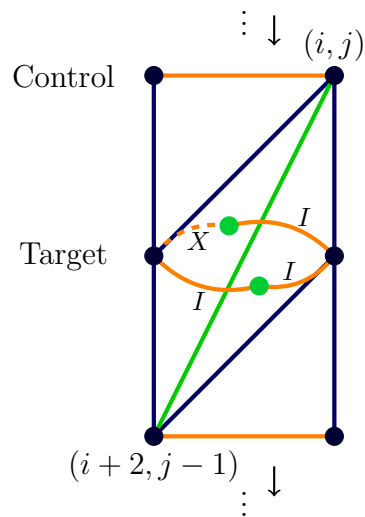
\begin{figure}[htb]
\centering
\begin{subfigure}[t]{0.51 \textwidth}
\centering
\begin{tikzpicture}[scale=0.4]
\draw[ultra thick, color=orange] (0,0)--(6,0);
\draw[ultra thick, color=orange] (0,-6)--(6,-6);
\draw[ultra thick, color=black!60!blue] (0,-6)--(0,0);
\draw[ultra thick, color=black!60!blue] (6,-6)--(6,0);
\draw[ultra thick, color=black!60!blue] (0,-3)--(3,0)--(6, -3);
\draw [ultra thick, color=green!80!black] (3,0)--(3,-3);
\draw [ultra thick, color=green!80!black] (3,-3)--(0,-6);
\draw [ultra thick, color=green!80!black] (3,-3)--(3,-6);
\draw [ultra thick, color=green!80!black] (3,-3)--(6,-6);
\draw [ultra thick, color=orange, dashed] (0,-3) to [bend left] (2.5,-2.5);
\draw [ultra thick, color=orange] (0,-3) to [bend right] (3.5, -4.2);
\draw [ultra thick, color=orange] (2.5,-2.5) to [bend left] (6,-3);
\draw [ultra thick, color=orange] (3.5, -4.2) to [bend right] (6,-3);
\foreach \x in {0,1,2}
\filldraw  [color=black!80!blue] (0+\x*3,0) circle (0.3cm);
\filldraw  [color=black!80!blue] (0,-3) circle (0.3cm);
\filldraw  [color=black!80!blue] (6,-3) circle (0.3cm);
\foreach \x in {0,1,2}
\filldraw  [color=black!80!blue] (0+\x*3,-6) circle (0.3cm);
\filldraw [color=green!80!blue]   (2.5, -2.5) circle (0.3cm);
\filldraw [color=green!80!blue]   (3.5, -4.2) circle (0.3cm);
\node at (-3,0) {Control};
\node at (-3,-3) {Target};
\node at (3, 2) {$\vdots$};
\node at (3, -8) {$\vdots$};
\node at (2, -2) {\footnotesize{$X$}};
\node at (1.3, -3.5) {\footnotesize{$I$}};
\node at (4.5, -3.7) {\footnotesize{$I$}};
\node at (4.2, -1.8) {\footnotesize{$I$}};
\draw [thick, ->] (1.5,2)--(1.5,1);
\draw [thick, ->] (1.5,-7)--(1.5,-8);
\draw [thick, <-] (4.5,2)--(4.5,1);
\draw [thick, <-] (4.5,-7)--(4.5,-8);
\end{tikzpicture}
\subcaption{Slow CNOT.}
\label{cnotNewa}
\end{subfigure}
\begin{subfigure}[t]{0.51 \textwidth}
\centering
\begin{tikzpicture}[scale=0.4]
\draw[ultra thick, color=orange] (0,0)--(6,0);
\draw[ultra thick, color=orange] (0,-12)--(6,-12);
\draw[ultra thick, color=black!60!blue] (0,-6)--(6,0);
\draw[ultra thick, color=black!60!blue] (6,-6)--(0,-12);
\draw[ultra thick, color=black!60!blue] (0,0)--(0,-12);
\draw[ultra thick, color=black!60!blue] (6,0)--(6,-12);
\draw [ultra thick, color=green!80!black] (6,0)--(0,-12);
\draw [ultra thick, color=orange, dashed] (0,-6) to [bend left] (2.5,-5);
\draw [ultra thick, color=orange] (0,-6) to [bend right] (3.5,-7);
\draw [ultra thick, color=orange] (3.5,-7) to [bend right] (6,-6);
\draw [ultra thick, color=orange] (2.5,-5) to [bend left] (6,-6);
\filldraw [color=green!80!blue]   (2.5, -5) circle (0.3cm);
\filldraw [color=green!80!blue]   (3.5, -7) circle (0.3cm);
\foreach \x in {0,1}
\foreach \y in {0,1,2}
\filldraw  [color=black!80!blue] (0+\x*6,0-\y*6) circle (0.3cm);
\node at (-3,0) {Control};
\node at (-3,-6) {Target};
\node at (1.5, -5.5) {\footnotesize{$X$}};
\node at (1.5, -7.5) {\footnotesize{$I$}};
\node at (5, -4.7) {\footnotesize{$I$}};
\node at (4.5, -6.5) {\footnotesize{$I$}};
\node at (3, 2) {$\vdots$};
\node at (6,1) {$(i,j)$};
\node at (0,-13) {$(i+2,j-1)$};
\node at (3, -14) {$\vdots$};
\draw [thick, ->] (4,-13)--(4,-14);
\draw [thick, <-] (4,1)--(4,2);
\end{tikzpicture}
\subcaption{Fast CNOT}
\label{cnotNewb}
\end{subfigure}
\caption{CNOT in the `single-clock snake' lattice model of Fig.~\ref{newLatticeFig}. The green edges are those who need to be modified similar to what we do in Subsec. \ref{subsecCNOT}.}
\label{cnotNew}
\end{figure}
The construction of the CNOT and Toffoli gates is similar to the construction discussed in Sec. \ref{sec:mql}. We will show that there are two ways of approaching the problem to realize a CNOT gate depicted in Fig. \ref{cnotNew} .

In Fig.~\ref{cnotNewa} we represent a slow CNOT. In this case we introduce a split-site in the lattice and we need to modify the green edges in the Hamiltonian (as compared to the standard case) and require the hopping terms to the split-sites to enact a possible $X$-gate. This is essentially the same gadget as the CNOT gadget in the Lloyd-Terhal scheme, described in Section \ref{subsecCNOT}. In this construction, the CNOT \emph{begins} in one column, but it is only terminated in the next one. In the meantime all the gates below the target in the first column and in the second column need to be executed before the CNOT can be finally terminated. This means that the control and the target particles are \emph{busy} for two rounds of the computation and cannot be involved in other gates. 

To avoid this problem, we might consider implementing the CNOT as in Fig.~\ref{cnotNewb}, which we call a fast CNOT since it can be accomplished in a single round. The idea is to add a new split-site for the CNOT in the middle of an orange hopping edge. The horizontal site coordinate $j$ can be given half-integer values at such a split-site, i.e. one uses annihilation operators $a_s[i,j+1/2,k]$ with $k=0,1$ labeling the two sites.
 We add new attractive green edges to this split-site which selectively depend on the state of the control qubit, e.g. the green edges in Fig.~\ref{cnotNewb} correspond to
\begin{equation*}
-\Delta \sum_{s=0,1} n_{s}[i,j] {\bf n}[i+1,j-1/2,s]
-\Delta \,{\bf n}[i+2,j-1] {\bf n}[i+1,j-1/2].
\end{equation*}
We can check that the CNOT is fully executed before the computation continues to run downwards as in Fig.~\ref{cnotNewb}. Note that the control particle can be either be above or below the target particle. This fast construction is advantageous since each CNOT will just occupy one site in the quantum walk on the line, while this conclusion cannot be drawn for the slow CNOT, as its effect in terms of overhead is algorithm-dependent. The disadvantage of the fast CNOT is that it requires a region which is more dense with interactions. An analogous construction can be imagined for the Toffoli gate. In the fast version each Toffoli adds two sites to the quantum walk. Even though the green split-sites only participate in two attractive edges each, the fast CNOT requires two normal sites which participate in 5 attractive edges in total which adds complexity.

\begin{figure}[htb]
\centering
\vspace{0.2 cm}
\includegraphics[scale=0.23]{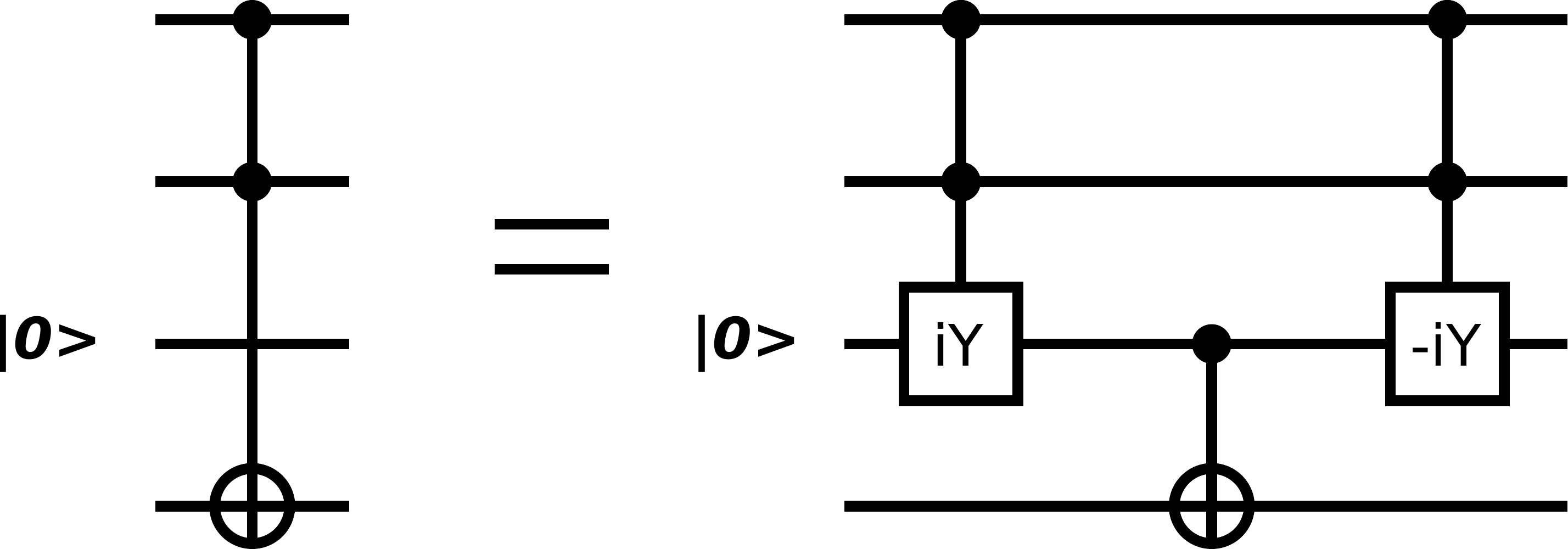}
\caption{Implementation of a Toffoli gate using an ancilla qubit, a CNOT and controlled-controlled-$iY$ and controlled-controlled-$(iY)^{\dagger}$ gates.}
\label{toffoliCminus}
\end{figure}
\begin{figure}[htb]
\centering
\includegraphics[scale=0.23]{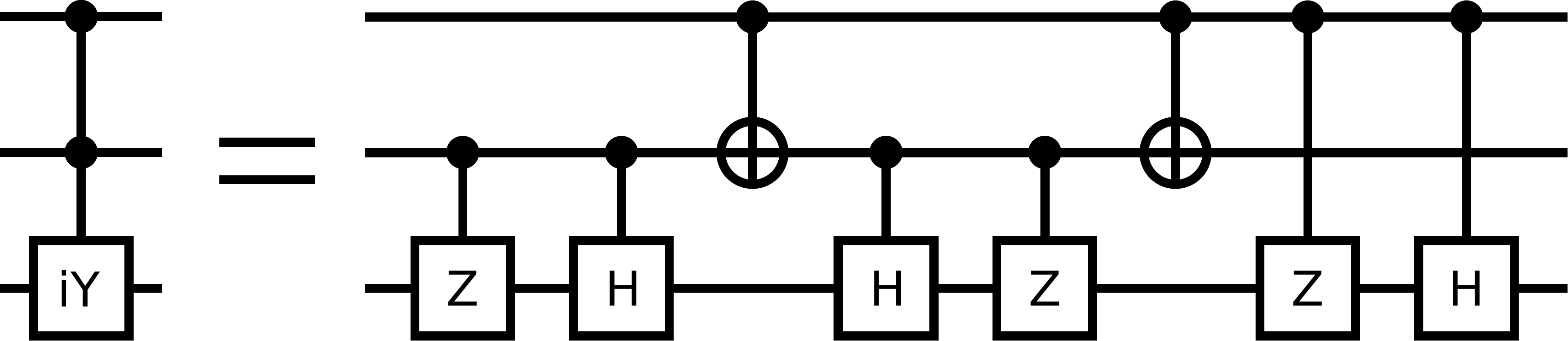}
\caption{Implementation of a controlled-controlled-$iY$ gate using CNOT, controlled-Hadamard and controlled-Z gates. The implementation of controlled-controlled-$(iY)^{\dagger}$ follows analogously.}
\label{cMinus}
\end{figure}
\section{Universality of Hadamard, CNOT and controlled-Hadamard}
\label{secControlledHadamard}

In this Appendix we show how the Toffoli gate can be constructed using Hadamard, CNOT and controlled-Hadamard gates with the help of an ancilla qubit initialized in $\ket{0}$. The key to this construction is shown in Fig. \ref{toffoliCminus}, where the Toffoli gate is obtained using a CNOT gate, a controlled-controlled-$iY$ and its Hermitian conjugate.

Now we show that the controlled-controlled-$iY$, and its Hermitian conjugate can be obtained from Hadamard, CNOT and controlled-Hadamard. To this end we follow the general construction of a controlled-controlled-$U$ described in Sec. 4.3 of Ref. \cite{nielsenChuang}. In order to use this method to construct the controlled-controlled-$iY$ operator, we have to find a unitary operator $V$ such that $V^2=iY$, which turns out to be the real-valued gate $V= Z H$.
\begin{figure}[htb]
\centering
\includegraphics[scale=0.23]{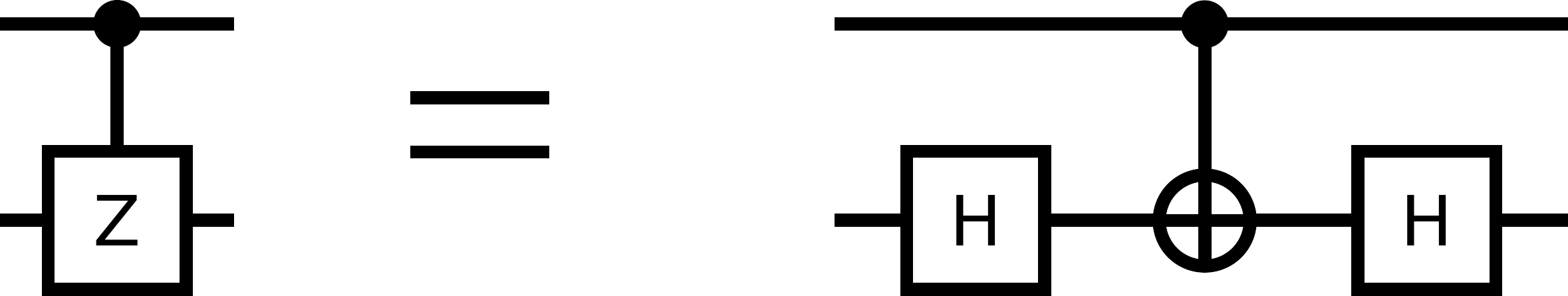}
\caption{Controlled-$Z$ gate from Hadamard and CNOT.}
\label{CZcnot}
\end{figure}
 As shown in Fig. \ref{cMinus} the controlled-controlled-$iY$ is thus obtained in terms of CNOT, controlled-Hadamard and controlled-Z gate (CZ). Finally, using the well-known relation between the CZ gate and the CNOT gate shown  in Fig. \ref{CZcnot}, we complete the synthesis of a Toffoli gate using Hadamard, CNOT and controlled-Hadamard, showing the universality of this set of quantum gates.

%BMT2 AC last Fig sits in the references...
% AC: it should be ok now

\clearpage
\bibliographystyle{ieeetr}
\bibliography{hamiltonianComputing.bib}

\begin{thebibliography}{10}

\bibitem{feynman1986}
R.~P. Feynman, ``Quantum mechanical computers,'' {\em Foundations of Physics},
  vol.~16, p.~507, 1986.

\bibitem{kitaevBook}
A.~Y. Kitaev, A.~H. Shen, and M.~N. Vyalyi, {\em Classical and Quantum
  Computation}.
\newblock Boston, MA, USA: American Mathematical Society, 2002.

\bibitem{ahronovAdiabatic2007}
D.~Aharonov, W.~van Dam, J.~Kempe, Z.~Landau, S.~Lloyd, and O.~Regev,
  ``Adiabatic quantum computation is equivalent to standard quantum
  computation,'' {\em SIAM J. Comput.}, vol.~37, pp.~166--194, Apr. 2007.

\bibitem{lidarmMizelProof}
A.~Mizel, D.~A. Lidar, and M.~Mitchell, ``Simple proof of equivalence between
  adiabatic quantum computation and the circuit model,'' {\em Phys. Rev.
  Lett.}, vol.~99, p.~070502, Aug 2007.

\bibitem{Margolus90parallelquantum}
N.~Margolus, ``Parallel quantum computation,'' in {\em Complexity, Entropy, and
  the Physics of Information,SFI Studies in the Sciences of Complexity},
  pp.~273--287, Addison-Wesley, 1990.

\bibitem{janzing2007}
D.~Janzing, ``Spin-1/2 particles moving on a two-dimensional lattice with
  nearest-neighbor interactions can realize an autonomous quantum computer,''
  {\em Phys. Rev. A}, vol.~75, p.~012307, Jan 2007.

\bibitem{mizel2001}
A.~Mizel, M.~W. Mitchell, and M.~L. Cohen, ``Energy barrier to decoherence,''
  {\em Phys. Rev. A}, vol.~63, p.~040302, Mar 2001.

\bibitem{mizel2004}
A.~Mizel, ``Mimicking time evolution within a quantum ground state:
  Ground-state quantum computation, cloning, and teleportation,'' {\em Phys.
  Rev. A}, vol.~70, p.~012304, Jul 2004.

\bibitem{breuckmannTerhal}
N.~P. Breuckmann and B.~M. Terhal, ``Space-time circuit-to-{H}amiltonian
  construction and its applications,'' {\em Journal of Physics A: Mathematical
  and Theoretical}, vol.~47, no.~19, p.~195304, 2014.

\bibitem{gossetTerhal}
D.~Gosset, B.~M. Terhal, and A.~Vershynina, ``Universal adiabatic quantum
  computation via the space-time circuit-to-{H}amiltonian construction,'' {\em
  Phys. Rev. Lett.}, vol.~114, p.~140501, Apr 2015.

\bibitem{terhalLloyd}
S.~Lloyd and B.~M. Terhal, ``Adiabatic and {H}amiltonian computing on a {2D}
  lattice with simple two-qubit interactions,'' {\em New Journal of Physics},
  vol.~18, no.~2, p.~023042, 2016.

\bibitem{CGW:walk}
A.~M. Childs, D.~Gosset, and Z.~Webb, ``Universal computation by multiparticle
  quantum walk,'' {\em Science}, vol.~339, no.~6121, pp.~791--794, 2013.

\bibitem{childs+:switch}
A.~Childs, D.~Gosset, D.~Nagaj, M.~Raha, and Z.~Webb, ``Momentum switches,''
  {\em Quant. Inf. Comp.}, vol.~15, pp.~0601--0621, 2015.

\bibitem{englund:walk}
Y.~Lahini, G.~R. Steinbrecher, A.~D. Bookatz, and D.~Englund, ``Quantum logic
  using correlated one-dimensional quantum walks,'' {\em npj Quantum
  Information}, vol.~4, no.~1, p.~2, 2018.

\bibitem{georgescuNori}
I.~Georgescu, S.~Ashhab, and F.~Nori, ``Quantum simulation,'' {\em Rev. Mod.
  Phys.}, vol.~86, pp.~152--185, Mar 2013.

\bibitem{koch2007}
J.~Koch, T.~M. Yu, J.~Gambetta, A.~A. Houck, D.~I. Schuster, J.~Majer,
  A.~Blais, M.~H. Devoret, S.~M. Girvin, and R.~J. Schoelkopf,
  ``Charge-insensitive qubit design derived from the {C}ooper pair box,'' {\em
  Phys. Rev. A}, vol.~76, p.~042319, Oct 2007.

\bibitem{richerPop}
S.~Richer, N.~Maleeva, S.~T. Skacel, I.~M. Pop, and D.~DiVincenzo,
  ``Inductively shunted transmon qubit with tunable transverse and longitudinal
  coupling,'' {\em Phys. Rev. B}, vol.~96, p.~174520, Nov 2017.

\bibitem{snail}
N.~E. Frattini, U.~Vool, S.~Shankar, A.~Narla, K.~M. Sliwa, and M.~H. Devoret,
  ``3-wave mixing {J}osephson dipole element,'' {\em Applied Physics Letters},
  vol.~110, no.~22, p.~222603, 2017.

\bibitem{Kounalakis2018}
M.~Kounalakis, C.~Dickel, A.~Bruno, N.~K. Langford, and G.~A. Steele,
  ``Tuneable hopping and nonlinear cross-{K}err interactions in a
  high-coherence superconducting circuit,'' {\em npj Quantum Information},
  vol.~4, 2018.

\bibitem{wallraff:observation}
M.~C. {Collodo}, A.~{Poto{\v c}nik}, S.~{Gasparinetti}, J.-C. {Besse},
  M.~{Pechal}, M.~{Sameti}, M.~J. {Hartmann}, A.~{Wallraff}, and C.~{Eichler},
  ``{Observation of the Crossover from Photon Ordering to Delocalization in
  Tunably Coupled Resonators},'' {\em ArXiv e-prints}, Aug. 2018.

\bibitem{nigg:bb}
S.~E. {Nigg}, H.~{Paik}, B.~{Vlastakis}, G.~{Kirchmair}, S.~{Shankar},
  L.~{Frunzio}, M.~H. {Devoret}, R.~J. {Schoelkopf}, and S.~M. {Girvin},
  ``{Black-Box Superconducting Circuit Quantization},'' {\em Physical Review
  Letters}, vol.~108, p.~240502, June 2012.

\bibitem{DOR:quantum-crys}
R.~{Dijkgraaf}, D.~{Orlando}, and S.~{Reffert}, ``{Quantum crystals and spin
  chains},'' {\em Nuclear Physics B}, vol.~811, pp.~463--490, Apr. 2009.

\bibitem{nielsenChuang}
M.~A. Nielsen and I.~L. Chuang, {\em Quantum Information and Quantum
  Computation}.
\newblock Cambridge, England: Cambridge University Press, 2000.

\bibitem{ShiToffoli}
Y.~Shi, ``Both {T}offoli and controlled-{NOT} need little help to do universal
  quantum computing,'' {\em Quantum Info. Comput.}, vol.~3, pp.~84--92, Jan.
  2003.

\bibitem{banchiToffoli}
L.~Banchi, N.~Pancotti, and S.~Bose, ``Quantum gate learning in qubit networks:
  {T}offoli gate without time-dependent control,'' {\em Npj Quantum
  Information}, vol.~2, p.~16019, 2016.

\bibitem{nagaj2010}
D.~Nagaj, ``Fast universal quantum computation with railroad-switch local
  {H}amiltonians,'' {\em Journal of Mathematical Physics}, vol.~51, no.~6,
  p.~062201, 2010.

\bibitem{zinnerSpinNature}
O.~V. Marchukov, A.~G. Volosniev, D.~Valiente, M.~Petrosyan, and N.~T. Zinner,
  ``Quantum spin transistor with a {H}eisenberg spin chain,'' {\em Nature
  Communications}, vol.~7, p.~13070, 2016.

\bibitem{zinnerSpinTrans}
N.~J.~S. {Loft}, L.~B. {Kristensen}, C.~K. {Andersen}, and N.~T. {Zinner},
  ``{Quantum spin transistors in superconducting circuits},'' {\em ArXiv
  e-prints}, Feb. 2018.

\bibitem{braumuller2016}
J.~Braumüller, M.~Sandberg, M.~R. Vissers, A.~Schneider, S.~Schlör,
  L.~Grünhaupt, H.~Rotzinger, M.~Marthaler, A.~Lukashenko, A.~Dieter, A.~V.
  Ustinov, M.~Weides, and D.~P. Pappas, ``Concentric transmon qubit featuring
  fast tunability and an anisotropic magnetic dipole moment,'' {\em Applied
  Physics Letters}, vol.~108, no.~3, p.~032601, 2016.

\bibitem{reiner2016}
J.-M. Reiner, M.~Marthaler, J.~Braum\"uller, M.~Weides, and G.~Sch\"on,
  ``Emulating the one-dimensional {F}ermi-{H}ubbard model by a double chain of
  qubits,'' {\em Phys. Rev. A}, vol.~94, p.~032338, Sep 2016.

\bibitem{neumeierLeib}
L.~Neumeier, M.~Leib, and M.~J. Hartmann, ``Single-photon transistor in circuit
  quantum electrodynamics,'' {\em Phys. Rev. Lett.}, vol.~111, p.~063601, Aug
  2013.

\bibitem{jinRossini}
J.~Jin, D.~Rossini, R.~Fazio, M.~Leib, and M.~J. Hartmann, ``Photon solid
  phases in driven arrays of nonlinearly coupled cavities,'' {\em Phys. Rev.
  Lett.}, vol.~110, p.~163605, Apr 2013.

\bibitem{MARCOS2014634}
D.~Marcos, P.~Widmer, E.~Rico, M.~Hafezi, P.~Rabl, U.-J. Wiese, and P.~Zoller,
  ``Two-dimensional lattice gauge theories with superconducting quantum
  circuits,'' {\em Annals of Physics}, vol.~351, pp.~634 -- 654, 2014.

\bibitem{leibZoller}
M.~Leib, P.~Zoller, and W.~Lechner, ``A transmon quantum annealer: decomposing
  many-body ising constraints into pair interactions,'' {\em Quantum Science
  and Technology}, vol.~1, no.~1, p.~015008, 2016.

\bibitem{pozar}
D.~M. Pozar, {\em {Microwave engineering; 3rd ed.}}
\newblock Hoboken, NJ: Wiley, 2005.

\bibitem{ManucharyanPhd}
V.~E. Manucharyan, {\em Superinductance}.
\newblock PhD thesis, Yale University, 2012.

\bibitem{manucharyanPhaseSlip}
V.~E. Manucharyan, N.~A. Masluk, A.~Kamal, J.~Koch, L.~I. Glazman, and M.~H.
  Devoret, ``Evidence for coherent quantum phase slips across a {J}osephson
  junction array,'' {\em Phys. Rev. B}, vol.~85, p.~024521, Jan 2012.

\bibitem{rastelliPop}
G.~Rastelli, I.~M. Pop, and F.~W.~J. Hekking, ``Quantum phase slips in
  {J}osephson junction rings,'' {\em Phys. Rev. B}, vol.~87, p.~174513, May
  2013.

\bibitem{nazarov_blanter_2009}
Y.~V. Nazarov and Y.~M. Blanter, {\em Quantum Transport: Introduction to
  Nanoscience}.
\newblock Cambridge University Press, 2009.

\bibitem{matveev2002}
K.~A. Matveev, A.~I. Larkin, and L.~I. Glazman, ``Persistent current in
  superconducting nanorings,'' {\em Phys. Rev. Lett.}, vol.~89, p.~096802, Aug
  2002.

\bibitem{Manucharyan113}
V.~E. Manucharyan, J.~Koch, L.~I. Glazman, and M.~H. Devoret, ``Fluxonium:
  Single {C}ooper-pair circuit free of charge offsets,'' {\em Science},
  vol.~326, no.~5949, pp.~113--116, 2009.

\bibitem{voolSnail}
U.~Vool, A.~Kou, W.~C. Smith, N.~E. Frattini, K.~Serniak, P.~Reinhold, I.~M.
  Pop, S.~Shankar, L.~Frunzio, S.~M. Girvin, and M.~H. Devoret, ``Driving
  forbidden transitions in the fluxonium artificial atom,'' {\em Phys. Rev.
  Applied}, vol.~9, p.~054046, May 2018.

\bibitem{blais2004}
A.~Blais, R.-S. Huang, A.~Wallraff, S.~M. Girvin, and R.~J. Schoelkopf,
  ``Cavity quantum electrodynamics for superconducting electrical circuits: An
  architecture for quantum computation,'' {\em Phys. Rev. A}, vol.~69,
  p.~062320, Jun 2004.

\bibitem{brito2015}
F.~Brito, F.~Rouxinol, M.~D. LaHaye, and A.~O. Caldeira, ``Testing time
  reversal symmetry in artificial atoms,'' {\em New Journal of Physics},
  vol.~17, no.~7, p.~075002, 2015.

\bibitem{kochTimeRev}
J.~Koch, A.~A. Houck, K.~L. Hur, and S.~M. Girvin, ``Time-reversal-symmetry
  breaking in circuit-{QED}-based photon lattices,'' {\em Phys. Rev. A},
  vol.~82, p.~043811, Oct 2010.

\bibitem{vogtCirculator}
C.~M\"uller, S.~Guan, N.~Vogt, J.~H. Cole, and T.~M. Stace, ``Passive on-chip
  superconducting circulator using a ring of tunnel junctions,'' {\em Phys.
  Rev. Lett.}, vol.~120, p.~213602, May 2018.

\bibitem{CLN:clock}
L.~Caha, Z.~Landau, and D.~Nagaj, ``Clocks in {F}eynman's computer and
  {K}itaev's local hamiltonian: Bias, gaps, idling, and pulse tuning,'' {\em
  Phys. Rev. A}, vol.~97, p.~062306, Jun 2018.

\bibitem{mariantoni2016}
J.~H. B\'ejanin, T.~G. McConkey, J.~R. Rinehart, C.~T. Earnest, C.~R.~H. McRae,
  D.~Shiri, J.~D. Bateman, Y.~Rohanizadegan, B.~Penava, P.~Breul, S.~Royak,
  M.~Zapatka, A.~G. Fowler, and M.~Mariantoni, ``Three-dimensional wiring for
  extensible quantum computing: The quantum socket,'' {\em Phys. Rev. Applied},
  vol.~6, p.~044010, Oct 2016.

\bibitem{metz:bsc}
F.~Metz, ``Space-time {C}ircuit-to-{H}amiltonian {C}onstruction applied to a
  {MERA} circuit.''
\newblock Bachelor Thesis, 2015, RWTH Aachen University,
  \url{http://www.quantuminfo.physik.rwth-aachen.de/global/show_document.asp?id=aaaaaaaaaanxrlq}.

\bibitem{peres1985}
A.~Peres, ``Reversible logic and quantum computers,'' {\em Phys. Rev. A},
  vol.~32, pp.~3266--3276, Dec 1985.

\bibitem{christandl2004}
M.~Christandl, N.~Datta, A.~Ekert, and A.~J. Landahl, ``Perfect state transfer
  in quantum spin networks,'' {\em Phys. Rev. Lett.}, vol.~92, p.~187902, May
  2004.

\bibitem{DKS:localization}
F.~Delyon, H.~Kunz, and B.~Souillard, ``One-dimensional wave equations in
  disordered media,'' {\em Journal of Physics A: Mathematical and General},
  vol.~16, no.~1, p.~25, 1983.

\bibitem{bravyiDiVincenzo}
S.~Bravyi, D.~P. DiVincenzo, and D.~Loss, ``Schrieffer-{W}olff transformation
  for quantum many-body systems,'' {\em Annals of Physics}, vol.~326, no.~10,
  pp.~2793 -- 2826, 2011.

\bibitem{bravyiHastings}
S.~Bravyi and M.~Hastings, ``On complexity of the quantum {I}sing model,'' {\em
  Communications in Mathematical Physics}, vol.~349, pp.~1--45, Jan 2017.

\bibitem{oliveiraTerhal}
R.~Oliveira and B.~M. Terhal, ``The complexity of quantum spin systems on a
  two-dimensional square lattice,'' {\em Quantum Info. Comput.}, vol.~8,
  pp.~900--924, Nov. 2008.

\bibitem{nagaj2012}
D.~Nagaj, ``Universal two-body-hamiltonian quantum computing,'' {\em Phys. Rev.
  A}, vol.~85, p.~032330, Mar 2012.

\bibitem{kayStateTransfer}
A.~Kay, ``Perfect, {E}fficient, {S}tate {Transfer} and its application as a
  constructive tool,'' {\em International Journal of Quantum Information},
  vol.~08, no.~04, pp.~641--676, 2010.

\end{thebibliography}
\end{document}